\documentclass[lettersize,journal]{IEEEtran}
\usepackage{amsmath,amsfonts}
\usepackage{algorithmic}
\usepackage{algorithm}
\usepackage{array}
\usepackage[caption=false,font=normalsize,labelfont=sf,textfont=sf]{subfig}
\usepackage{textcomp}
\usepackage{stfloats}
\usepackage{url}
\usepackage{verbatim}
\usepackage{graphicx}
\usepackage[table]{xcolor}
\usepackage{tikz}
\usepackage[edges]{forest}
\usepackage[export]{adjustbox}
\definecolor{myred}{RGB}{217,83,79}
\definecolor{myblue}{RGB}{66,139,202}
\definecolor{myorange}{RGB}{240,173,78}
\definecolor{mygreen}{RGB}{92,184,92}
\definecolor{mygrey}{RGB}{119,119,119}
\definecolor{hidden-draw}{RGB}{20,68,106}
\usepackage{cite}
\usepackage{booktabs}
\usepackage{makecell} 
\usepackage{tabularx}
\usepackage{hyperref}
\newcolumntype{L}[1]{>{\raggedright\arraybackslash}m{#1}}

\begin{document}

\title{Video Generation Models as World Models: Efficient Paradigms, Architectures and Algorithms}




\author{Muyang He, Hanzhong Guo, Junxiong Lin, 
Yizhou Yu,~\IEEEmembership{Fellow,~IEEE}

\thanks{Muyang He and Yizhou Yu are with School of Computing and Data Science, The University of Hong Kong, Hong Kong SAR, China and Hong Kong Generative AI Research and Development Center, Hong Kong SAR, China (E-mail: muyanghe@connect.hku.hk; yizhouy@acm.org).}
\thanks{Hanzhong Guo and Junxiong Lin are with School of Computing and Data Science, The University of Hong Kong, Hong Kong SAR, China (E-mail: hanzhong@connect.hku.hk; junxionglin26@outlook.com).}
\thanks{(M. He, H. Guo and J. Lin contributed equally.)}
\thanks{(Corresponding author: Yizhou Yu.)}

}

\markboth{}%
{ \MakeLowercase{\textit{}}: Video Generation Models as World Models}


\maketitle

\begin{abstract}
The rapid evolution of video generation has enabled models to simulate complex physical dynamics and long-horizon causalities, positioning them as potential world simulators. However, a critical gap still remains between the theoretical capacity for world simulation and the heavy computational costs of spatiotemporal modeling. To address this, we comprehensively and systematically review video generation frameworks and techniques that consider efficiency as a crucial requirement for practical world modeling. We introduce a novel taxonomy in three dimensions: efficient modeling paradigms, efficient network architectures, and efficient inference algorithms. We further show that bridging this efficiency gap directly empowers interactive applications such as autonomous driving, embodied AI, and game simulation. Finally, we identify emerging research frontiers in efficient video-based world modeling, arguing that efficiency is a fundamental prerequisite for evolving video generators into general-purpose, real-time, and robust world simulators. A curated GitHub repository of the reviewed literature can be found at \href{https://github.com/Isaachhh/Efficient-VWM-Survey}{Efficient-VWM-Survey}.
\end{abstract}


\section{Introduction}

In the rapidly evolving landscape of generative artificial intelligence, video generation has received remarkable attention due to its potential to simulate complex world dynamics. This field has undergone a transformative journey, progressing from early generative adversarial networks (GANs)~\cite{goodfellow2014generative, saito2017temporal} and pixel-level auto-regressive (AR) models~\cite{kalchbrenner2017video, yan2021videogpt} to high-fidelity diffusion-based approaches~\cite{ramesh2021zero,rombach2022high,saharia2022photorealistic,ho2022video,song2020score, blattmann2023align,betker2023improving,podellsdxl,esser2024scaling}, and more recently to large-scale architectures that function as ``World Simulators" capable of modeling physical laws and long-horizon causalities~\cite{videoworldsimulators2024, peebles2023scalable}. This progression marks a substantial leap in generative capabilities, enabling models not only to synthesize visual content but to understand and predict the underlying physics of the environment, thereby paving the way for AGI~\cite{ha2018world, yang2024learning}.

To fully appreciate this leap, it is essential to understand video generation has the potential to achieve world modeling. The concept of \textit{world modeling} seeks to move beyond simple pattern matching toward a fundamental understanding of environmental dynamics. A world model is generally defined as an internal representation of environmental dynamics that enables the prediction of future states based on historical contexts and, optionally, actions~\cite{ha2018world}. In the context of visual synthesis, \textit{video-based world models} treat the generative process as a simulation of the physical world, where the objective is to model the underlying causal mechanisms such as gravity, collision, and object permanence rather than just pixel transitions. Mathematically, this can be viewed as learning the transition function $\mathcal{P}(s_{t+1} | s_t, a_t)$, where $s$ represents the state (video frames or latents) and $a$ represents the conditions or actions (e.g., text prompts or camera trajectories). As emphasized in the development of {Sora}~\cite{videoworldsimulators2024}, scaling video generation models leads to the emergence of simulation capabilities, where the model demonstrates an initial comprehension of physical laws without explicit hard-coding. 

This alignment between video generation and world modeling offers several advantages:

\textbf{Emergent Physics:} Large-scale training on diverse video data allows models to learn complex interactions, such as agent-environment interactions or fluid dynamics, which are difficult to model via traditional analytical engines.

\textbf{Latent Imagination:} Modern world models often operate in compact latent spaces~\cite{ha2018world, yang2024learning}, allowing the imagination of future scenarios to occur at a lower computational cost than high-resolution pixel rendering. This inherently links the concept of world modeling to computational efficiency.

\textbf{Unified Reasoning:} By treating video generation as world modeling, the same architecture can be applied to diverse domains ranging from media production to autonomous driving~\cite{hu2023gaia, wang2024drivedreamer} and robotic manipulation~\cite{du2023learning}, where the model acts as a general-purpose simulator for decision-making.

To function as effective world simulators, video generators must ensure long-term consistency, physical accuracy, and high-resolution interactivity~\cite{du2023learning, deepmind_genie3_models}. However, high-dimensional video data and complex dynamics impose severe computational and memory bottlenecks. For example, autoregressive models face key-value (KV) cache explosion during long-sequence generation~\cite{li2025stable, oshima2025worldpack}, and diffusion models suffer from high latency due to iterative denoising. In addition, the vast redundancy in video frames must be reduced without losing semantics~\cite{li2024vidtome, chen2025dc}. To prevent these efficiency limitations from hindering scalability, the development of \textit{efficient architectures and algorithms} is crucial for enabling real-time deployment.

\tikzstyle{my-box}=[
    rectangle,
    draw=hidden-draw,
    rounded corners,
    align=center,
    text opacity=1,
    minimum height=1.2em,
    inner sep=2pt,
    fill opacity=.8,
    line width=0.8pt,
]
\tikzstyle{leaf-head}=[my-box, draw=gray!80, fill=gray!15, text=black, inner xsep=4pt, inner ysep=2pt]
\tikzstyle{leaf-datasets}=[my-box, draw=myred!80, fill=myred!15, text=black, inner xsep=4pt, inner ysep=2pt]
\tikzstyle{leaf-methods}=[my-box, draw=myblue!80, fill=myblue!15, text=black, inner xsep=4pt, inner ysep=2pt]
\tikzstyle{leaf-metrics}=[my-box, draw=myorange!80, fill=myorange!15, text=black, inner xsep=4pt, inner ysep=2pt]
\tikzstyle{leaf-app}=[my-box, draw=mygreen!80, fill=mygreen!15, text=black, inner xsep=4pt, inner ysep=2pt]
\tikzstyle{leaf-chall}=[my-box, draw=mygrey!80, fill=mygrey!15, text=black, inner xsep=4pt, inner ysep=2pt]
\tikzstyle{modelnode-datasets}=[my-box, draw=myred!100, fill=white, text=black, inner xsep=4pt, inner ysep=2pt]
\tikzstyle{modelnode-methods}=[my-box, draw=myblue!100, fill=white, text=black, inner xsep=4pt, inner ysep=2pt]
\tikzstyle{modelnode-metrics}=[my-box, draw=myorange!100, fill=white, text=black, inner xsep=4pt, inner ysep=2pt]
\tikzstyle{modelnode-app}=[my-box, draw=mygreen!100, fill=white, text=black, inner xsep=4pt, inner ysep=2pt]
\tikzstyle{modelnode-chall}=[my-box, draw=mygrey!100, fill=white, text=black, inner xsep=4pt, inner ysep=2pt]

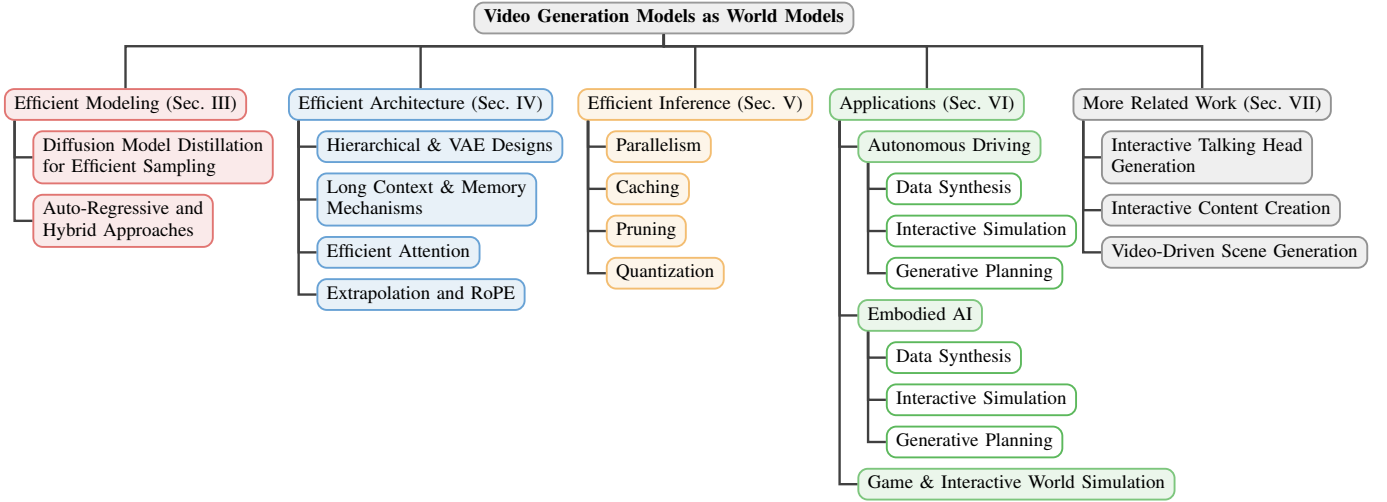
\begin{figure*}[!th]
    \centering
    \begin{adjustbox}{max width=\textwidth, max height=0.9\textheight}
\begin{forest}
forked edges,
for tree={
    font=\footnotesize,
    rectangle,
    draw=hidden-draw,
    rounded corners,
    align=left,
    edge+={darkgray, line width=1pt},
    inner xsep=4pt,
    inner ysep=2pt,
    line width=0.8pt,
},
where level=0{
    grow=south, 
    parent anchor=south, 
    child anchor=north
}{},
where level>=1{
    folder,
    grow'=0,
    s sep=4pt,
    l sep=12pt
}{},
[
\textbf{Video Generation Models as World Models}, align=center, leaf-head, calign=edge midpoint
    [
        Efficient Modeling (Sec.~\ref{sec:efficient_modeling}), leaf-datasets
        [Diffusion Model Distillation\\ for Efficient Sampling, leaf-datasets]
        [Auto-Regressive and\\ Hybrid Approaches, leaf-datasets]
    ]
    [
        Efficient Architecture (Sec.~\ref{sec:efficient_arch}), leaf-methods
        [Hierarchical \& VAE Designs, leaf-methods]
        [Long Context \& Memory\\ Mechanisms, leaf-methods]
        [Efficient Attention, leaf-methods]
        [Extrapolation and RoPE, leaf-methods]
    ]
    [
        Efficient Inference (Sec.~\ref{sec:efficient_inference}), leaf-metrics
        [Parallelism, leaf-metrics]
        [Caching, leaf-metrics]
        [Pruning, leaf-metrics]
        [Quantization, leaf-metrics]
    ]
    [
        Applications (Sec.~\ref{sec:applications}), leaf-app
        [
            Autonomous Driving, leaf-app
        ]
        [
            Embodied AI, leaf-app
        ]
        [Game \& Interactive\\ World Simulation, leaf-app]
    ]
    [
        More Related Work (Sec.~\ref{sec:more_related}), leaf-chall
        [Interactive Talking Head\\ Generation, leaf-chall]
        [Interactive Content Creation, leaf-chall]
        [Video-Driven Scene Generation, leaf-chall]
    ]
]
\end{forest}
    \end{adjustbox}
    \caption{A taxonomy of representative topics related to efficiency improvement for video generation-based world models.}
    \label{fig:taxonomy}
\end{figure*}

\textbf{Taxonomy.} As shown in Figure~\ref{fig:taxonomy}, this article systematically investigates the role of efficiency in the aspects of modeling, architectures, and algorithms for video-based world models, covering the spectrum between AR-based and Diffusion-based paradigms. Our discussion is structured around three core dimensions: \textit{Efficient Modeling} (covering efficiency-oriented modeling paradigms), \textit{Efficient Architectures} (designs such as VAEs, memory mechanisms, and efficient attention), and \textit{Efficient Inference} (system deployment considerations including parallelism, caching, pruning, and quantization). Furthermore, this article also explores how these efficient models are used in downstream application scenarios, such as autonomous driving, embodied AI and games/interactive simulations. By reviewing comprehensive insights in this rapidly evolving field, we aim to catalyze new advances in video-based world models that leverage efficient computing to tackle increasingly sophisticated simulation challenges.

Within the existing literature, previous studies have primarily explored general video generation or specific diffusion model based techniques. More recently, amidst the significant advances in Sora-like models~\cite{videoworldsimulators2024}, some works have begun to address the computational demands of video generation. However, a systematic review specifically elucidating how \textit{efficiency improvement techniques} can benefit a video-based world model is notably absent. To the best of our knowledge, this article presents the first systematic exploration dedicated to the intersection of efficiency improvement techniques and the multiple facets of video-based world models. The main contributions of this paper are summarized as follows:

\begin{itemize}
\item We provide the first comprehensive review of the critical intersection between efficiency improvement techniques and video-based world models.
\item We introduce a novel taxonomy that provides a structured perspective on efficiency across three dimensions: modeling paradigms, architectural designs, and inference optimizations.
\item We detail how these efficiency improvement techniques empower critical applications such as autonomous driving, embodied AI, and interactive simulation.
\item We further discuss key challenges and future opportunities in efficient video-based world modeling.
\end{itemize}

In the remainder of this paper, we introduce background knowledge in Section~\ref{sec:background}, efficient modeling paradigms in Section~\ref{sec:efficient_modeling}, efficient architectures in Section~\ref{sec:efficient_arch}, and efficient inference algorithms in Section~\ref{sec:efficient_inference}. In addition, promising applications and more related works are discussed in Section~\ref{sec:applications} and Section~\ref{sec:more_related}. Finally, Section~\ref{sec:conclusions} concludes this paper.

\section{Background}
\label{sec:background}

\subsection{Generative Paradigms}
Modern video generation models are largely built upon paradigms established in image synthesis. We introduce the mathematical formulations of these generative models, focusing on Diffusion Models and Flow Matching as the current dominant approaches, followed by Auto-regressive models.

\subsubsection{Denoising Diffusion Probabilistic Models (DDPM)}
Diffusion models~\cite{ho2020denoising} formulate generation as a denoising process. To improve efficiency, most state-of-the-art models operate in the latent space of a pre-trained variational autoencoder (VAE), known as Latent Diffusion Models (LDMs)~\cite{rombach2022high}.

\noindent \textbf{Forward Process.} Given a data sample $x_0 \sim q(x_0)$ (or its latent representation $z_0$), the forward process is a fixed Markov chain that gradually adds Gaussian noise according to a variance schedule $\beta_t \in (0,1)$. The transition probability is defined as:
\begin{equation}
q(x_t | x_{t-1}) = \mathcal{N}(x_t; \sqrt{1-\beta_t}x_{t-1}, \beta_t \mathbf{I})
\end{equation}
Using the notation $\alpha_t = 1 - \beta_t$ and $\bar{\alpha}_t = \prod_{s=1}^t \alpha_s$, we can sample $x_t$ at any timestep $t$ directly from $x_0$:
\begin{equation}
\label{fwd_ddpm}
x_t = \sqrt{\bar{\alpha}_t}x_0 + \sqrt{1-\bar{\alpha}_t}\epsilon, \quad \text{where } \epsilon \sim \mathcal{N}(0, \mathbf{I})
\end{equation}
\noindent \textbf{Reverse Process and Training.} The generative process reverses this noise addition. Since the true posterior $q(x_{t-1}|x_t)$ is intractable, we approximate it with a parameterized distribution $p_\theta(x_{t-1}|x_t)$. In practice, the model is trained to predict the added noise $\epsilon$ or the velocity $v$. The simplified training objective is often the mean squared error (MSE) between the actual noise $\epsilon$ and the predicted noise $\epsilon_\theta$:
\begin{equation}
\mathcal{L}_{\text{simple}} = \mathbb{E}_{x_0, t, \epsilon} \left[ \| \epsilon - \epsilon_\theta(x_t, t) \|^2 \right]
\end{equation}
Once trained, the model generates data by iteratively denoising pure Gaussian noise $x_T$ to $x_0$.

\subsubsection{Flow Matching}
While DDPMs rely on a pre-defined forward process in Eq.~\eqref{fwd_ddpm}, which transports samples through a fixed and typically curved noising trajectory, Flow Matching (FM)~\cite{lipman2023flow,liu2024instaflow} instead models generation as a continuous-time probability path governed by ordinary differential equations (ODEs). FM defines a probability density path $p_t$ that transforms a simple prior distribution into the data distribution through a time-dependent vector field $v_t(x)$:
\begin{equation}
\frac{d}{dt} \phi_t(x) = v_t(\phi_t(x)), \quad \phi_0(x) = x
\end{equation}
where $\phi_t$ maps samples from $t=0$ to $t$. The goal is to learn a parameterized vector field $v_\theta(x,t)$ that matches the target velocity field associated with the chosen probability path.

Since directly regressing the marginal target velocity field is generally intractable for complex data distributions, flow matching is commonly implemented in a conditional form. Given a source sample $z_0$ and a target data sample $x_1$, one defines a conditional probability path $p_t(x \mid z_0, x_1)$ together with a tractable conditional target vector field $u_t(x \mid z_0, x_1)$. The resulting conditional flow matching (CFM) objective is
\begin{equation}
\small
\mathcal{L}_{\text{CFM}}
=
\mathbb{E}_{t, z_0, x_1, x_t \sim p_t(\cdot \mid z_0,x_1)}
\left[
\| v_\theta(x_t,t)-u_t(x_t \mid z_0,x_1)\|^2
\right].
\end{equation}
In common straight-line path formulations, the conditional path is chosen as a linear interpolation between noise $z_0$ and data $x_1$, namely $x_t = t x_1 + (1-t) z_0$. In this case, the target velocity becomes a constant, i.e., $u_t = x_1 - z_0$, and the objective reduces to
\begin{equation}
\mathcal{L}_{\text{CFM}} = \mathbb{E}_{t,z_0,x_1} \left[ \| v_\theta(x_t, t) - (x_1 - z_0) \|^2 \right].
\end{equation}



\subsubsection{Auto-regressive (AR) Models}
AR models decompose the joint probability distribution of a sequence $x$ into a product of conditional probabilities. In a canonical visual generation formulation, $x$ represents a flattened sequence of discrete visual tokens derived from a VQ-VAE-style tokenizer~\cite{esser2021taming}, where an encoder maps patches or frames to continuous latents that are snapped to a learned finite codebook via nearest-neighbor vector quantization (VQ), although more general autoregressive video models may also operate on other compressed latent token sequences. For a sequence of length $N$:
\begin{equation}
p(x) = \prod_{i=1}^{N} p(x_i \mid x_{<i})
\end{equation}
Training maximizes the log-likelihood of the next token given the previous context. While training is efficient due to parallel teacher forcing, inference is inherently sequential and can become computationally expensive for long videos ($O(N)$). 

\subsection{From Image to Video Generation}
Transitioning from image to video generation involves extending 2D spatial modeling to the 3D spatiotemporal domain ($T \times H \times W$). Efficient techniques largely focus on how to manage the cubic growth in complexity.

\noindent \textbf{Inflation} Early approaches directly inflated 2D kernels into 3D kernels (e.g., $3 \times 3 \to 3 \times 3 \times 3$)~\cite{ho2022video}. While preserving spatial priors from pre-trained image models, this drastically increases parameter count and computational load.

\noindent \textbf{Factorization} To improve efficiency, modern architectures factorize 3D operations into separate 2D spatial and 1D temporal operations. For instance, Video LDM~\cite{blattmann2023align} inserts temporal attention layers after spatial blocks in a pre-trained image U-Net. This allows the model to learn motion dynamics without catastrophic forgetting of spatial concepts and reduces computational complexity in attention mechanisms from $\mathcal{O}((THW)^2)$ to $\mathcal{O}(T(HW)^2 + HW(T)^2)$.

\noindent \textbf{Spacetime Tokenization} Emerging Transformer-based video models (e.g., Latte~\cite{ma2025latte}) treat video as a unified volumetric sequence. Instead of processing frames individually, they extract 3D spacetime cubes (``tubelets'') as tokens, utilizing a spatial and temporal downsampling mechanism by encapsulating a local spatial region across multiple consecutive frames into a single token. Consequently, the model allows for jointly capturing spatial semantics and temporal evolution within a unified attention layer, although this necessitates sophisticated positional embeddings (e.g., 3D RoPE) to accurately preserve spatiotemporal geometry.

\subsection{Architectures}




Modern video generative frameworks typically follow a modular pipeline consisting of three core components.

\noindent \textbf{Latent Compression Module (usually a VAE)} To mitigate the high dimensionality of video, VAEs compress pixel data into a latent space~\cite{rombach2022high}. Modern video generators often utilize {3D causal VAEs}~\cite{yu2023magvit, yang2025cogvideox} to jointly reduce spatial and temporal redundancy.

\noindent \textbf{Generative Backbone} The central component performs denoising or next-step prediction within the latent space. This backbone is primarily implemented using either a convolutional {U-Net}~\cite{ronneberger2015u} or a {Diffusion Transformer (DiT)}~\cite{peebles2023scalable}. DiT adopts 3D patchification and self-attention to capture long-range spatiotemporal dependencies.

\noindent \textbf{Conditioning Module} Modern video generators, especially video-based world models, are no longer conditioned by text alone, but increasingly support multimodal inputs such as reference images, video clips, audio, actions, trajectories, layouts, and other structured control signals. 
Textual guidance is commonly encoded by CLIP~\cite{radford2021learning} or T5-XXL~\cite{raffel2020exploring} and other vision-language models (VLMs)~\cite{kong2024hunyuanvideo,gao2025seedance,wan2025wan,wu2025hunyuanvideo}.
Beyond text prompts, structured conditions such as bounding boxes, road layouts, and ego trajectories can be injected to constrain scene geometry and motion, as demonstrated in driving-oriented models such as MagicDrive-V2~\cite{gao2025magicdrive}. 
In interactive world models, action signals can be represented as discrete tokens, latent actions, or control embeddings, and integrated into generation to obtain action-conditioned rollouts, as in Genie~\cite{bruce2024genie}, Matrix-Game 2.0~\cite{he2025matrix}, and Cosmos-Predict~\cite{ali2025world}. 
Audio conditions are typically encoded by a speech or motion encoder and used to guide temporal dynamics such as lip motion, facial expression, or speech rhythm~\cite{tian2024emo,li2025ditto, shen2025soulx, yang2025infinitetalkaudiodrivenvideogeneration,zheng2024memo,yi2025magicinfinite}. 
These conditions are injected into the generative backbone through cross-attention, adaptive normalization, or token merging. For example, autoregressive frameworks such as iVideoGPT~\cite{wu2024ivideogpt} serialize heterogeneous conditions into a unified sequence, whereas diffusion-based models more often fuse them through cross-attention layers or a token merging mechanism~\cite{kong2024hunyuanvideo,gao2025seedance}. 
Overall, the conditioning module determines not only \emph{what} should be generated, but also \emph{how} the generated world should evolve under external instructions or interactions.

\section{Efficient Modeling}
\label{sec:efficient_modeling}

Efficient modeling is central to scaling video generation from short clips to long-horizon, high-resolution sequences under practical latency and memory constraints. This section reviews two major directions: (i) diffusion model distillation, which reduces the number of sampling steps required for high-fidelity generation, and (ii) long-horizon interactive modeling paradigms, including autoregressive, hybrid AR-diffusion, and streaming causal diffusion approaches that aim to support real-time interaction and persistent world simulation.

\subsection{Diffusion Model Distillation for Efficient Sampling}
\label{subsec:diffusion_distillation}
While architectural and system optimizations reduce wall-clock latency per step, a complementary direction is \emph{post-training acceleration} that directly reduces the number of denoising steps. In diffusion-based video generation, the sampling cost scales linearly with the step count $K$. Distillation aims to train a \emph{student} model that matches the teacher diffusion model's sampling behavior with significantly fewer steps---down to few-step or even one-step generation.
\subsubsection{Step-Reduction Distillation}
A direct approach distills a $K$-step teacher sampler into a $K'$-step student sampler ($K'\ll K$)~\cite{salimans2022progressive,frans2024one}. 
Let $\mathcal{T}$ denote a fixed teacher solver. Starting from $x_t$, the teacher produces a target $x_{t-\Delta}$ after $\Delta$ steps. The student is trained to match this result in one macro-step:
\begin{equation}
\mathcal{L}_{\text{step}}(\theta)=
\mathbb{E}\left[\left\| \hat{x}^{(S)}_{t-\Delta}-\hat{x}^{(T)}_{t-\Delta}\right\|_2^2\right],
\label{eq:step_distill}
\end{equation}
where $\hat{x}^{(T)}_{t-\Delta}=\mathcal{T}^\Delta(x_t,c)$ is the teacher rollout target. Progressive variants halve the step count iteratively. In video generation, GPD~\cite{liang2026gpd} provides a representative example of this direction by progressively guiding the student model to operate with larger step sizes, reducing the sampling steps of Wan~\cite{wan2025wan} from 48 to 6 while maintaining competitive quality.
\subsubsection{Consistency Distillation}
Consistency-style objectives learn a mapping $f_\theta(x_t, t)$ that maps any point on the trajectory to the origin $x_0$. Consistency training enforces that predictions from two timepoints along the same trajectory agree~\cite{song2023consistency,luo2023latent}:
\begin{equation}
\mathcal{L}_{\text{cons}}(\theta)=
\mathbb{E}\left[\left\|
f_\theta(x_t,t,c)-f_\theta(x_s,s,c)
\right\|_2^2\right], \quad s<t,
\label{eq:consistency_loss}
\end{equation}
where $x_s$ is obtained by advancing from $x_t$. This enables one-step generation. VideoLCM~\cite{wang2024videolcm} and AnimateLCM~\cite{wang2024animatelcm} extend this to latent video models, enabling real-time synthesis. 
TurboDiffusion~\cite{zhang2025turbodiffusion} introduces a unified framework that combines consistency models with reward-guided distillation, significantly enhancing the visual quality of one-step outputs. Similarly, open-source initiatives like FastVideo~\cite{fastvideo2025} provide optimized pipelines for distilling large-scale video models into few-step or one-step variants, democratizing real-time video generation capabilities.


\subsubsection{Adversarial Distillation}

To maintain perceptual fidelity under extremely small step budgets, recent distillation methods increasingly optimize the student at the \emph{distribution level} rather than relying only on pointwise regression targets. A generic objective can be written as

\begin{equation}
\min_\theta \; D\!\left(p_S(\cdot|c)\,\|\,p_T(\cdot|c)\right),
\label{eq:dmd_generic}
\end{equation}

where $D$ denotes a generic discrepancy between the student distribution $p_S$ and the teacher distribution $p_T$. Such discrepancies can be instantiated in three ways. First, $D$ can be an explicit statistical divergence or its score-based surrogate, such as approximate KL divergence or Fisher-type score matching. Second, $D$ can be an implicitly learned discrepancy induced by a discriminator, as in GAN-style adversarial training. Third, practical systems often combine the two, using distribution/score matching to preserve teacher alignment while introducing adversarial supervision to improve realism and perceptual sharpness.

Representative examples of the first direction include DMD~\cite{yin2024one} and related DMD-style methods, which match the student to the teacher at the distribution level without enforcing a strict one-to-one correspondence with the teacher's sampling trajectory. In the hybrid regime, DMD2~\cite{yin2024improved} further augments distribution matching with a GAN loss on real data, and AVDM2D~\cite{zhu2024accelerating} can also be viewed within this broader family of perceptually enhanced distribution-matching distillation.

For video generation, recent work increasingly moves toward pure adversarial post-training. Seaweed-APT~\cite{lin2025seaweedapt} applies adversarial post-training against real data after diffusion pre-training, together with an approximated R1 regularization, enabling real-time one-step video generation.


However, these distillation-based approaches primarily improve step efficiency and wall-clock latency. They are usually insufficient to support persistent, long-horizon generation, which requires explicit mechanisms for causal inference, memory retention, and error control over long horizons.

\subsection{Auto-Regressive and Hybrid Approaches}
\label{subsec:hyperapproach}
Autoregressive and hybrid approaches aim to overcome the limitation of traditional video diffusion models as mainly clip-based generators. By combining autoregressive temporal rollout with efficient video synthesis, these methods move toward persistent, interactive, and long-horizon world modeling.
These methods focus on infinite-length generation with real-time interactivity by strategically combining AR scalability with diffusion fidelity.

\subsubsection{Auto-Regressive Modeling}
Treating video generation as a discrete token prediction problem allows models to inherit the scalability of autoregressive language models. A representative early work is VideoGPT~\cite{yan2021videogpt}, which employs a VQ-VAE to learn discrete spatiotemporal latent tokens from raw videos and then uses a GPT-like transformer to autoregressively model these tokens. This formulation establishes a clean and reproducible baseline for transformer-based video generation. More recent work extends this idea to large-scale multimodal and long-horizon generation. VideoPoet~\cite{kondratyuk2023videopoet} adopts a decoder-only transformer architecture that processes multimodal inputs, including text, images, videos, and audio, in a unified autoregressive framework. By following an LLM-style pretraining and adaptation pipeline, it demonstrates strong capability in zero-shot video generation and high-fidelity motion synthesis. 
Loong~\cite{wang2024loong} further pushes autoregressive generation toward minute-level long videos by modeling text tokens and video tokens as a unified sequence for autoregressive language models, together with progressive short-to-long training and inference strategies to mitigate error accumulation.

Pure autoregressive modeling is also highly relevant to interactive world modeling. Genie~\cite{bruce2024genie} introduces a generative interactive environment model composed of a spatiotemporal video tokenizer, an autoregressive dynamics model, and a latent action model. This design enables frame-by-frame controllable generation and shows that autoregressive world models can support interactive environment simulation. Along a similar direction, iVideoGPT~\cite{wu2024ivideogpt} formulates world modeling as next-token prediction over a unified sequence of visual observations, actions, and rewards. Its scalable autoregressive transformer architecture supports action-conditioned video generation, visual planning, and model-based reinforcement learning, making it a strong representative of autoregressive video-based world models.

\subsubsection{Hybrid AR-Diffusion Modeling}
Hybrid AR-diffusion modeling aims to combine the long-horizon rollout capability of autoregressive generation with the high-fidelity synthesis ability of diffusion models. In this paradigm, temporal dependencies are modeled autoregressively along the time dimension, such that previously generated frames or chunks serve as the context for predicting subsequent content, while the current frame or chunk is still generated by a diffusion model rather than directly decoded by a pure AR backbone. Progressive Autoregressive Video Diffusion Models~\cite{xie2024progressive} is a representative work in this direction. It revisits autoregressive long video generation by assigning progressively increasing noise levels across frames and performing denoising together with temporal shifting in small intervals, thereby improving information propagation and generation fidelity over long horizons. FramePack~\cite{zhang2025frame} further develops this paradigm for next-frame or next-frame-section prediction by compressing historical frame contexts according to frame-wise importance, allowing much longer effective contexts under a fixed context length while introducing drift prevention strategies to reduce long-horizon error accumulation. Overall, this hybrid paradigm provides a practical compromise between temporal scalability and visual fidelity, making it an important direction for efficient long-horizon video world modeling.



\subsubsection{Streaming Causal Diffusion Modeling}
Streaming causal diffusion modeling can be viewed as a complementary line of work to token-level autoregressive models and hybrid AR-diffusion pipelines. Instead of explicitly predicting discrete future tokens or introducing a separate autoregressive module or algorithm, it causalizes the diffusion model itself---typically through temporal causal attention or block-causal design, so that frames or chunks can be generated incrementally without relying on future context. In this way, conventional offline clip-wise diffusion is transformed into a streaming-friendly generation paradigm. For example, CausVid~\cite{yin2025slow} reformulates video diffusion with causal attention masks, enabling frame-by-frame generation and making diffusion models more suitable for autoregressive streaming scenarios.

However, causal attention alone is insufficient for stable long-horizon rollout, since continuous autoregressive generation still suffers from severe train-test mismatch and error accumulation. Diffusion Forcing~\cite{chen2024diffusionforcing} provides a representative training paradigm for this setting by combining causal next-token prediction with full-sequence diffusion and allowing different tokens to be denoised under independent noise levels. Building on this idea, Self Forcing~\cite{huang2025selfforcing} explicitly bridges the train-test gap by conditioning training on self-generated histories rather than only on ground-truth contexts, thereby improving long-horizon stability. Rolling Forcing~\cite{rollingforcing2025} further extends this principle with a rolling-window denoising strategy that jointly generates multiple future frames, substantially reducing long-range drift while enabling real-time multi-minute video generation.
Follow-up work pushes the same forcing-based streaming line toward longer horizons and sharper training objectives. Self-Forcing++~\cite{cui2025selfforcingplus} scales self-generated guidance to minute-scale video generation far beyond the original teacher horizon, while Reward Forcing~\cite{lu2025rewardforcing} enhances motion dynamics through Rewarded Distribution Matching Distillation, biasing the model toward high-reward dynamic regions and thereby allocating limited generation capacity more effectively to behaviorally important events.
Closely related to few-step interactive streaming, Causal Forcing~\cite{zhu2026causalforcing} studies autoregressive diffusion distillation from pretrained bidirectional video diffusion models. It argues that initializing the causal student via ODE distillation from a \emph{bidirectional} teacher can violate frame-level injectivity under the probability-flow ODE, leading to a biased conditional-expectation solution rather than the teacher's flow map, and instead performs ODE initialization using an \emph{autoregressive} teacher to align the distillation objective with causal generation.
Orthogonal to further training-time refinements, Rolling Sink~\cite{li2026rollingsink} targets the mismatch between \emph{finite} clip-length training and \emph{open-ended} test-time horizons, and proposes a training-free test-time procedure for autoregressive cache maintenance that scales autoregressive video diffusion models trained with Self Forcing to minute-scale generation while improving long-horizon fidelity and temporal consistency.
Beyond likelihood- or distillation-based causal diffusion training, AAPT~\cite{lin2025aapt} further extends adversarial post-training to autoregressive streaming generation, reducing error accumulation through student-forcing and demonstrating that diffusion-based video generators can move closer to interactive real-time rollout.

\subsection{Discussion}
\begin{figure*}[t] 
    \centering

    \setlength{\tabcolsep}{1.5pt} 

    \renewcommand{\arraystretch}{1.0}

    \newcommand{\vcenterimg}[1]{\includegraphics[width=0.145\textwidth, valign=c]{#1}}

    \newcommand{\rowname}[2]{%
        \scriptsize
        \begin{tabular}{@{}r@{}}
            #1 \\
            #2
        \end{tabular}%
    }

    \begin{tabular}{r cccccc} 
        & \textbf{0s} & \textbf{15s} & \textbf{40s} & \textbf{2m} & \textbf{5m} & \textbf{10m} \\
        \midrule[0.8pt]

        \rowname{Self Forcing}{\cite{huang2025selfforcing}} & 
        \vcenterimg{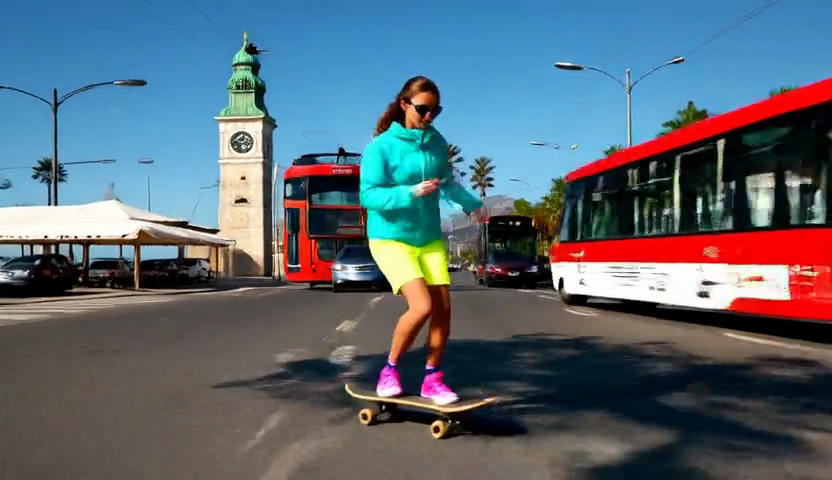} &
        \vcenterimg{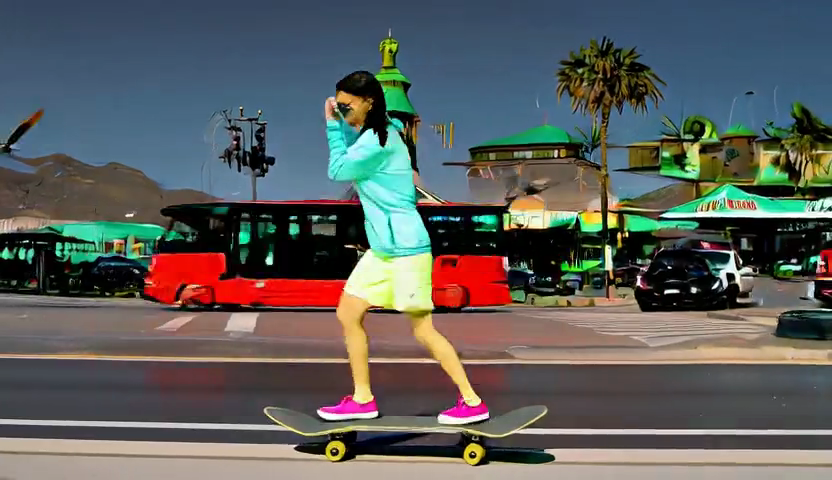} &
        \vcenterimg{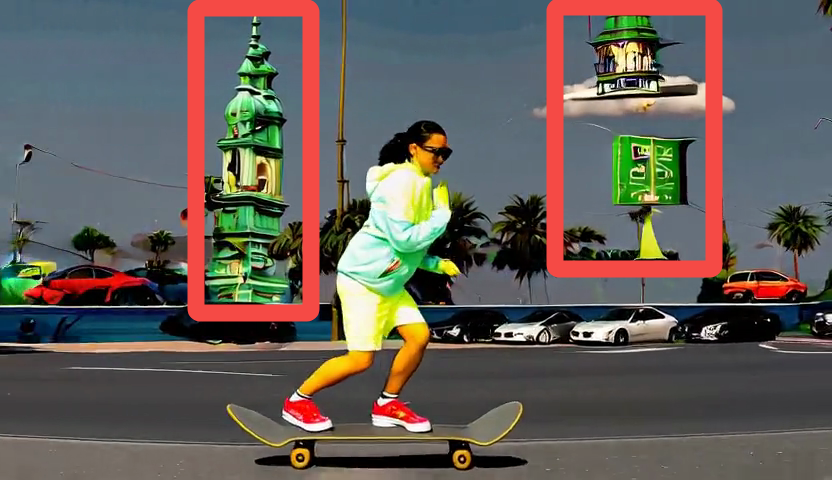} &
        \vcenterimg{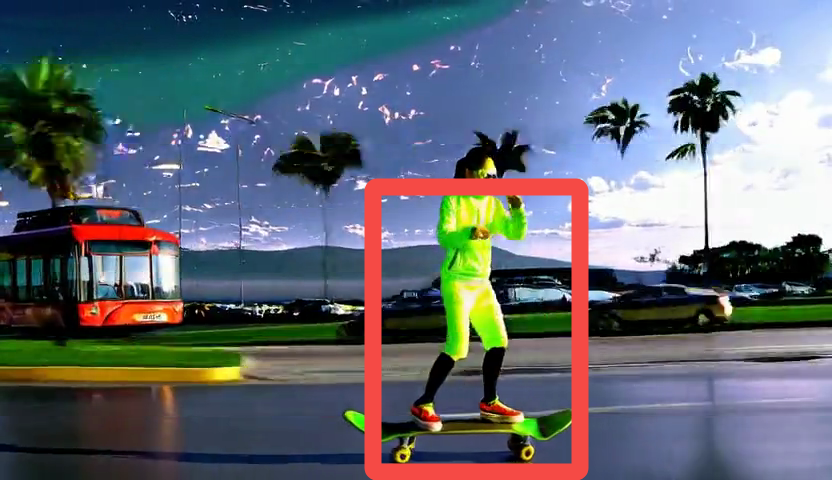} &
        \vcenterimg{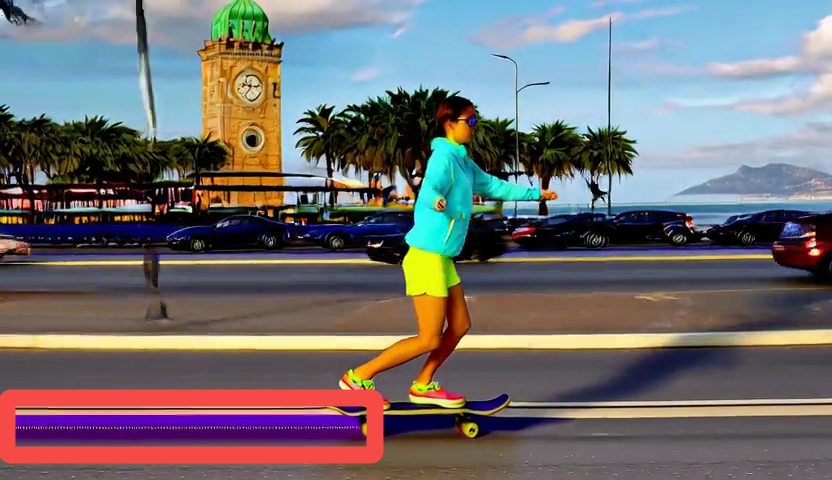} &
        \vcenterimg{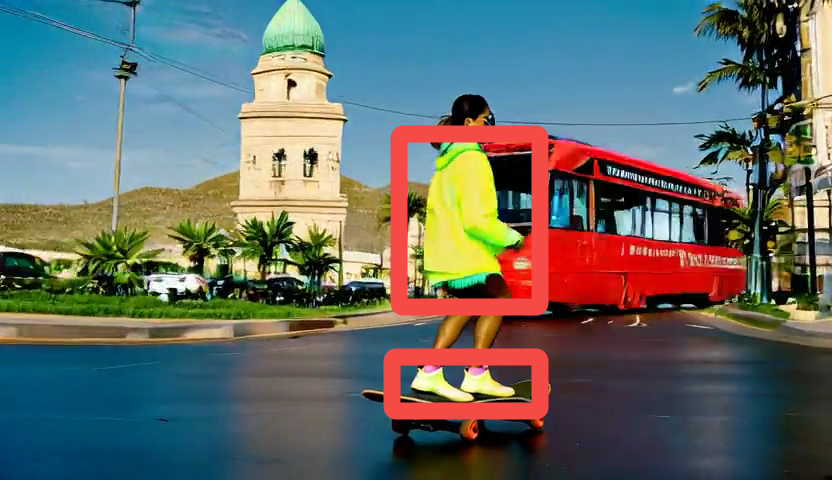} \\[3pt]

        \rowname{LongLive}{\cite{yang2025longlive}} & 
        \vcenterimg{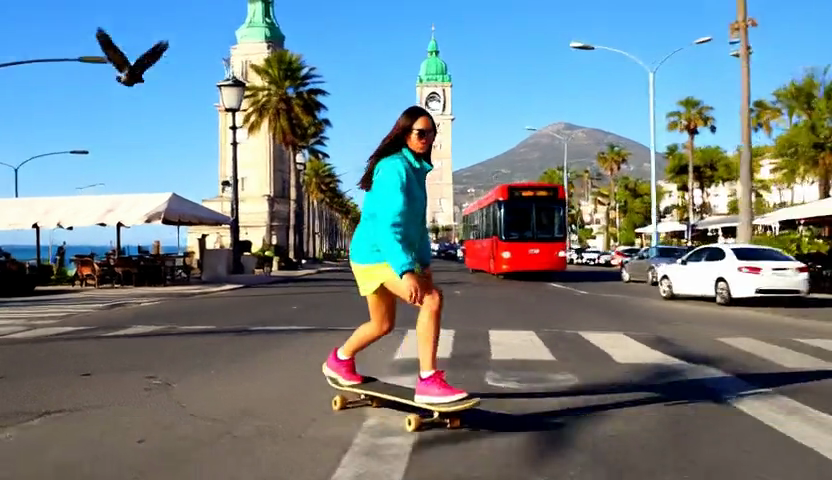} &
        \vcenterimg{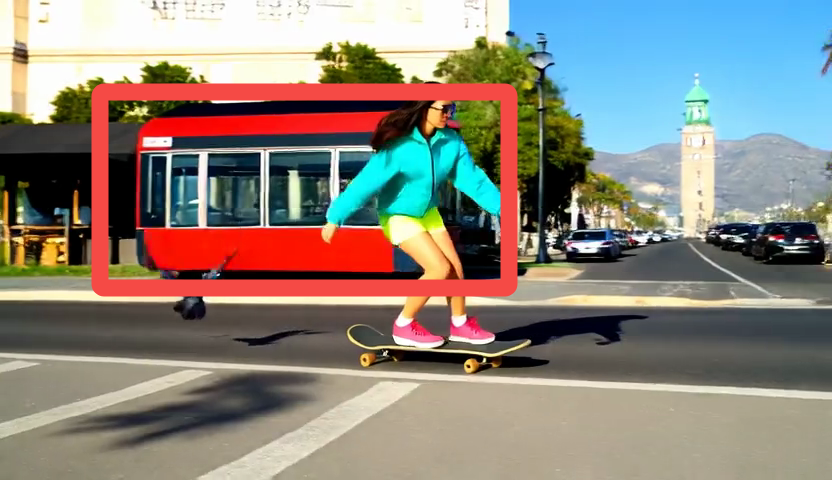} &
        \vcenterimg{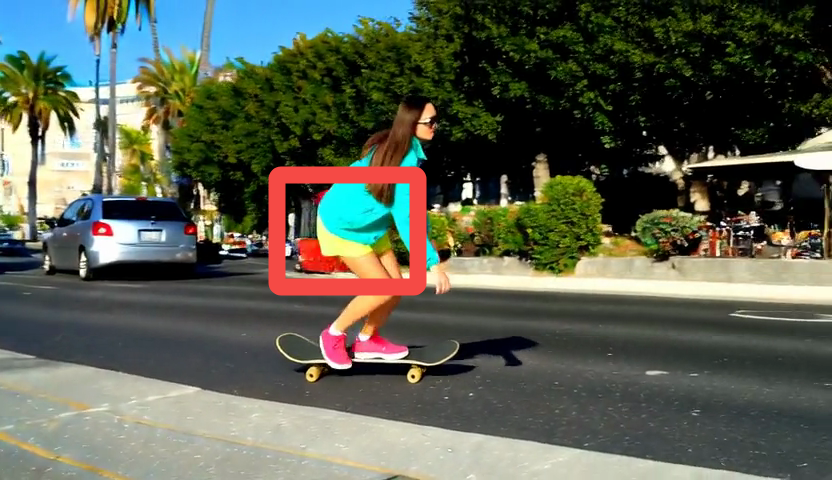} &
        \vcenterimg{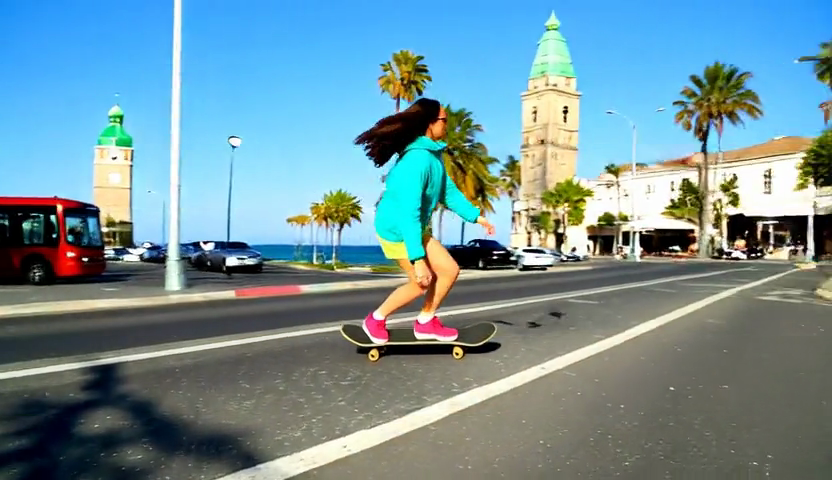} &
        \vcenterimg{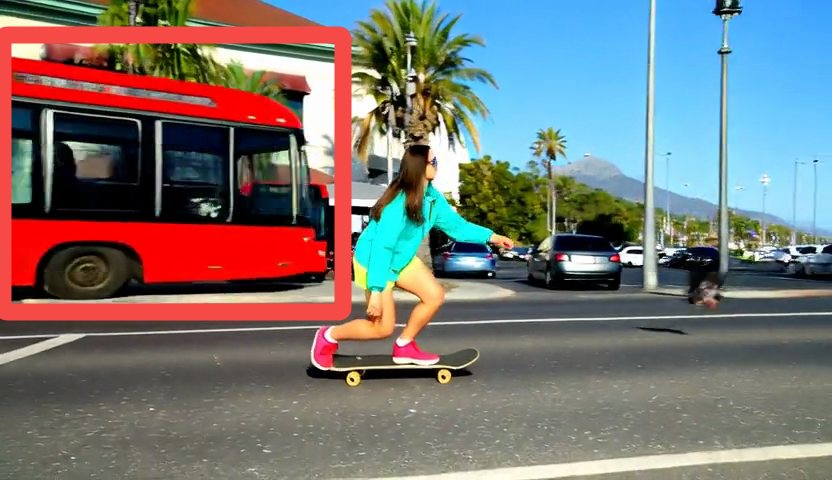} &
        \vcenterimg{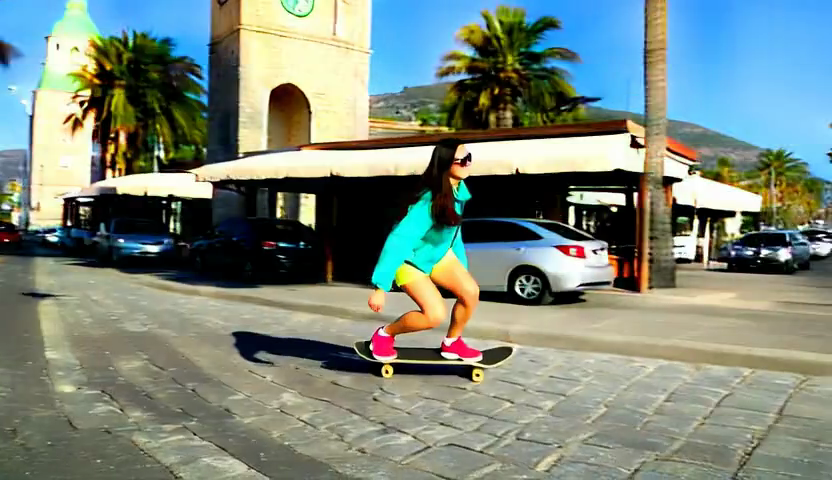} \\[3pt]

        \rowname{Rolling Forcing}{\cite{rollingforcing2025}} & 
        \vcenterimg{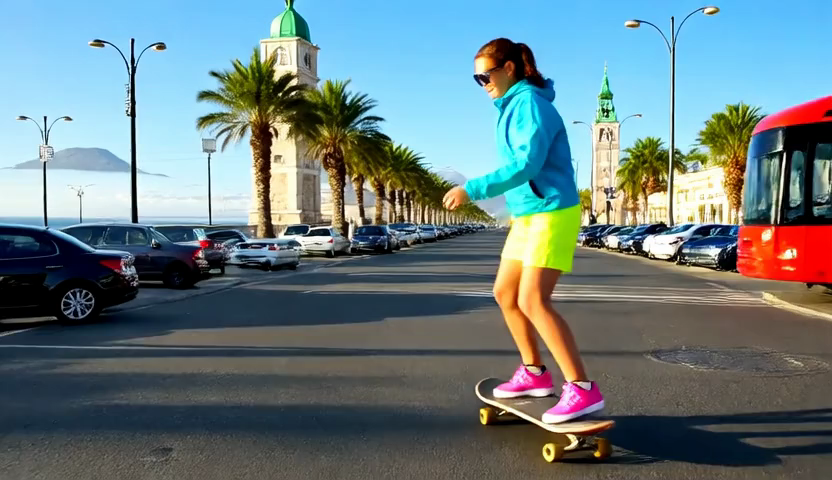} &
        \vcenterimg{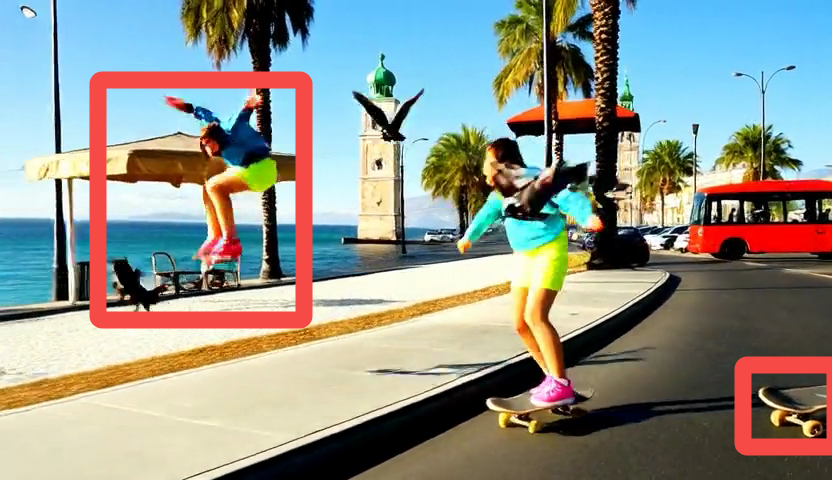} &
        \vcenterimg{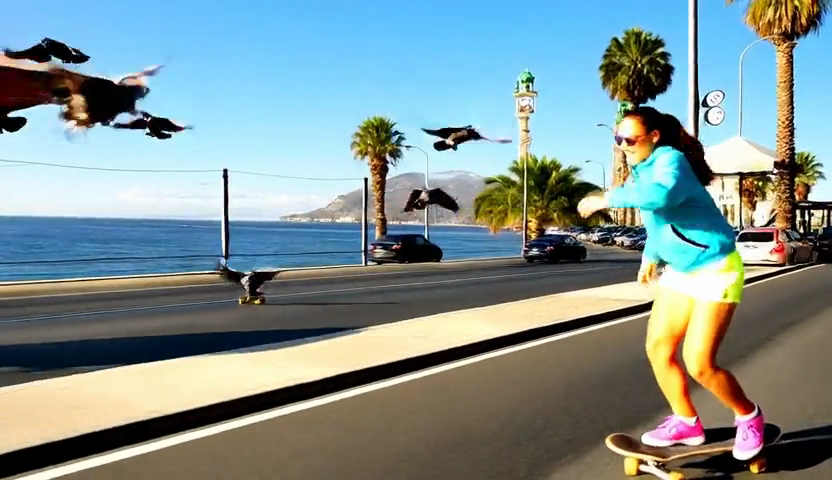} &
        \vcenterimg{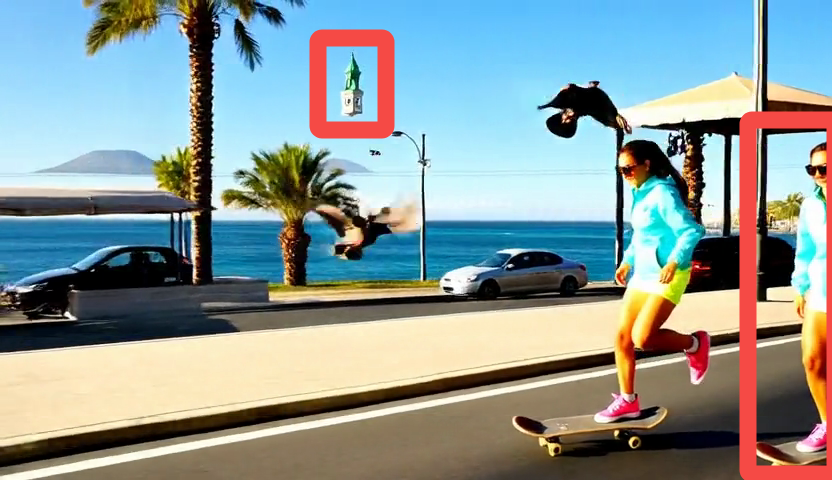} &
        \vcenterimg{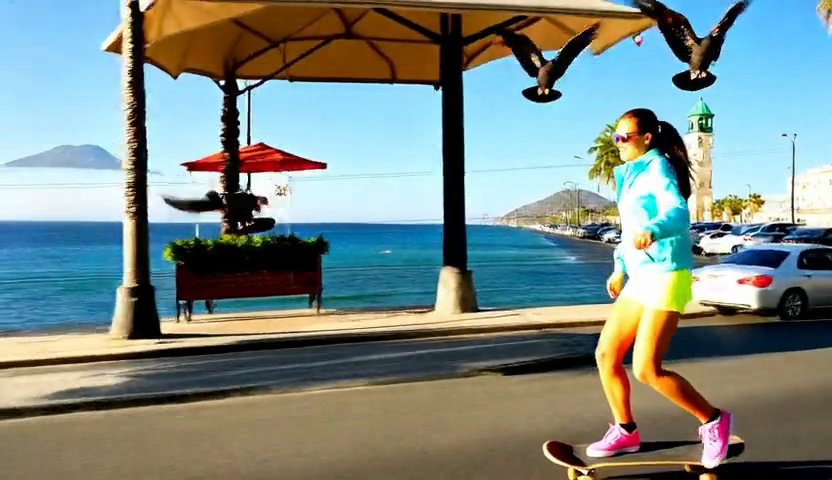} &
        \vcenterimg{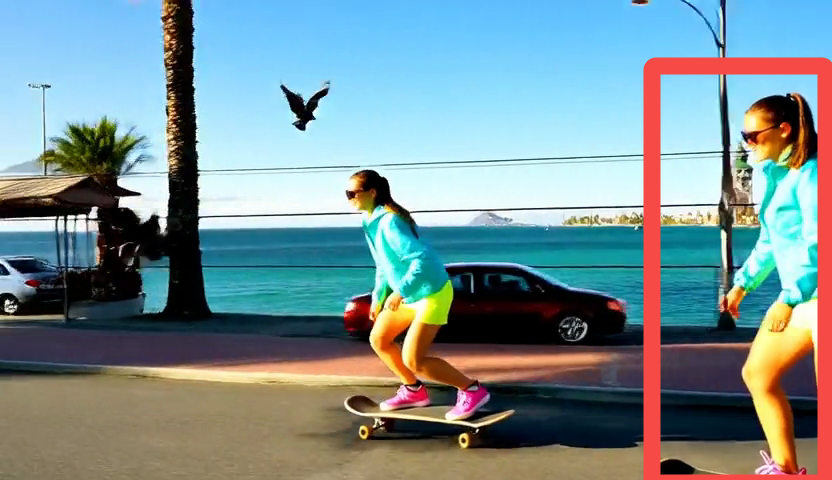} \\[3pt]

        \rowname{Deep Forcing}{\cite{yi2025deep}} & 
        \vcenterimg{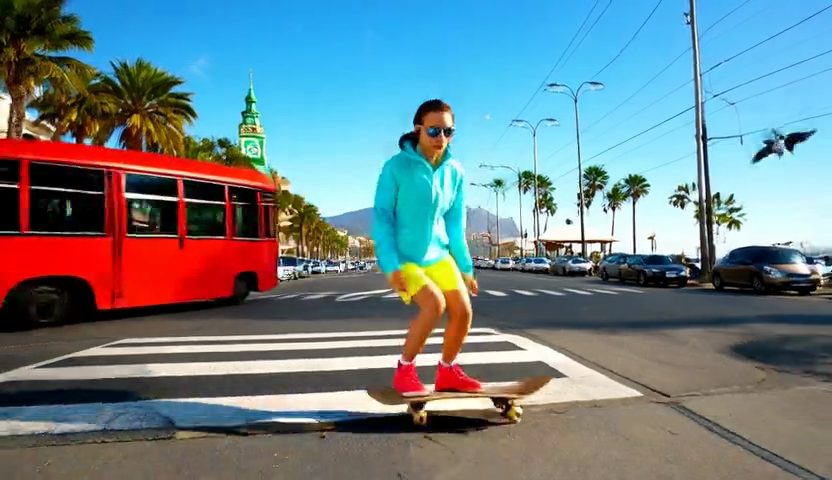} &
        \vcenterimg{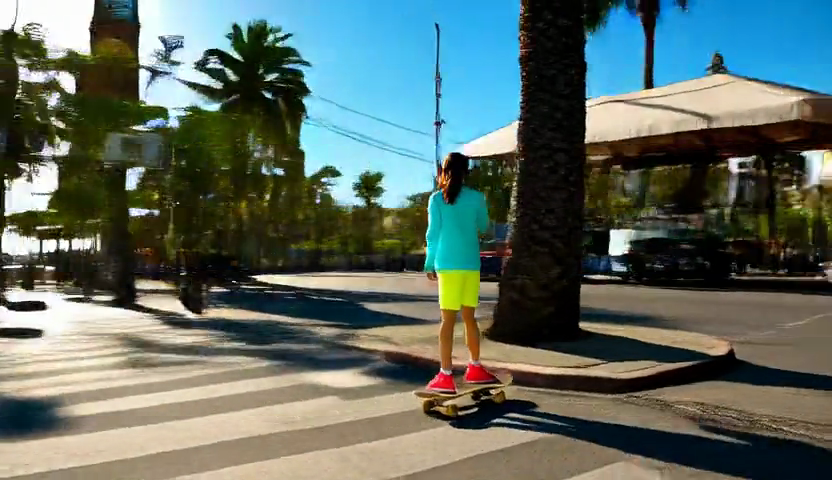} &
        \vcenterimg{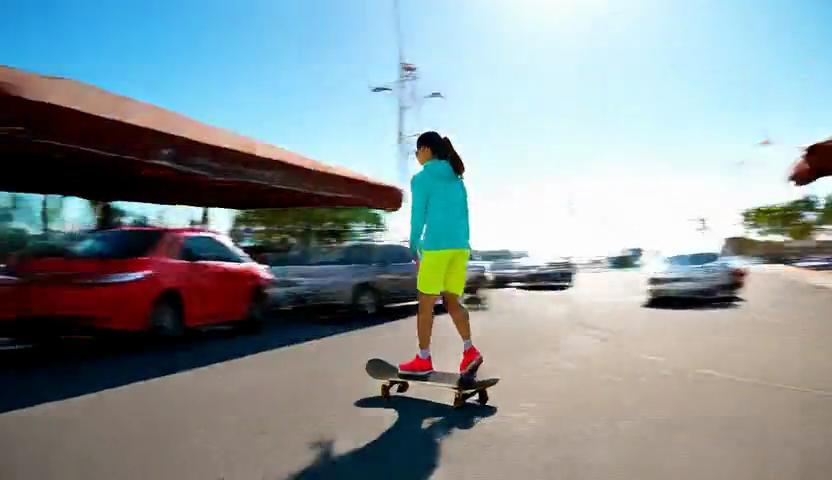} &
        \vcenterimg{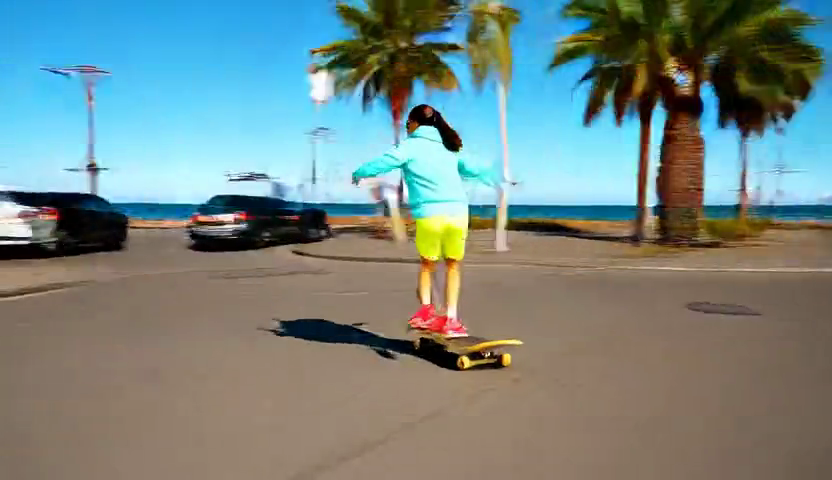} &
        \vcenterimg{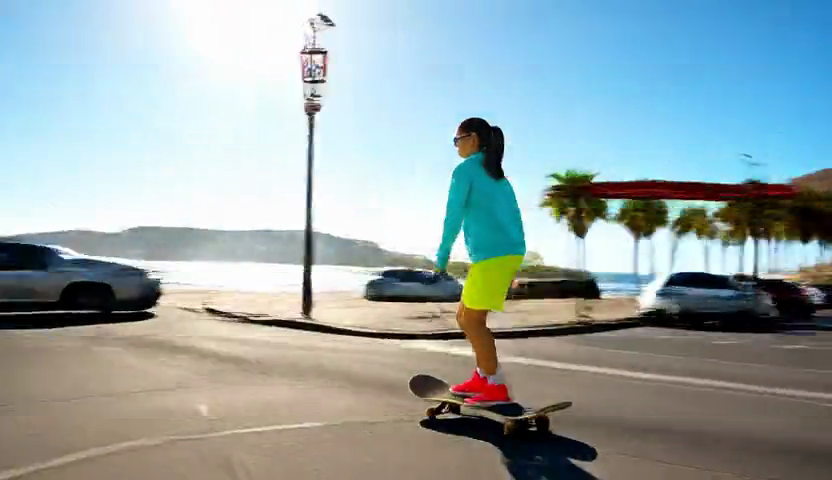} &
        \vcenterimg{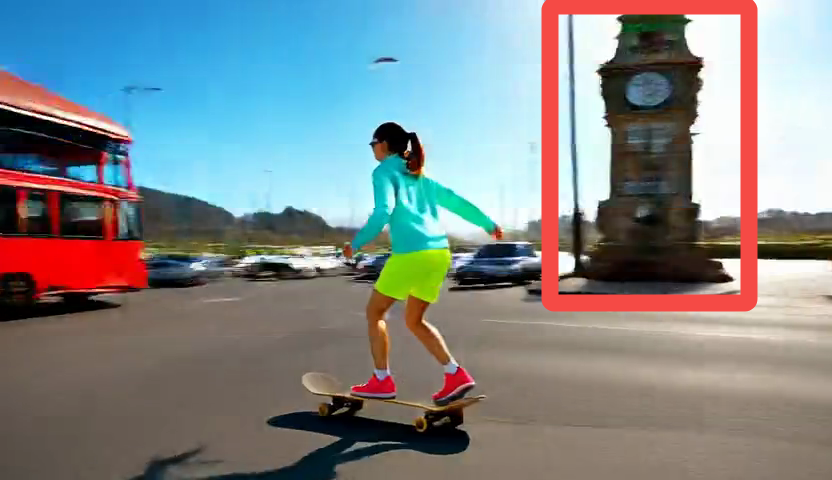} \\[3pt]

        \rowname{Causal Forcing}{\cite{zhu2026causalforcing}} & 
        \vcenterimg{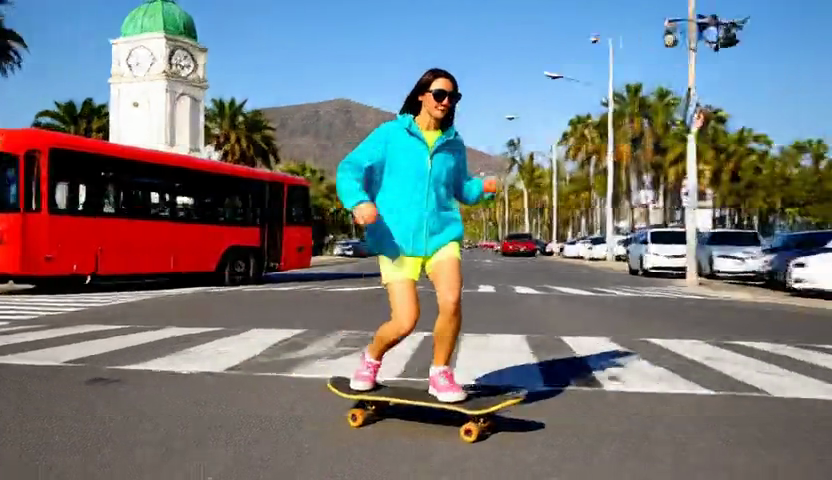} &
        \vcenterimg{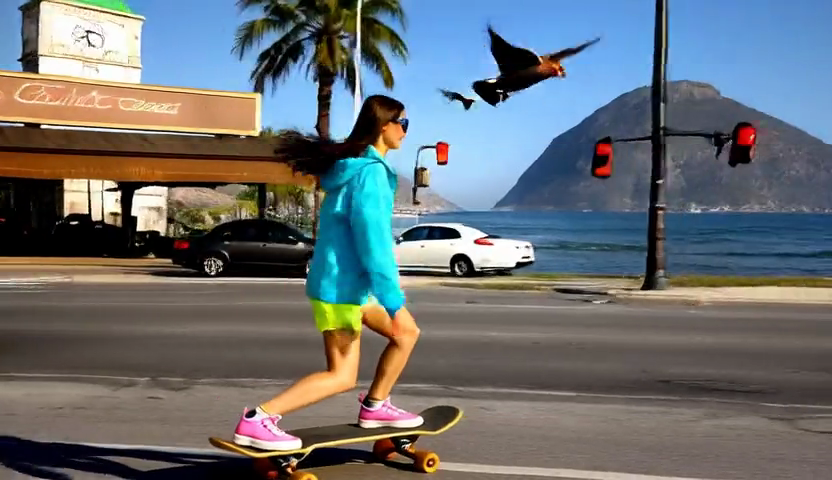} &
        \vcenterimg{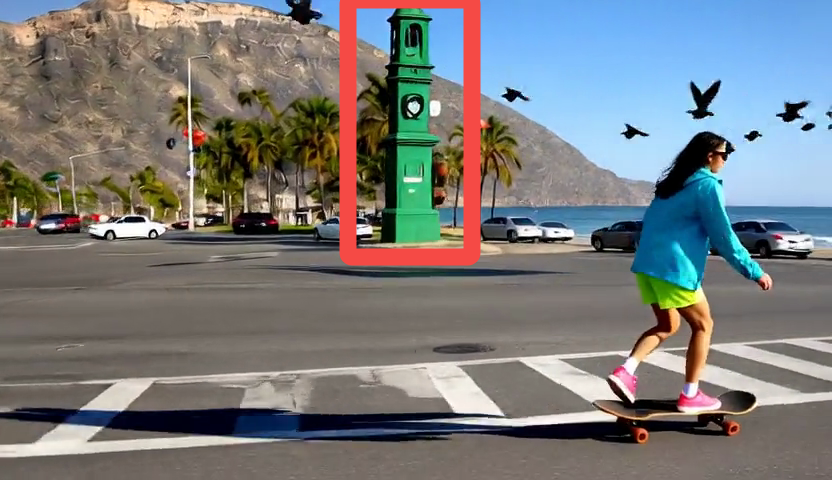} &
        \vcenterimg{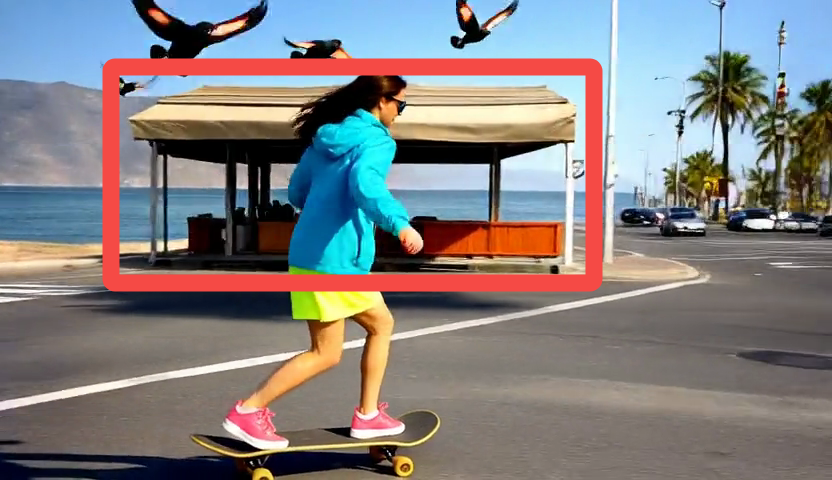} &
        \vcenterimg{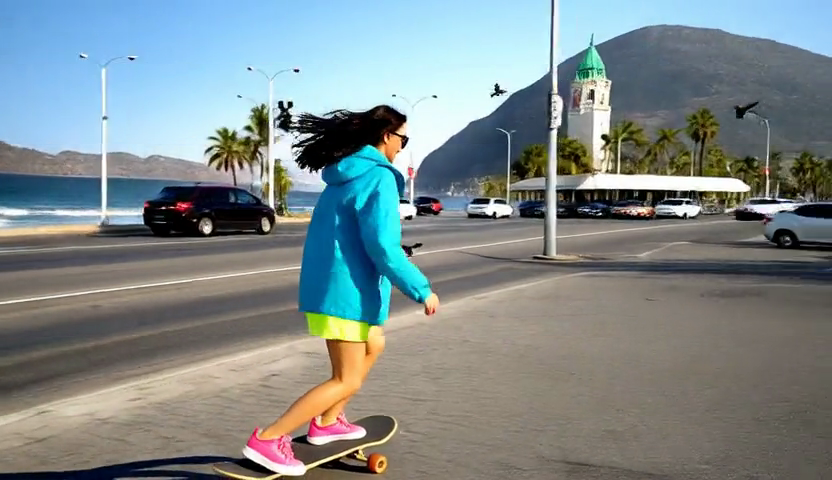} &
        \vcenterimg{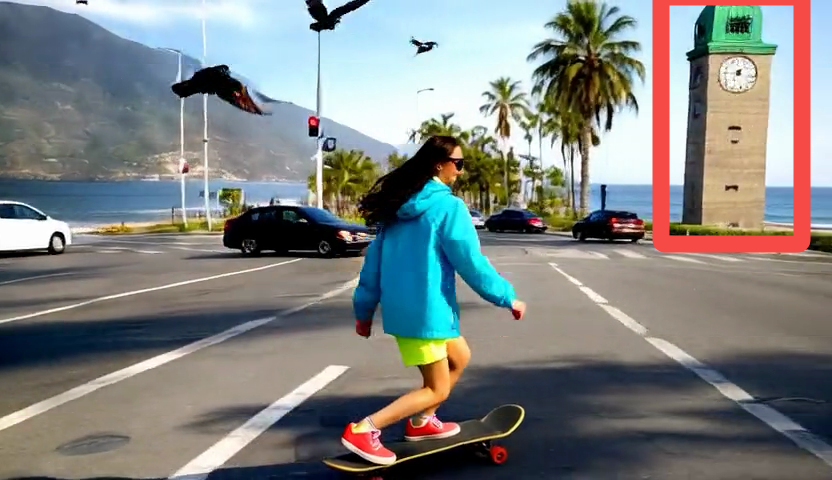} \\[3pt]

        \rowname{Rolling Sink}{\cite{li2026rollingsink}} & 
        \vcenterimg{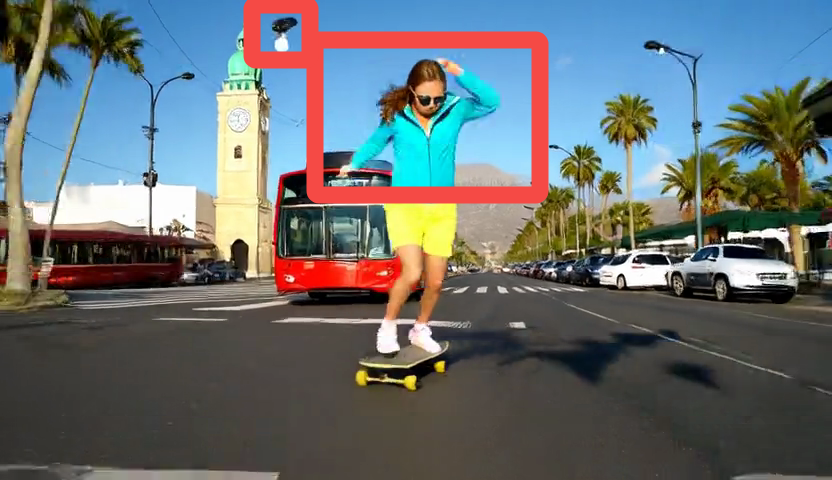} &
        \vcenterimg{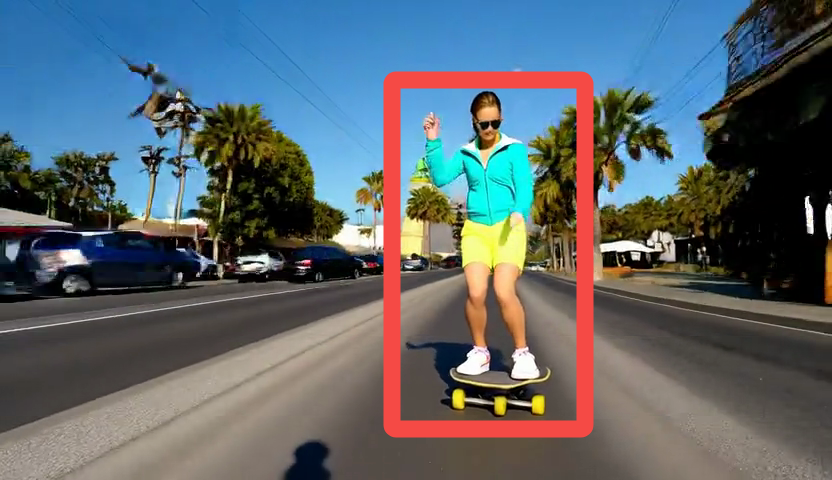} &
        \vcenterimg{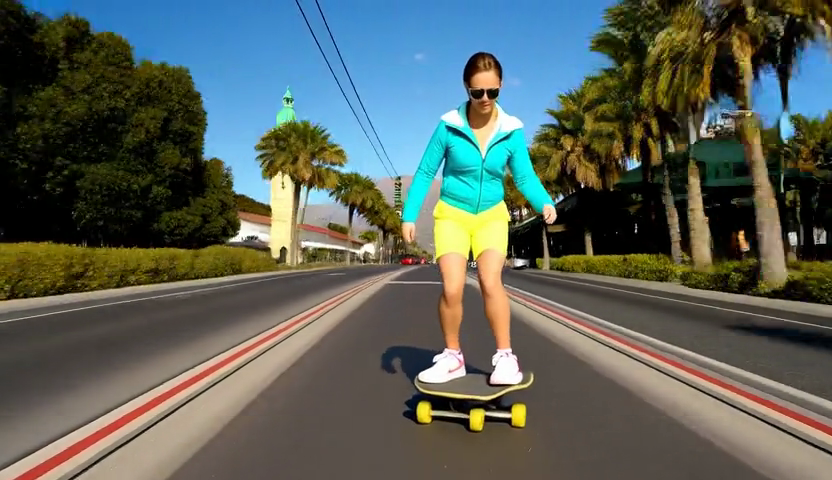} &
        \vcenterimg{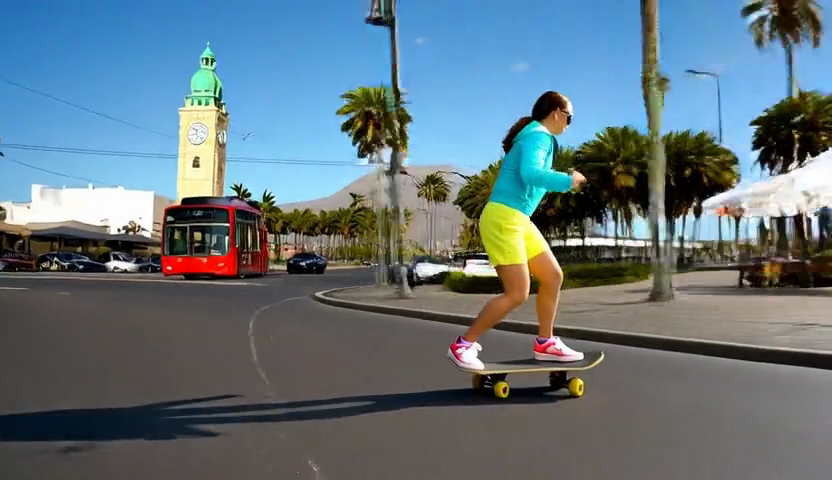} &
        \vcenterimg{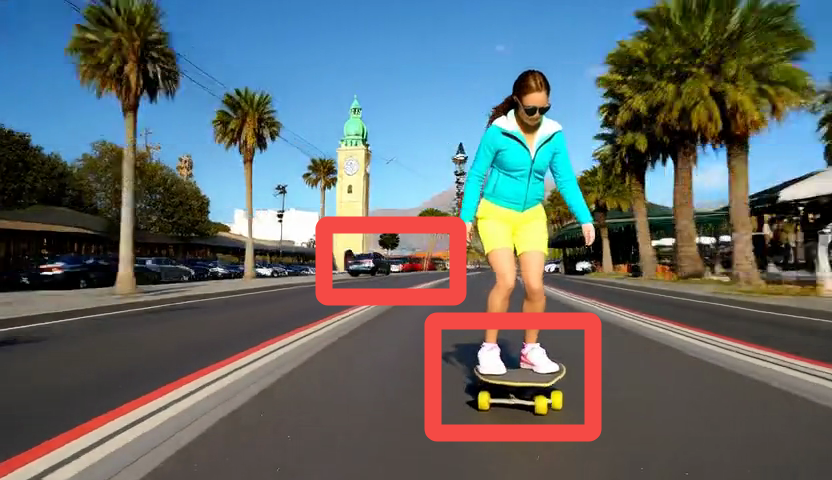} &
        \vcenterimg{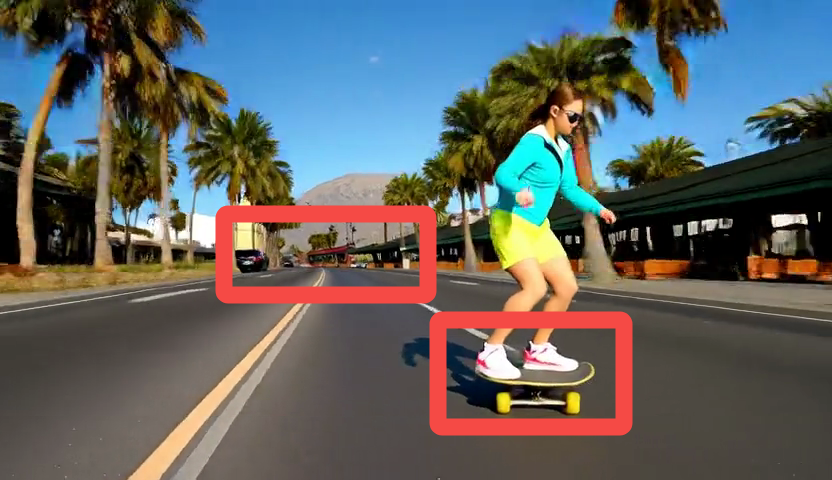} \\
        
        \bottomrule 
    \end{tabular}
    
    \vspace{2mm} 

    \caption{\textbf{Qualitative comparison of long-video generation.} We evaluate six methods built upon Wan2.1-T2V-1.3B~\cite{wan2025wan} across a 10-minute timeline. {\setlength{\fboxrule}{1pt}\setlength{\fboxsep}{2pt}\color{red}\fbox{\textcolor{black}{Red boxes}}} are added to highlight specific failure modes for each method. 
    \textbf{Self Forcing}~\cite{huang2025selfforcing}: inherently limited to a maximum generation length of 5 seconds (81 frames), leading to catastrophic structural collapse (e.g., at 40s), outfit inconsistencies (e.g., at 2m and 10m), and color artifacts (e.g., at 5m).
    \textbf{LongLive}~\cite{yang2025longlive}: severe structural hallucination and geometric distortion of the red bus (e.g., at 15s, 40s, and 5m).
    \textbf{Rolling Forcing}~\cite{rollingforcing2025}: inconsistent spatial structures, such as detached skateboards or duplicated characters (e.g., at 15s, 2m, and 10m).
    \textbf{Deep Forcing}~\cite{yi2025deep}: severe background hallucinations, with the clock tower disappearing or completely transforming (e.g., at 10m).
    \textbf{Causal Forcing}~\cite{zhu2026causalforcing}: illogical 3D geometry and inconsistent backgrounds, such as a kiosk in the middle of the road (e.g., at 2m) and a mutating clock tower. 
    \textbf{Rolling Sink}~\cite{li2026rollingsink}: human body distortions (e.g., at 0s and 15s), scene flattening and incorrect spatial relationships (e.g., at 5m and 10m).
    Overall, maintaining complex interactions and persistent 3D structures over minutes remains a significant challenge.}

    \label{fig:generation_comparison}
\end{figure*}

Efficient modeling methods are driven by two equally fundamental objectives: \emph{per-step efficiency} and \emph{long-horizon interactivity}.
Distillation-based approaches are highly effective in straightening the denoising trajectory and reducing latency from tens of steps to only a few or even a single step, making real-time generation increasingly practical. However, most of these methods remain centered on accelerating fixed-length generation, and therefore do not by themselves resolve the challenges of persistent rollout and long-term error accumulation. In contrast, autoregressive, hybrid AR-diffusion, and causal streaming diffusion paradigms, based on efficient per-step inference, explicitly target these world-model requirements by introducing causal generation interfaces.

While per-step efficiency and interactivity form the foundation of video-based world models, a more critical and unresolved goal is optimizing consistency, stability, and spatial understanding in long-duration scenarios. As illustrated in Figure~\ref{fig:generation_comparison}, current state-of-the-art models still exhibit severe degradation over extended timelines. Over a 10-minute generation window, we observe that most methods struggle with object permanence and structural consistency in later stages. Furthermore, accurately synthesizing complex human-object interactions (e.g. skateboarding) and maintaining coherent relative spatial positioning within a 3D environment remains a significant obstacle.

Addressing these limitations requires a novel modeling paradigm. This involves designing algorithms to mitigate cumulative errors in long-term interactions and memory mechanisms to ensure spatial, logical, and physical consistency.

\section{Efficient Architecture}
\label{sec:efficient_arch}

To overcome spatiotemporal redundancy and the quadratic cost of attention in long-horizon video generation, efficient architectural design is the most direct and effective method for enhancing video generation from short clips to persistent, high-fidelity world models. This section reviews four structural optimization paradigms: (i) Hierarchical and VAE Designs, which compress the world’s state into compact or coarse-to-fine representations; (ii) Long Context and Memory Mechanisms, offering scalable alternatives for long-term consistency; (iii) Efficient Attention, accelerating computation via sparse, windowed, or linear mechanisms; and (iv) Extrapolation and RoPE, providing cost-effective methods to extend generation beyond training horizons.

\subsection{Hierarchical \& VAE Designs}
The common framework for efficient video-based world modeling involves decomposing the high-dimensional spatiotemporal signal into a coarse-to-fine hierarchy or a compact latent space, reducing the state complexity that the model must simulate.
\subsubsection{Hierarchical and Pyramidal Generation}
\begin{figure}
    \centering
    \includegraphics[width=1\linewidth]{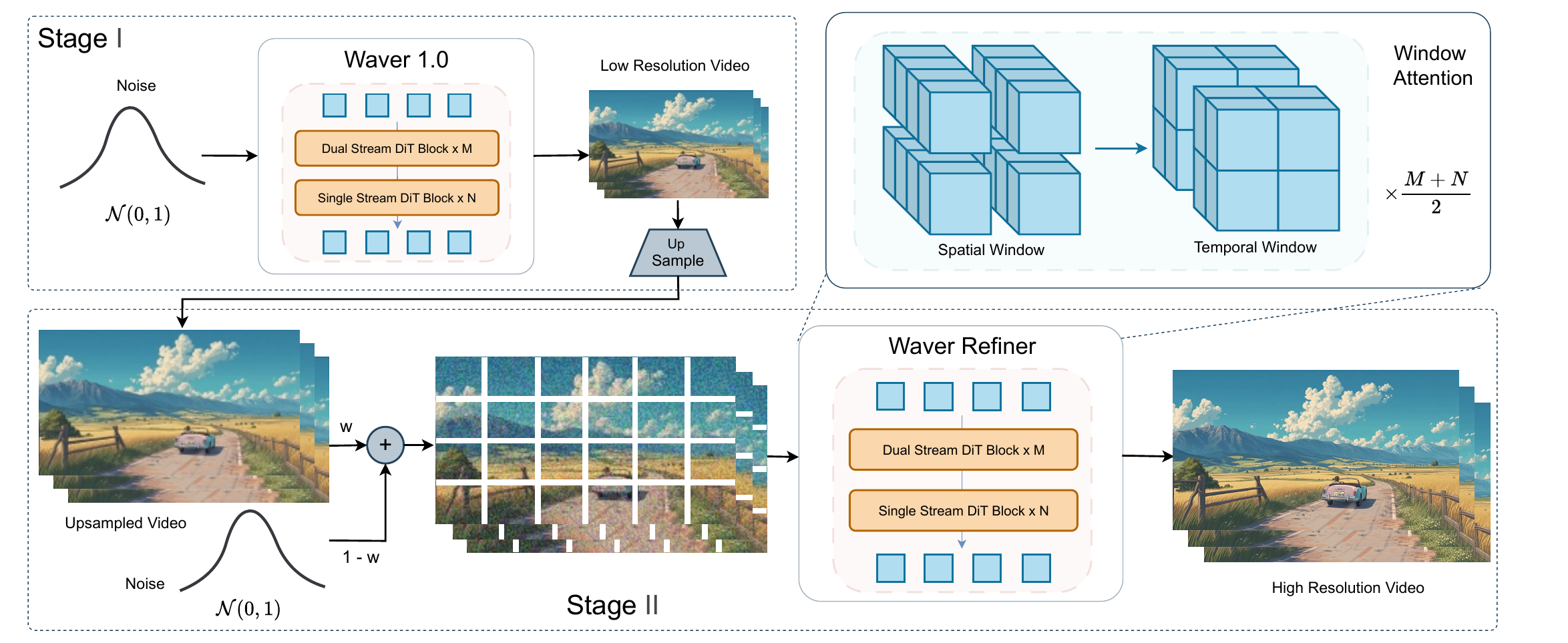}
    \caption{Pipeline of cascaded video generation. Figure courtesy of~\cite{zhang2025waver}.}
    \label{fig:cascade_arch}
\end{figure}

This approach is a multi-stage refinement process where a base module establishes a general world model followed by specialized modules for detail enhancement. Pyramidal Flow Matching~\cite{jin2025pyramidal} and TPDiff~\cite{ran2025tpdiff} establish multiple stages with increasing frame rate; the former extends the flow matching algorithm to an efficient spatial pyramid representation and proposes a temporal pyramid design to further improve training efficiency, while the latter introduces a temporal pyramid to increase frame rate along the diffusion process. Waver~\cite{zhang2025waver} (Figure~\ref{fig:cascade_arch}) and FlashVideo~\cite{zhang2026flashvideo} adopt a cascade paradigm to upsample low-resolution videos generated by the DiT to a final resolution of 1080p. Compared to flat architectures, these hierarchical designs significantly reduce the redundant computation of details during the early semantic planning phase. SUPERGEN~\cite{ye2025supergen} also follows this two-stage paradigm, where a sketch provides an overview and iterative fine-grained tile-based refinement enriches details. PatchVSR~\cite{du2025patchvsr} introduces inter-patch modulation during the detail generation process. In addition, on the parameter dimension, SRDiffusion~\cite{cheng2025srdiffusion} uses a large model during the early high-noise steps to generate higher-quality structures and motion while using a small model during the late low-noise steps to generate finer details, thus accelerating the overall diffusion process. This method offers a more flexible efficiency-quality trade-off than fixed hierarchical cascades, and switches models during different stages of the diffusion process. Conversely, on the spatial dimension, FlexiDiT~\cite{anagnostidis2025flexidit} applies a reversed compute strategy. Observing that the early denoising steps focus on low-frequency structures, it uses larger patch sizes (lower compute) for the initial denoising steps and switches to smaller patches (higher compute) for later refinement.
\subsubsection{Efficient VAE and Latent Compression}
To model a persistent world, the world state must be compressed into a manageable latent representation. DC-VideoGen~\cite{chen2025dc} introduces a deep compression video autoencoder with a chunk-causal design to achieve up to 64$\times$ spatial compression and 4$\times$ temporal compression. REGEN~\cite{zhang2025regen} further expands this by relaxing the criterion for decoding from exact reproduction to plausible reconstruction. The decoder itself is generative, allowing the encoder to store ultra-compact semantic tokens only, achieving a temporal compression ratio of up to 32$\times$. Considering that VAE's fixed compression rates cannot capture the temporal non-uniformity of real-world video contents, DLFR-VAE~\cite{yuan2025dlfr} proposes a dynamic VAE that dynamically adjusts the optimal latent frame rate according to the content complexity. In addition, VGDFR~\cite{yuan2025vgdfr} adaptively merges frames in the latent space, allowing subsequent denoising steps to be executed in a smaller latent space, significantly reducing computational costs.
Recent works such as Turbo-VAED~\cite{zou2025turbo} distill heavy decoders into lightweight versions.

\subsection{Long Context \& Memory Mechanisms}
Video-based world modeling relies on maintaining consistency over long horizons. The common method augments the generative backbone with an external or implicit memory that serves as a persistent storage of the simulated world.

\subsubsection{Visual Memory}
Visual Memory retains raw or semi-compressed keyframes as distinct reference points to anchor the generation. FramePack~\cite{zhang2025frame} compresses historical frames according to their relative importance to encode more frames within a fixed context length limit. Following this, WorldPack~\cite{oshima2025worldpack} combines trajectory packing, which enables efficient utilization of long-term history within a fixed-length context by hierarchically compressing and allocating frames, and memory retrieval, which selectively recalls past scenes that share substantial visual overlap with the target scene. This design allows recent frames and frames recalled by memory retrieval to be stored at a high resolution, and the remaining frames to be stored at a lower resolution, enabling the model to retain long-term history while keeping computation efficient.
Related works such as StoryMem~\cite{zhang2025storymem} maintain a compact and dynamically updated memory bank of keyframes from historically generated shots. Astra~\cite{zhu2026astra} also adopts frame packing and uses a noise-augmented history memory to avoid over-reliance on past frames.

\subsubsection{Spatial Memory}
Spatial Memory utilizes explicit geometric representations (e.g., point clouds, meshes) to enforce strict physical consistency. EvoWorld~\cite{wang2025evoworld} maintains an evolving panoramic world using a spherical 3D memory, while VMem~\cite{li2025vmem} constructs a surfel-indexed global map for camera-pose queries. These works shift the model's role from pixel generation to rendering from a consistent memory. Following these, HunyuanWorld-Voyager~\cite{huang2025voyager} and~\cite{wu2025video} align frames with persistent point clouds for spatial consistency. Memory Forcing~\cite{huang2025memory}  combines spatial and temporal memory for maintaining long-term geometric consistency.

\subsubsection{Compressed Contexts}
\begin{figure}
    \centering
    \includegraphics[width=1\linewidth]{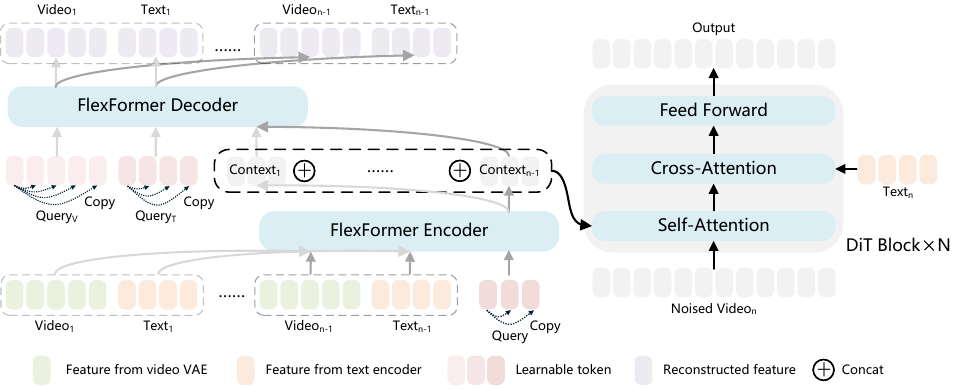}
    \caption{LoViC~\cite{jiang2025lovic} introduces FlexFormer, a flexible encoder that compresses context of arbitrary length under an adaptive compression ratio. The resulting compressed context features are fed into a DiT-based decoder to generate the current video chunk. Figure courtesy of~\cite{jiang2025lovic}.}
    \label{fig:lovic}
\end{figure}

These approaches focus on reducing historical information into compact latent vectors. SVI~\cite{li2025stable} utilizes an Error Replay Memory that dynamically banks and selectively resamples self-generated diffusion errors across discretized timesteps to simulate and correct accumulation artifacts during fine-tuning. MemFlow~\cite{ji2025memflow} also dynamically updates the memory bank by retrieving the most relevant historical frames with the text prompt of the current chunk. VideoSSM~\cite{yu2025videossm} introduces a global memory to absorb tokens evicted from the local window and relies on a state space model to recurrently compress them into a compact, fixed-size state. Other works such as Context as Memory~\cite{yu2025context}, LoViC~\cite{jiang2025lovic} (Figure~\ref{fig:lovic}), and Mixture of Contexts~\cite{cai2025mixture} refine how contexts are retrieved. Compared to spatial maps, compressed contexts are more flexible, but may struggle with precise geometric grounding.

\subsubsection{Implicit Model Memory}
Implicit Model Memory embeds historical contexts directly into the model's weights via online updates (test-time training, TTT). TTT\cite{sun2025learning} has emerged as a promising approach for efficient sub-quadratic sequence modeling, which extends the concept of recurrent states in RNNs to a small, adaptive sub-network. The weights of this sub-network are rapidly adapted online via self-supervised objectives to memorize in-context information.~\cite{dalal2025one} incorporate TTT layers into DiT to capture global narrative dependencies for minute-long generation. Addressing the hardware inefficiency of frequent updates, LaCT~\cite{zhang2025test} performs weight updates for massive token blocks rather than individual steps, enabling scalable autoregressive modeling with contexts exceeding 50k tokens. Although this paradigm offers the memorization of very long contexts, it incurs higher inference latency due to the cost of online optimization.

\subsection{Efficient Attention}
Full attention accounts for major end-to-end runtime in video generation. Due to the quadratic computational complexity with respect to context length, attention can be much more dominant as the resolution and number of frames increase.
To handle this quadratic complexity of high-resolution or long-horizon world simulation, efficient architectures approximate full attention with sparse attention, window attention, or linear attention, or even replace the attention mechanism with linear-complexity alternatives (e.g., SSMs).
\subsubsection{Sparse Attention}
\begin{figure}
    \centering
    \includegraphics[width=1\linewidth]{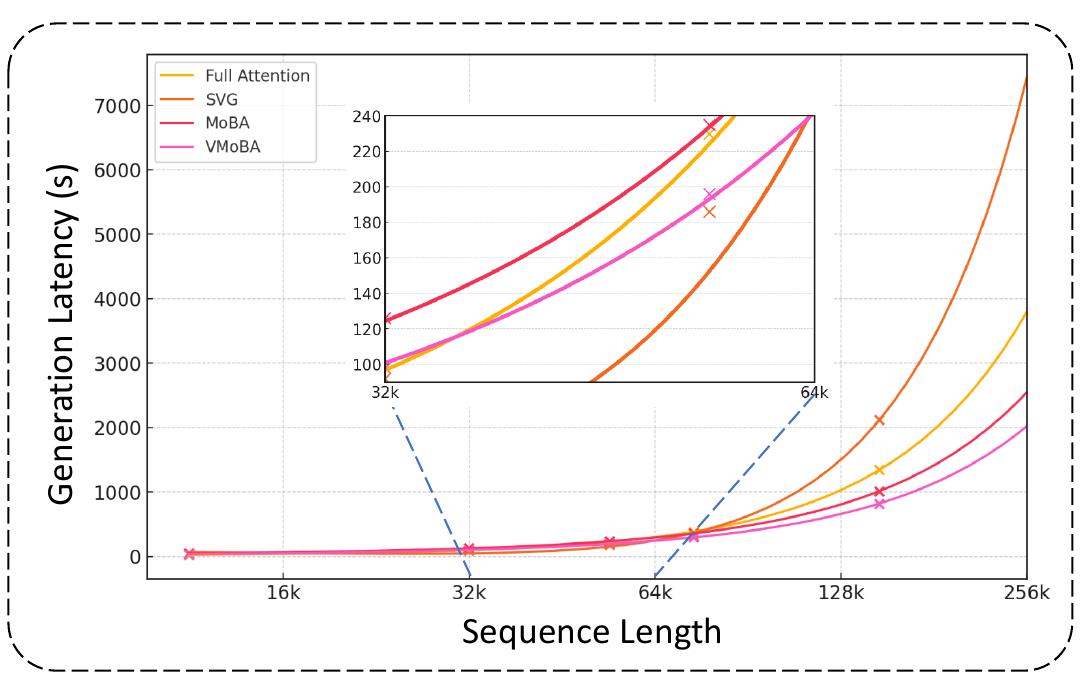}
    \caption{Comparison of inference time among full attention, SVG\cite{xi2025sparse}, MoBA~\cite{lu2025moba} and VMoBA\cite{wu2025vmoba} as sequence length increases. Figure courtesy of~\cite{wu2025vmoba}.}
    \label{fig:vmoba_compare}
\end{figure}

Attention in transformers is inherently sparse~\cite{zhang2023ho}, which offers a great opportunity to reduce computation. Sparse attention selectively restricts computation to highly relevant or local token pairs. SVG~\cite{xi2025sparse} and SVG2~\cite{yang2025sparse} pioneer this direction; the former reveals inherent sparse patterns (e.g., temporal and spatial heads focusing on critical tokens), while the latter leverages semantics-aware permutation to maximize efficiency. Following this paradigm, several works identify similar structural sparsity patterns~\cite{chen2025sparse, li2025compact}, or directly unify SVG's heads into scalable radial attention~\cite{li2025radial}. To address the costly dynamic detection overhead of SVG, VMoBA~\cite{wu2025vmoba} adapts text-centric MoBA~\cite{lu2025moba} to capture spatiotemporal locality via layer-wise recurrent block partitions. Figure~\ref{fig:vmoba_compare} compares the inference time of some of the aforementioned methods. The aforementioned methods, along with several other variants~\cite{gu2025blade, zhang2025spargeattention, zhang2026spargeattention2, zhang2025trainingfree}, rely primarily on standard block sparse attention, which partitions queries/keys into blocks with a fixed size and computes attention for selected blocks only, skipping entire blocks to leverage hardware-efficient GPU kernels. However, recent advances explore new dimensions. For instance, FG-Attn~\cite{durvasula2025fg} challenges the block paradigm with finer-grained slice-level sparsity; to address overlooked query-side redundancy, BSA~\cite{zhan2025bidirectional} and Astraea~\cite{liu2025astraea} propose mechanisms to selectively prune queries alongside or instead of key-value pairs.

\subsubsection{Windowed Attention}
\begin{figure}
    \centering
    \includegraphics[width=1\linewidth]{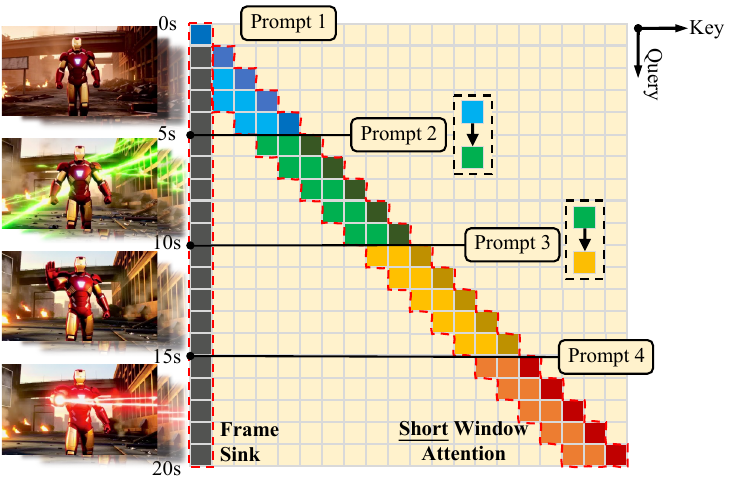}
    \caption{LongLive~\cite{yang2025longlive} processes sequential user prompts and generates a corresponding long video using efficient short window attention and frame sink. Figure courtesy of~\cite{yang2025longlive}.}
    \label{fig:longlive}
\end{figure}

Exploiting inherent spatiotemporal locality, window-based attention restricts computation to local neighborhoods to mitigate the complexity of full 3D attention. To address potential quality degradation caused by limited receptive fields, various works propose combining local windows with global context mechanisms. For instance, DiTFastAttn~\cite{yuan2024ditfastattn} caches the residual difference between full and windowed attention, while LongLive~\cite{yang2025longlive} combines short window attention with a frame sink mechanism as shown in Figure~\ref{fig:longlive} that permanently caches initial frames as global anchors to maintain long-range coherence during infinite-length streaming generation. Alternatively, several approaches explicitly decompose attention into local and global branches. Specifically, UltraGen~\cite{hu2025ultragen} enables native 4K generation via spatially compressed global attention alongside hierarchical cross-window local attention. Similarly, VORTA~\cite{sun2025vorta} and VideoNSA~\cite{song2025videonsa} employ learnable routing or gating mechanisms to dynamically fuse local sliding windows with global sparse or compressed tokens. Beyond receptive field limitations, standard sliding windows often suffer from hardware inefficiencies. To resolve this, STA~\cite{zhang2025fast} proposes a hardware-aware Sliding Tile Attention that aligns window strides with GPU tile sizes, successfully translating the theoretical FLOP reduction into significant wall-clock speedup.

\subsubsection{Linear Attention}
Linear attention mitigates the $O(N^2)$ complexity of standard self-attention by employing a kernel feature map $\phi(\cdot)$ and the associative property of matrix multiplication: $O = \phi(Q) (\phi(K)^\top V)$. This decoupling avoids the explicit $N \times N$ matrix, reducing complexity to $O(N)$. Recent video generation models integrate linear attention through various structural paradigms. At the global level, {SANA-Video}~\cite{chen2026sanavideo} entirely replaces vanilla attention with ReLU-based linear attention, enabling efficient block-wise autoregressive generation. At the layer level (serial integration), {LinVideo}~\cite{huang2025linvideo} adapts pre-trained models by selectively substituting a subset of quadratic attention layers with linear ones, optimizing via distribution matching. At the token level (parallel routing), {SLA}~\cite{zhang2025sla} decomposes attention weights to apply exact sparse attention to critical tokens and linear attention to the marginal majority. Building on these efficient formulations, models like {Yume-1.5}~\cite{mao2025yume} further apply linear attention to interactive long-video generation.

\subsubsection{State Space Models (SSMs)}
SSMs, particularly Mamba~\cite{gu2024mamba}, offer a linear-complexity $O(N)$ alternative to Transformers by modeling sequences through recurrent state transitions. LaMamba-Diff~\cite{Fu_2025_BMVC} designs a novel backbone for diffusion models for image generation.
LinGen~\cite{Wang_2025_CVPR} introduces a hybrid linear-complexity block that couples a bidirectional Mamba2~\cite{dao2024transformers} branch with a Temporal Swin Attention branch, achieving stable minute-length video generation with strictly linear scaling.

\subsection{Extrapolation and RoPE}
A true world model must simulate the future beyond its seen horizon, requiring modifications to Rotary Positional Embeddings (RoPE) to prevent distribution drift.
\subsubsection{Frequency-Based Extrapolation}
Early adaptations focused on RoPE frequency scaling. RIFLEx~\cite{zhao2025riflex} identifies that high-frequency components cause temporal repetition and proposes frequency shifting to enable 3$\times$ length extrapolation. This remains a simple, training-free baseline for extending temporal horizons.
\subsubsection{Mitigating Attention Dispersion}
UltraViCo~\cite{zhao2025ultravico} identifies ``attention dispersion"—where distant tokens dilute learned patterns—as the root cause of quality decay, introducing a constant decay factor to suppress distant scores. Compared to frequency-only scaling, this maintains better imaging quality at larger extrapolation limits.
\subsubsection{From Long to Infinite}
To enable effectively infinite simulation, Infinity-RoPE~\cite{yesiltepe2025infinity} proposes Block-Relativistic RoPE, rotating new latent blocks relative to a moving local reference frame. This shifts from ``extending a window" to a ``sliding world" paradigm.
Related works like FreeNoise~\cite{qiu2024freenoise} and Align your Latents~\cite{blattmann2023align} explore complementary tuning-free noise and attention rescheduling strategies.



\subsection{Discussion}
Despite significant advances, existing efficient architectures face fundamental trade-offs between computational cost and spatiotemporal/causal integrity. Specifically, hierarchical compression often sacrifices long-term semantic consistency for visual refinement, while training-free extrapolation techniques fail to maintain causal progression over long horizons, inevitably leading to motion decay or temporal loops. Furthermore, memory mechanisms face the stability-plasticity dilemma: how to retain a persistent global map (stability) while rapidly adapting to new, unexpected environmental changes (plasticity).  At the operational level, efficient attention variants frequently fail to translate theoretical complexity reduction into wall-clock acceleration. To overcome these bottlenecks, future research must shift toward physics-aware and adaptive paradigms. Promising directions include exploring physically constrained latent spaces, designing hybrid memory hierarchies that couple slowly updated global maps with agile working memories, and using interactive causal chains to replace absolute frame indices. Ultimately, realizing real-time, physics-compliant generation will necessitate hardware-software co-designed mechanisms that dynamically allocate compute based on semantic importance and motion dynamics.

\section{Efficient Inference}
\label{sec:efficient_inference}

As video generation models scale to billions of parameters and become capable of generating videos with extended durations (e.g., Seedance 1.0~\cite{gao2025seedance} with 30B parameters), running inference on a single GPU often faces severe memory bottlenecks and unacceptable latency. To address these challenges, 
efficient inference strategies focus on distributing the computational load, reducing redundant calculations and quantization. This section reviews four critical strategies: (i) Parallelism, which distributes inference across multiple devices via spatial, sequence, and pipeline partitioning; (ii) Caching, which exploits spatial and temporal redundancy to accelerate generation; (iii) Pruning, which directly mitigates sequence length explosion and architectural redundancy by merging tokens and streamlining networks; and (iv) Quantization, which lowers the precision of weights and activations to reduce computational resource and memory cost.

\subsection{Parallelism}
Parallel inference is critical for scaling video generation to high resolution, long duration, and real-time inference, since the computational and memory costs of diffusion transformers grow rapidly with sequence length. In practice, existing systems mainly improve inference efficiency through sequence-level partition, pipeline-style execution, and hybrid parallel frameworks, making it possible to generate in real-time with multi-GPUs.
A straightforward strategy is to split spatial or temporal tokens across multiple devices so that memory and computation can be distributed. In diffusion inference, DistriFusion~\cite{li2024distrifusion} shows that patch-wise distributed inference can be made efficient by reusing features from the previous denoising step, thereby overlapping communication with computation. For long-form video generation, Video-Infinity~\cite{chen2024videoinfinity} further extends this idea with clip parallelism and dual-scope attention, enabling distributed long-video generation across multiple GPUs.


Another complementary strategy is to partition model execution into a pipeline so that different devices process different parts of the workload concurrently. Rather than fully relying on one type of parallelism, recent DiT inference systems increasingly exploit pipeline-style execution at the patch level to improve device utilization and reduce end-to-end latency. Related system-oriented designs also appear in streaming avatar generation. For example, LiveAvatar~\cite{huang2025live} introduces timestep-forcing pipeline parallelism, which assigns different denoising timesteps to different devices and converts the diffusion chain into a high-throughput streaming pipeline.

Since no single parallel strategy is optimal under all hardware and model settings, unified frameworks have recently emerged to combine multiple forms of parallelism. xDiT~\cite{fang2024xdit} is a representative example, which integrates sequence parallelism, PipeFusion-style pipeline parallelism, and classifier-free guidance (CFG)~\cite{ho2022classifier} parallelism into a scalable inference engine for diffusion transformers.

\subsection{Caching}
\label{subsec:caching}

Caching methods accelerate video generation by exploiting redundancy across adjacent denoising steps. As diffusion inference proceeds through a sequence of timesteps, intermediate activations often evolve gradually, making it unnecessary to recompute all features from scratch at every step. In recent video generation systems, this direction has rapidly evolved from coarse feature reuse to more adaptive and fine-grained caching strategies.

Representative recent methods include PAB~\cite{zhao2025pab}, TeaCache~\cite{liu2024teacache}, FasterCache~\cite{lv2025fastercache}, and PreciseCache~\cite{wang2026precisecache}. PAB~\cite{zhao2025pab} accelerates video generation by broadcasting attention outputs in a pyramid manner across timesteps, based on the observation that attention redundancy varies across different stages and attention types. TeaCache~\cite{liu2024teacache} instead adopts a timestep-aware policy for video diffusion models rather than using a fixed cache interval. It estimates output variation from timestep-related signals and selectively reuses cached outputs when the predicted change is sufficiently small. FasterCache~\cite{lv2025fastercache} further improves training-free acceleration by combining dynamic feature reuse with classifier-free guidance (CFG)~\cite{ho2022classifier}-aware caching, reducing redundancy both across timesteps and between conditional and unconditional branches. More recently, PreciseCache~\cite{wang2026precisecache} combines step-wise and block-wise caching to skip only truly redundant computations, using low-frequency difference to identify step-level redundancy and then performing additional block-level reuse within non-skipped steps.
Table~\ref{tab:cache_perf} summarizes a compact comparison under the unified 4 A800 GPU setting.

While Table~\ref{tab:cache_perf} focuses on cache-based acceleration results for general video generation models, recent work has also begun to explore caching mechanisms for video-based world models. 
HERO~\cite{song2025hero} proposes a hybrid acceleration scheme for video generation based on multimodal data, such as depth and RGB views. It figures out that shallow layers, which exhibit larger variation, should be recomputed more frequently, whereas deeper and more stable layers can be accelerated through linear extrapolation from preceding timesteps, effectively reducing attention computation. 
More recently, WorldCache~\cite{feng2026worldcache} explicitly targets video-based world models and identifies two world-model-specific obstacles for caching, namely heterogeneous token behavior caused by multimodal coupling and non-uniform temporal dynamics where a small subset of hard tokens dominates error accumulation. To address these issues, it introduces curvature-guided heterogeneous token prediction together with chaotic-prioritized adaptive skipping, achieving up to 3.7$\times$ end-to-end speedup while maintaining 98\% rollout quality. 
\begin{table}[t]
\centering
\renewcommand{\arraystretch}{1.15}
\setlength{\tabcolsep}{4pt}
\caption{Compact comparison of representative cache-based acceleration methods under the unified 4 A800 GPU setting, reported by PreciseCache~\cite{wang2026precisecache}.}
\resizebox{0.48\textwidth}{!}{
\begin{tabular}{L{0.14\textwidth}|L{0.13\textwidth}|L{0.10\textwidth}|L{0.10\textwidth}|L{0.10\textwidth}}
\toprule
Benchmark Block & Method & VBench~\cite{Huang_2024_VBench} $\uparrow$ & Speedup $\uparrow$ & Latency (s) $\downarrow$ \\
\midrule
Open-Sora 1.2~\cite{zheng2024open}
& Original & 78.79\% & 1.00$\times$ & 47.23 \\
& PAB~\cite{zhao2025pab} & 78.15\% & 1.26$\times$ & 38.40 \\
& TeaCache~\cite{liu2024teacache} & 78.23\% & 1.95$\times$ & 24.73 \\
& FasterCache~\cite{lv2025fastercache} & 78.46\% & 1.67$\times$ & 29.15 \\
& PreciseCache-Flash~\cite{wang2026precisecache} & 78.19\% & 2.60$\times$ & 18.38 \\
\midrule
HunyuanVideo~\cite{kong2024hunyuanvideo}
& Original & 80.66\% & 1.00$\times$ & 73.64 \\
& PAB~\cite{zhao2025pab} & 79.37 & 1.35$\times$ & 54.54 \\
& TeaCache~\cite{liu2024teacache} & 80.51\% & 1.64$\times$ & 44.90 \\
& FasterCache~\cite{lv2025fastercache} & 80.59\% & 1.43$\times$ & 51.50 \\
& PreciseCache-Turbo~\cite{wang2026precisecache} & 80.49\% & 1.95$\times$ & 37.76 \\
\bottomrule
\end{tabular}}
\label{tab:cache_perf}
\end{table}

\subsection{Pruning}
Pruning techniques in video diffusion models aim to reduce computational burden by eliminating redundant information at the token, channel, or layer level. To tackle the huge computational overhead introduced by the spatial resolution and temporal depth of video data, recent approaches exploit redundancies across both the video content and the diffusion generation process. We categorize these techniques into two primary paradigms: token-level reduction, which addresses sequence length explosion by merging or dropping redundant visual tokens, and structural pruning, which streamlines the model architecture by removing network components or reallocating compute based on layer roles.

\subsubsection{Token-Level Reduction}
\begin{figure}
    \centering
    \includegraphics[width=1\linewidth]{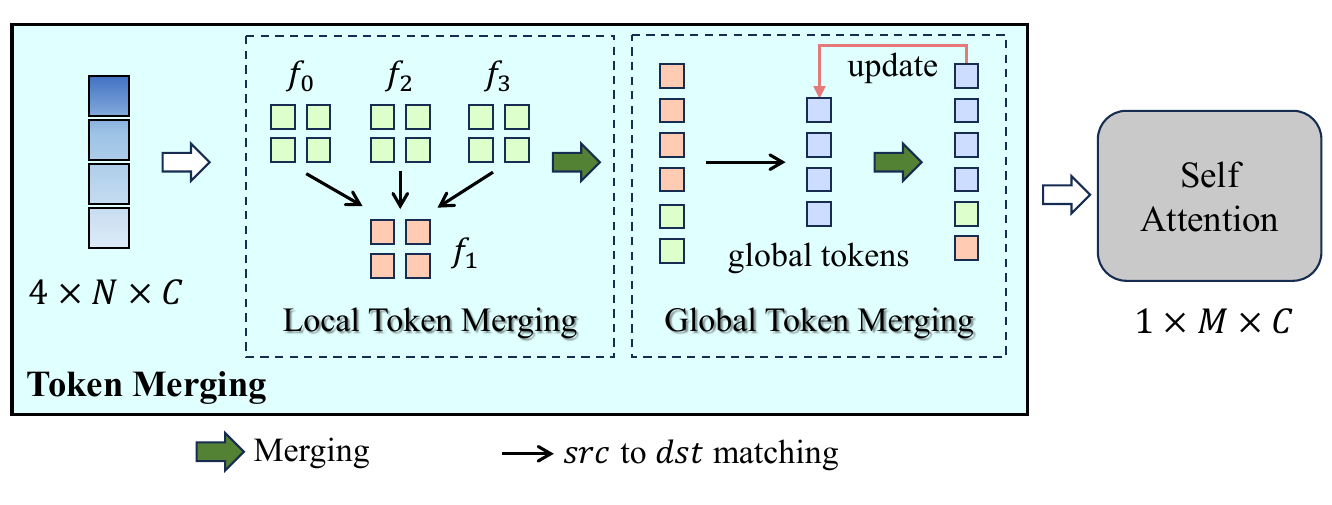}
    \caption{VidToMe~\cite{li2024vidtome} first merges tokens locally and then combines the merged tokens with the maintained global tokens. Figure courtesy of~\cite{li2024vidtome}.}
    \label{fig:vidtome}
\end{figure}

The high resolution and temporal depth of video data result in an explosion of tokens in diffusion models. To address this, early representative work such as {VidToMe}~\cite{li2024vidtome} utilizes ToMe~\cite{bolya2023token}'s bipartite matching algorithm to merge redundant self-attention tokens across video frames (Figure~\ref{fig:vidtome})). The algorithm partitions the tokens into a pair of source (\textit{src}) and destination (\textit{dst}) sets, and the source tokens are linked to their most similar tokens in the destination set. Specifically, VidToMe divides the video into frame chunks, mapping temporally correlated tokens from adjacent frames into a single target frame (intra-chunk local merging), and further links them with a persistent set of tokens across different chunks (inter-chunk global merging). Building on this foundation, importance-based token merging~\cite{wu2025importance} argues that \textit{dst} tokens are chosen randomly within predefined regions in existing methods, degrading generation quality due to the elimination of critical semantic details. To explicitly preserve informative regions, it leverages classifier-free guidance (CFG)~\cite{ho2022classifier} magnitudes as a cost-free indicator to construct a dynamic pool of important tokens and samples destination tokens from this pool.
To overcome the distortion and pixelation caused by extending ToMe methods, {AsymRnR}~\cite{sun2025asymrnr} introduces a training-free asymmetric reduction strategy that independently merges Q,K,V tokens at adaptive rates tailored for specific network blocks and denoising timesteps. Designed for resource-constrained mobile environments, On-device Sora~\cite{kim2025device} introduces temporal dimension token merging (TDTM) that explicitly averages consecutive tokens along the temporal axis to effectively halve the sequence length for attention computation.

\subsubsection{Structural Pruning}
Beyond reducing token counts, structural pruning directly targets the model architecture to reduce computation.
Early representative strategies target macro-level depth and temporal redundancy. For static architectural pruning, MobileVD~\cite{yahia2024mobile} introduces a learnable gating mechanism to prune redundant temporal blocks alongside a channel funneling strategy to compress layer widths during inference. {VDMini}~\cite{wu2025individual} empirically observes that shallow layers primarily focus on individual frame content, while deeper layers dictate temporal motion dynamics; consequently, it selectively prunes redundant shallow blocks and restores generation quality through a loss based on individual content and motion dynamics (ICMD). SnapGen-V~\cite{wu2025snapgen} also performs a lot of architecture searches.
Contrasting with VDMini's macro-level block removal, {Mobile Video DiT}~\cite{wu2025taming} executes a sensitivity-aware tri-level static pruning, simultaneously targeting blocks, attention heads, and FFN channels. Concurrently, {UniCP}~\cite{sun2025unicp} introduces finer dynamic pruning at the attention-matrix level. 


\subsection{Quantization}
Quantization reduces the precision of weights and activations to accelerate inference and lower memory usage, which is critical for deploying large-scale video-based world models. We categorize recent advances into attention-centric optimization, post-training quantization, quantization-aware training, and dynamic scheduling strategies.
\begin{table*}[t]
\footnotesize
\vspace{-2.5mm}
\centering
\caption{Performance comparison of quantization methods on VBench~\cite{Huang_2024_VBench} across multiple base models. Reported by DVD-Quant~\cite{li2025dvd}, {LRQ-DiT}~\cite{yang2025lrq} and {QVGen}~\cite{huang2026qvgen}.}
\vspace{-2mm}
\renewcommand{\arraystretch}{0.98} 
\resizebox{\textwidth}{!}{%
\begin{tabular}{l c c c c c c c c c} 
\toprule
Method & \makecell{Bit-width \\ (W/A)} & \makecell{Aesthetic \\ Quality} $\uparrow$ & \makecell{Imaging \\ Quality} $\uparrow$ & \makecell{Overall \\ Consistency} $\uparrow$ & \makecell{Scene \\ Consistency} $\uparrow$ & \makecell{Background \\ Consistency} $\uparrow$ & \makecell{Subject \\ Consistency} $\uparrow$ & \makecell{Dynamic \\ Degree} $\uparrow$ & \makecell{Motion \\ Smoothness } $\uparrow$ \\

\midrule \midrule
\multicolumn{10}{c}{\cellcolor{gray!15}{Base Model: HunyuanVideo~\cite{kong2024hunyuanvideo}}} \\
\midrule
BF16 Baseline    & 16/16 & 62.53 & 64.78 & 25.86 & 42.81 & 97.01 & 96.05 & 51.39 & 99.30 \\
\midrule
ViDiT-Q~\cite{zhao2025viditq}     & 4/8  & 57.01 & 59.74 & 24.77 & 27.11 & 97.37 & 95.16 & 48.61 & 99.06 \\
DVD-Quant~\cite{li2025dvd}   & 4/6  & 62.27 & 64.22 & 25.83 & 33.07 & 97.89 & 96.57 & 58.33 & 99.05 \\
\midrule
ViDiT-Q~\cite{zhao2025viditq}     & 4/4  & 45.36 & 40.10 & 19.66 & 7.85 & 97.19 & 97.29 & 0.00 & 99.43 \\
DVD-Quant~\cite{li2025dvd}                  & 4/4  & 61.96 & 61.82 & 25.68 & 29.94 & 97.82 & 96.61 & 56.94 & 99.15 \\

\midrule \midrule
\multicolumn{10}{c}{\cellcolor{gray!15}{Base Model: Open-Sora 1.2~\cite{zheng2024open}}} \\
\midrule
ViDiT-Q~\cite{zhao2025viditq}    & 4/6  & 50.89 & 55.57 & 25.98 & 36.77 & 96.52 & 94.83 & 52.77 & 98.66 \\
LRQ-DiT~\cite{yang2025lrq}                 & 4/6  & 52.25 & 56.57 & 26.68 & 41.28 & 96.90 & 95.28 & 48.62 & 98.85 \\
\midrule
ViDiT-Q~\cite{zhao2025viditq}        & 4/4  & 47.30 & 51.60 & 25.84 & 35.61 & 95.27 & 92.01 & 54.16 & 98.14 \\
LRQ-DiT~\cite{yang2025lrq}                 & 4/4  & 47.95 & 51.79 & 25.87 & 37.80 & 95.56 & 92.87 & 55.56 & 98.34 \\

\midrule \midrule
\multicolumn{10}{c}{\cellcolor{gray!15}{Base Model: CogVideoX-2B~\cite{yang2025cogvideox}}} \\
\midrule
BF16 Baseline    & 16/16 & 54.49 & 59.15 & 25.06 & 36.24 & 94.79 & 92.82 & 67.78 & 97.43 \\
\midrule
ViDiT-Q~\cite{zhao2025viditq}     & 4/6  & 43.01 & 54.72 & 20.41 & 26.25 & 90.76 & 81.02 & 43.22 & 92.18 \\
QVGen~\cite{huang2026qvgen}   & 4/4  & 54.61 & 60.16 & 24.61 & 31.42 & 94.38 & 93.01 & 67.22 & 98.06 \\

\bottomrule
\end{tabular}%
}
\label{tab:ptq_comparison}
\end{table*}

\subsubsection{Attention-Centric Quantization}
Attention mechanisms dominate the computational cost in video generation models, driving a rapid evolution from 8-bit to 4-bit precision. 
Early representative works such as {SageAttention}~\cite{zhang2025sageattention} and {FPSAttention}~\cite{liu2025fpsattention} established the baseline; SageAttention employs 8-bit quantization with smooth matrix to handle outliers, while FPSAttention co-designs FP8 quantization with sparsity constraints.
Building on this, the \textit{SageAttention} series has pushed the limits of low-bit inference: {SageAttention2}~\cite{zhang2025sageattention2} achieves INT4 precision by introducing per-thread quantization and thorough outlier smoothing; {SageAttention2++}~\cite{zhang2025sageattention2++} further optimizes kernel performance by utilizing faster FP8 matrix multiplication instructions accumulated in FP16.
The most recent member in this series, {SageAttention3}~\cite{zhang2025sageattention3}, introduces FP4 microscaling attention tailored for next-generation hardware (e.g., RTX 5090), effectively achieving extreme compression with negligible quality loss.

\subsubsection{Post-Training Quantization}
Quantizing DiTs is challenging due to significant outliers in activations, requiring robust post-training quantization (PTQ) techniques.
{ViDiT-Q}~\cite{zhao2025viditq} serves as a representative method in this domain, which addresses oscillating activations in video models through specialized metric-aware rounding.
Subsequent work further refines these methods: {DVD-Quant}~\cite{li2025dvd} extends quantization to a data-free setting by reconstructing calibration data to handle temporal dependencies, and {LRQ-DiT}~\cite{yang2025lrq} tackles the dual challenges of long-tailed, Gaussian-like weight distributions and diverse activation outliers by introducing twin-log quantization along with an adaptive rotation scheme. 
These methods collectively pave the way for the deployment of INT4-level DiTs in production environments.

\subsubsection{Quantization-Aware Training}
While PTQ methods offer efficient deployment, they often suffer from severe performance degradation when pushing video generation models to ultra-low precision (e.g., $\le 4$-bit). To bridge this performance gap, quantization-aware training (QAT) has emerged as a promising direction.
As a pioneering work in this paradigm, {QVGen}~\cite{huang2026qvgen} introduces a novel QAT framework tailored for video diffusion models under extremely low-bit settings (e.g., W4A4 and W3A3). Table~\ref{tab:ptq_comparison} shows a performance comparison of some PTQ and QAT methods.

\subsubsection{Dynamic and Temporal Quantization Strategies}
Video generation involves temporal redundancy and multi-step denoising, offering opportunities for adaptive precision.
Focusing on the temporal dimension, {QuantCache}~\cite{Wu_2025_ICCV} advances this concept by implementing an adaptive importance-guided quantization specifically for the KV cache and hierarchical latents, effectively exploiting the similarity between video frames to reduce memory bandwidth.
Conversely, addressing the temporal heterogeneity across denoising timesteps, {AdaTSQ}~\cite{zhang2026adatsq} introduces a timestep-dynamic quantization framework. By leveraging Fisher information to evaluate the varying sensitivity of different diffusion phases, AdaTSQ dynamically allocates bit-widths via Pareto-aware beam search. Coupled with a Fisher-guided temporal calibration mechanism, this strategy pushes the Pareto frontier of efficiency and quality for video generation models.


\subsection{Discussion}
Efficient video inference focuses on reducing per-step latency and scaling to high-fidelity, long-horizon generation. 
The four directions reviewed in this section address this problem from complementary perspectives: parallelism distributes computation across devices, caching reuses intermediate features across denoising steps, pruning removes redundant tokens or network components, and quantization reduces the precision cost of weights and activations. 
However, these techniques are not independent. For instance, aggressive caching or pruning may amplify approximation or accumulation errors in dynamic regions, while low-bit quantization can further destabilize activations already altered by token reduction or feature reuse. 
For video-based world models, this problem is more challenging because inference must support not only short clips, but also long-horizon and interactive generation. In practice, this means that parallelism, caching, pruning, and quantization should work together rather than be applied separately. Future methods should therefore improve both efficiency and stability, especially for long-duration interactive scenarios.


\begin{table*}[t]
\centering
\renewcommand{\arraystretch}{1.5}
\setlength{\tabcolsep}{10pt}      
\caption{Applications of Video-Based World Models}
\resizebox{0.95\textwidth}{!}{
\begin{tabular}{L{0.14\textwidth}|L{0.27\textwidth}|L{0.27\textwidth}|L{0.27\textwidth}}
\toprule
Application & Data Synthesis & Interactive Simulation & Generative Planning \\
\midrule
Autonomous Driving &
GAIA~\cite{hu2023gaia,russell2025gaia,wayve_gaia3_2025}, DriveDreamer4D~\cite{zhao2025drivedreamer4d}, InfinityDrive~\cite{guo2024infinitydrive}, Glad~\cite{xie2025glad}, STAGE~\cite{wang2025stage}, UniScene~\cite{li2025uniscene}, WorldSplat~\cite{zhu2025worldsplat}, EOT-WM~\cite{zhu2025other}, WoVoGen~\cite{lu2024wovogen},Cosmos-Drive-Dreams~\cite{ren2025cosmosdrivedreamsscalablesyntheticdriving} &
Drive-WM~\cite{wang2024driving}, Vista~\cite{gao2024vista}, MiLA~\cite{wang2025mila}, ADriver-I~\cite{jia2023adriverigeneralworldmodel},~\cite{santana2016learning}, Drivedreamer~\cite{wang2024drivedreamer}, MagicDrive-V2~\cite{gao2025magicdrive}, DriveArena~\cite{yang2025drivearena}, MAD~\cite{rahimi2026mad} &
Epona~\cite{zhang2025epona}, GenAD~\cite{yang2024genadgeneralizedpredictivemodel}, DriveLaW~\cite{xia2025drivelaw}, DrivingGPT~\cite{chen2025drivinggpt}, VaVAM~\cite{bartoccioni2025vavim}\\
\midrule
Embodied AI &
Vidar~\cite{feng2025vidar}, DreamGen~\cite{jang2025dreamgen},  GenMimic~\cite{ni2025generated}, RBench~\cite{deng2026rethinking}, GigaWorld-0~\cite{team2025gigaworld}, RIGVid~\cite{patel2025robotic}, LuciBot~\cite{qiu2025lucibot}, Gen2Act~\cite{bharadhwaj2024gen2act}, Dreamitate~\cite{liang2024dreamitate} &
World-Env~\cite{xiao2025world}, EVAC~\cite{jiang2025enerverse}, Ctrl-World~\cite{guo2025ctrl}, VideoAgent~\cite{soni2024videoagent}, VIPER~\cite{escontrela2023video}, WorldEval~\cite{li2025worldeval}, Genie Envisioner~\cite{ge2025}, World-Gymnast~\cite{sharma2026world}, DreamDojo~\cite{gao2026dreamdojogeneralistrobotworld} &
GR-1~\cite{wu2023unleashing}, VILP~\cite{xu2025vilp}, UVA~\cite{li2025unified}, RoboEnvision~\cite{yang2025roboenvision}, GEVRM~\cite{zhang2025gevrm},  EnerVerse~\cite{huang2025enerverseenvisioningembodiedfuture}, LingBot-VA~\cite{lingbotva2026}, Cosmos Policy~\cite{kim2026cosmospolicyfinetuningvideo},Fast-WAM~\cite{yuan2026fastwam},LeWorldModel~\cite{maes2026leworldmodel},DreamZero~\cite{ye2026dreamzero}\\
\midrule
Game \& Interactive World Simulation &
\multicolumn{3}{L{0.81\textwidth}}{%
GameGen-X~\cite{che2024gamegen}, GameFactory~\cite{yu2025gamefactory},
MineWorld~\cite{guo2025mineworld}, Matrix-Game~\cite{zhang2025matrixgame,he2025matrix},
GenieRedux-G~\cite{savov2025exploration}, Hunyuan-GameCraft
\cite{li2025hunyuangamecrafthighdynamicinteractivegame,tang2025hunyuan}, PlayGen~\cite{yang2024playablegamegeneration}, WorldPlay~\cite{sun2025worldplay}, Yume1.5~\cite{mao2025yume}, LingBot-World~\cite{robbyantteam2026advancingopensourceworldmodels}, Cosmos-Predict2.5~\cite{ali2025world}, Dreamer 4~\cite{hafner2025dreamerv4}, Genie 3~\cite{deepmind_genie3_models}
}\\
\bottomrule
\end{tabular}%
}
\vspace{-5mm}
\label{tab:application}
\end{table*}
\section{Applications}
\label{sec:applications}
World modeling via efficient video generation has been widely applied to domains including autonomous driving, embodied AI, and interactive game simulation, supporting tasks such as data synthesis, interactive simulation, and generative planning (Table~\ref{tab:application}). In such applications, online generation is often used to facilitate reinforcement learning, while offline data are more commonly used for supervised training.

\subsection{Autonomous Driving}
\subsubsection{Data Synthesis}
\begin{figure}
    \centering
    \includegraphics[width=1\linewidth]{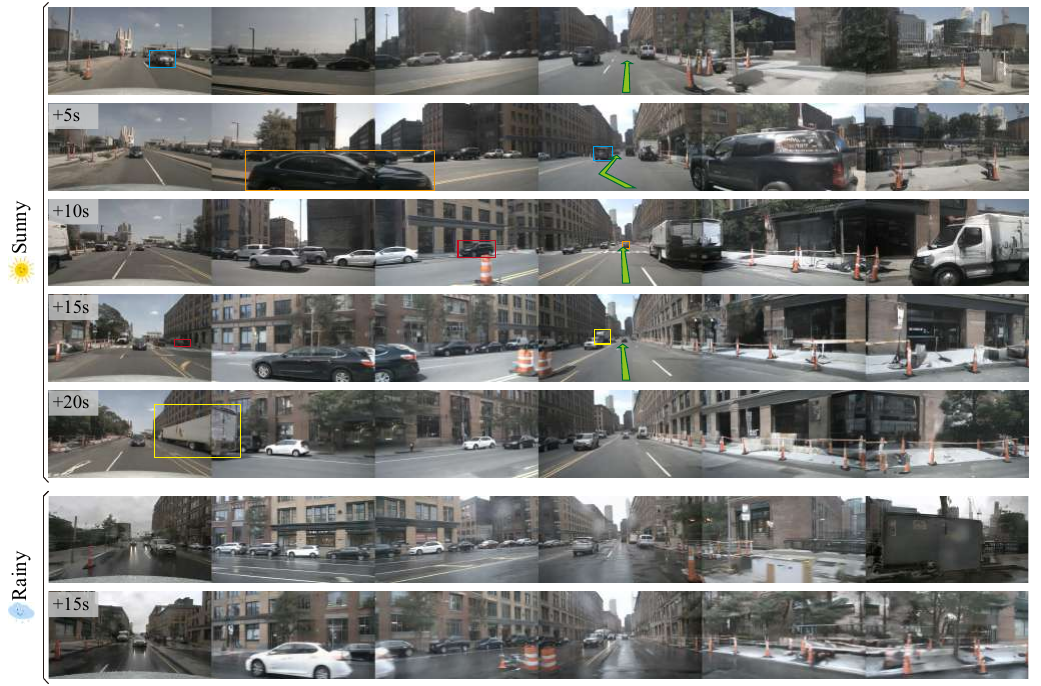}
    \caption{Examples of street-scene videos generated by MagicDrive-V2, which supports conditional generation with multiple types of control signals (e.g., road maps, object boxes, ego trajectories, and text). Figure courtesy of~\cite{gao2025magicdrive}.}
    \label{fig:MagicDrive}
\end{figure}

In autonomous driving, video-based world models improve coverage of long-tail and safety-critical scenarios by generating realistic, controllable driving videos that can be used as synthetic training data for perception, prediction, and planning, as well as evaluation data for testing robustness and safety under rare or hazardous conditions (Figure~\ref{fig:MagicDrive}). The GAIA series~\cite{hu2023gaia,russell2025gaia,wayve_gaia3_2025} advances generative world modeling for autonomous driving: GAIA-1 demonstrated that models can learn from video, text, and actions to generate realistic driving scenarios; GAIA-2 added stronger controllability, broader geographic coverage, and multi-camera scene generation across diverse vehicle embodiments. GAIA-3 combines the realism of real-world driving data with the controllability of simulation, allowing authentic driving sequences to be replayed with modifications-—for example, altering the trajectory of the ego vehicle while making every other element in the scene consistent. DriveDreamer4D~\cite{zhao2025drivedreamer4d} leverages world-model priors to enhance 4D driving scene representations. InfinityDrive~\cite{guo2024infinitydrive} introduces a spatiotemporal co-modeling module and an extended temporal training strategy, producing high-resolution spatiotemporally consistent videos.

\subsubsection{Interactive Simulation}
Some works integrate video-based world models into closed-loop interaction pipelines, rolling out action-conditioned generation for interactive simulation and evaluation. Vista~\cite{gao2024vista} generates realistic and temporally continuous videos at high spatiotemporal resolution and supports diverse behavior-conditioned control. MiLA~\cite{wang2025mila} adopts a coarse-to-fine approach to stabilize video generation and correct distortions in dynamic objects. ADriver-I~\cite{jia2023adriverigeneralworldmodel} enables infinite autonomous driving within a virtual world created by a video generation model.

\subsubsection{Generative Planning}
Some works explore generative planning by using video-based world models to assist action selection during inference, while others leverage them as an auxiliary training objective. Drive-WM~\cite{wang2024driving} can roll out multiple trajectories under different driving actions and select the best trajectory using image-based rewards. As an auxiliary training signal, Epona~\cite{zhang2025epona} explicitly integrates trajectory prediction into a video generation framework, using a dual-branch diffusion model to separately generate trajectories and video frames, and supports trajectory-only planning to improve real-time performance. GenAD~\cite{yang2024genadgeneralizedpredictivemodel} can generalize in a zero-shot manner to diverse unseen driving datasets, and be adapted into a motion planner or an action-conditioned generation model for future frames, highlighting its great value for real-world autonomous driving applications. By sharing a latent space, DriveLAW~\cite{xia2025drivelaw} treats the video model as a feature generator by directly injecting the latent representation produced by the Video DiT into the Action DiT. This chained design allows the planner (Action DiT) to better exploit the world modeling capability of the video model.
\begin{figure}
    \centering
    \includegraphics[width=0.99\linewidth]{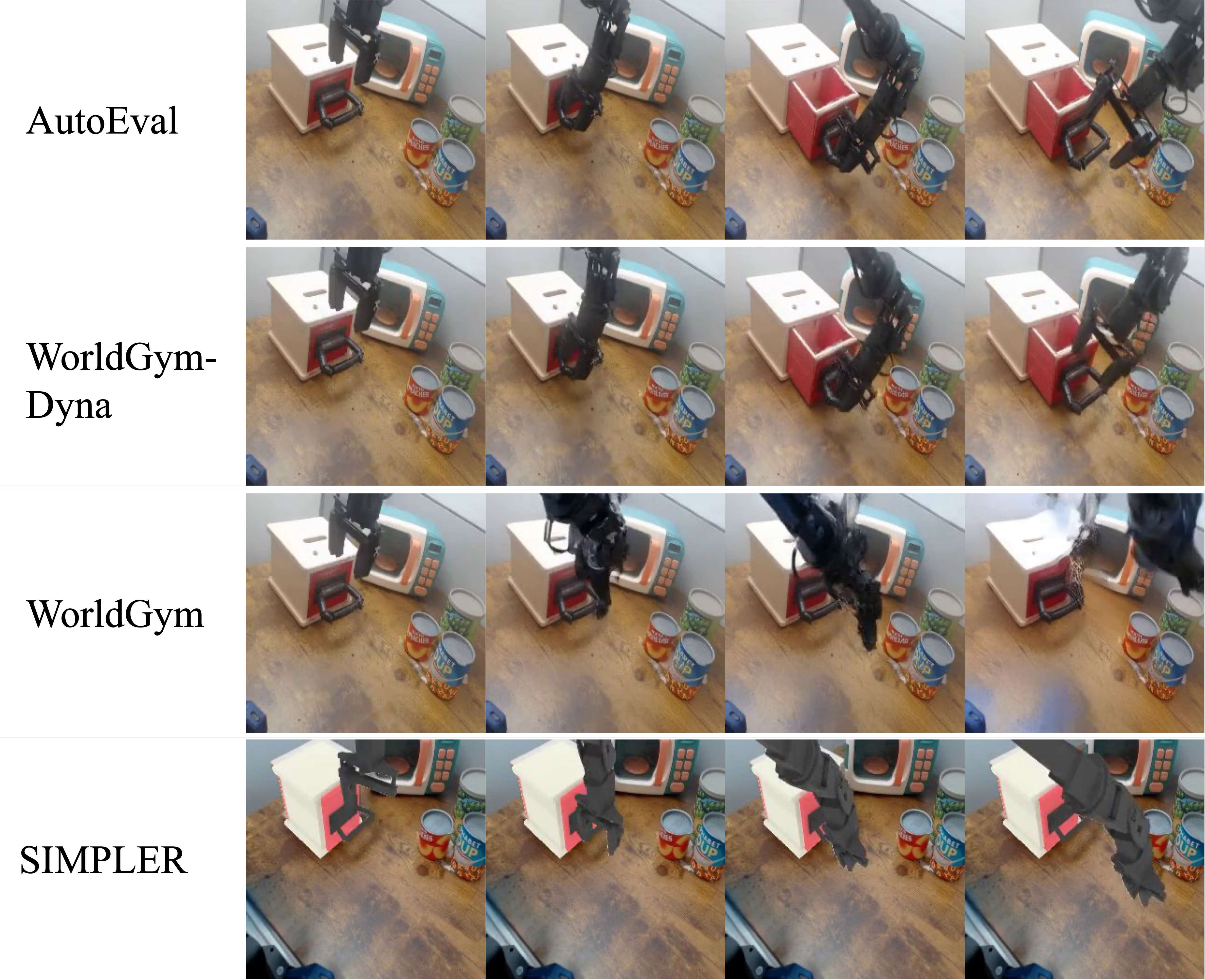}
    \caption{Comparison of videos generated by video-based world models on the same robot action sequence. Figure courtesy of~\cite{sharma2026world}.}
    \label{fig:rollouts}
\end{figure}
\subsection{Embodied AI}
\subsubsection{Data Synthesis}
In embodied AI, video world models can serve as a data engine to augment training data, covering broader distributions and rare cases to improve policy generalization in dynamic and long-tail tasks (Figure~\ref{fig:rollouts}). GigaWorld-0~\cite{team2025gigaworld} modifies real-world videos through text-guided editing and promotes sim-to-real transfer, helping bridge the simulation-to-reality gap. DreamGen~\cite{jang2025dreamgen} forms a synthetic-data loop by turning world-model rollouts into trajectory-style supervision, enabling diverse sample generation based on data from a single real environment. To mitigate the sim-to-real gap, GenMimic~\cite{ni2025generated} first lifts videos of human movements to 4D reconstructions, then retargets the extracted human motion to humanoid embodiments, and finally trains reinforcement learning policies for robust motion imitation. 

\subsubsection{Interactive Simulation}
As an interactive environment simulator, a video-based world model can support stable action-conditioned rollout generation for reinforcement learning and facilitate real-time evaluation of generated trajectories, allowing safe and reproducible policy testing and improvement. World-Env~\cite{xiao2025world} couples a video simulator with VLM-guided reflection to provide dense rewards and completion-based termination; EVAC~\cite{jiang2025enerverse} generates multi-view, controllable observations as a low-cost evaluation proxy. Ctrl-World~\cite{guo2025ctrl} provides a world simulator that enables robots to evaluate and improve their manipulation skills in a virtual environment. DreamDojo~\cite{gao2026dreamdojogeneralistrobotworld} is pretrained on 44k hours of human videos to learn physical dynamics without explicit action labels, and is then post-trained on robot data for downstream adaptation.

\subsubsection{Generative Planning}
Video world models can be extended to world action models (WAM) that support robot policy learning and action generation. One line of work jointly models future video frames and actions, leveraging shared representations between video generation and action prediction to improve policy learning and enhance scene dynamics understanding. DreamZero~\cite{ye2026dreamzero} builds a large-scale WAM based on a pretrained video diffusion backbone and jointly predicts future video frames and actions, showing strong zero-shot generalization, real-time closed-loop control, and cross-embodiment transfer. UVA~\cite{li2025unified} models video frames and actions in a shared latent space with two lightweight diffusion decoders, enabling action-only inference and flexible task switching via random masking. Fast-WAM~\cite{yuan2026fastwam} provides evidence through controlled ablation studies that the gains of video world models stem primarily from the video co-training objective shaping physical representations during training, rather than from explicit future imagination at test time. Another line of work first generates future visual trajectories and then predicts actions according to the generated trajectories. VILP~\cite{xu2025vilp} learns action prediction from generated videos via imitation learning, while enabling real-time receding-horizon control. RoboEnvision~\cite{yang2025roboenvision} generates keyframes via instruction decomposition plus interpolation for long-horizon consistency, then regresses joint controls with a lightweight policy. Based on JEPA~\cite{lecun2022path}, LeWorldModel~\cite{maes2026leworldmodel} proposes a stable end-to-end latent world model that avoids representation collapse using only a prediction loss and a SIGReg regularizer enforcing Gaussian-distributed embeddings, enabling efficient latent planning with only 15M parameters.

\subsection{Game \& Interactive World Simulation}
Efficient video generation provides critical support for interactive world simulation, and games have become a common deployment setting because of their well-defined interaction interfaces and controllable closed-loop evaluation. 

Among representative works, GameGen-X~\cite{che2024gamegen} targets open-world game videos, injecting keyboard actions and multimodal instructions into the generation process to improve interactive responsiveness over long sequences. GameFactory~\cite{yu2025gamefactory} models action control independently of the game genre to enable action-conditioned interactive video generation for diverse open-world scenarios. Focusing on Minecraft, MineWorld~\cite{guo2025mineworld} increases interactive frame rates by alleviating the throughput bottleneck of autoregressive tokens via parallel decoding. Matrix-Game 2.0~\cite{he2025matrix}, trained on data from GTA5 and Unreal Engine, reports interactive generation at around 25 frames per second and supports minute-level long rollouts. 
DreamerV4~\cite{hafner2025dreamerv4} uses a video-based world model as an interactive environment for reinforcement learning, allowing the agent to practice complex long-horizon tasks.

Toward more general interactive world generation, existing methods typically combine streaming generation with contextual memory to support long-term exploration, and rely on architectural choices and inference acceleration to meet real-time requirements. WorldPlay~\cite{sun2025worldplay} emphasizes high-resolution real-time generation and long-term consistency under action conditioning. Yume1.5~\cite{mao2025yume} focuses on text controllability and event editing, reducing long-context latency through context compression and distillation. LingBot-World~\cite{robbyantteam2026advancingopensourceworldmodels} is an open-source world simulator that combines a hierarchical semantic data engine with multi-stage training for low-latency interaction and long-term memory. 

\subsection{Discussion}
Video generation models, empowered by strong world modeling capabilities, can predict future observations conditioned on actions or instructions. In autonomous driving and embodied AI, a clear trend is the gradual convergence of data generation and interactive simulation: during closed-loop interaction and rollout-based prediction, the model continuously produces new samples and hard cases, forming a “generate–evaluate–retrain” loop for policy training and shifting data provisioning from offline augmentation to online iteration. Meanwhile, video-based world models are also moving toward supporting the full pipeline of data, model, and evaluation. Representative works such as Genie Envisioner~\cite{ge2025}, Cosmos~\cite{ali2025world}, and LingBot~\cite{lingbotva2026,robbyantteam2026advancingopensourceworldmodels} attempt to integrate data generation, interactive simulation, feedback, and policy optimization within a single generative framework, reducing cross-platform adaptation costs and enabling more systematic and reproducible evaluation and training paradigms.

\begin{figure}[]
    \centering
    \includegraphics[width=0.99\linewidth]{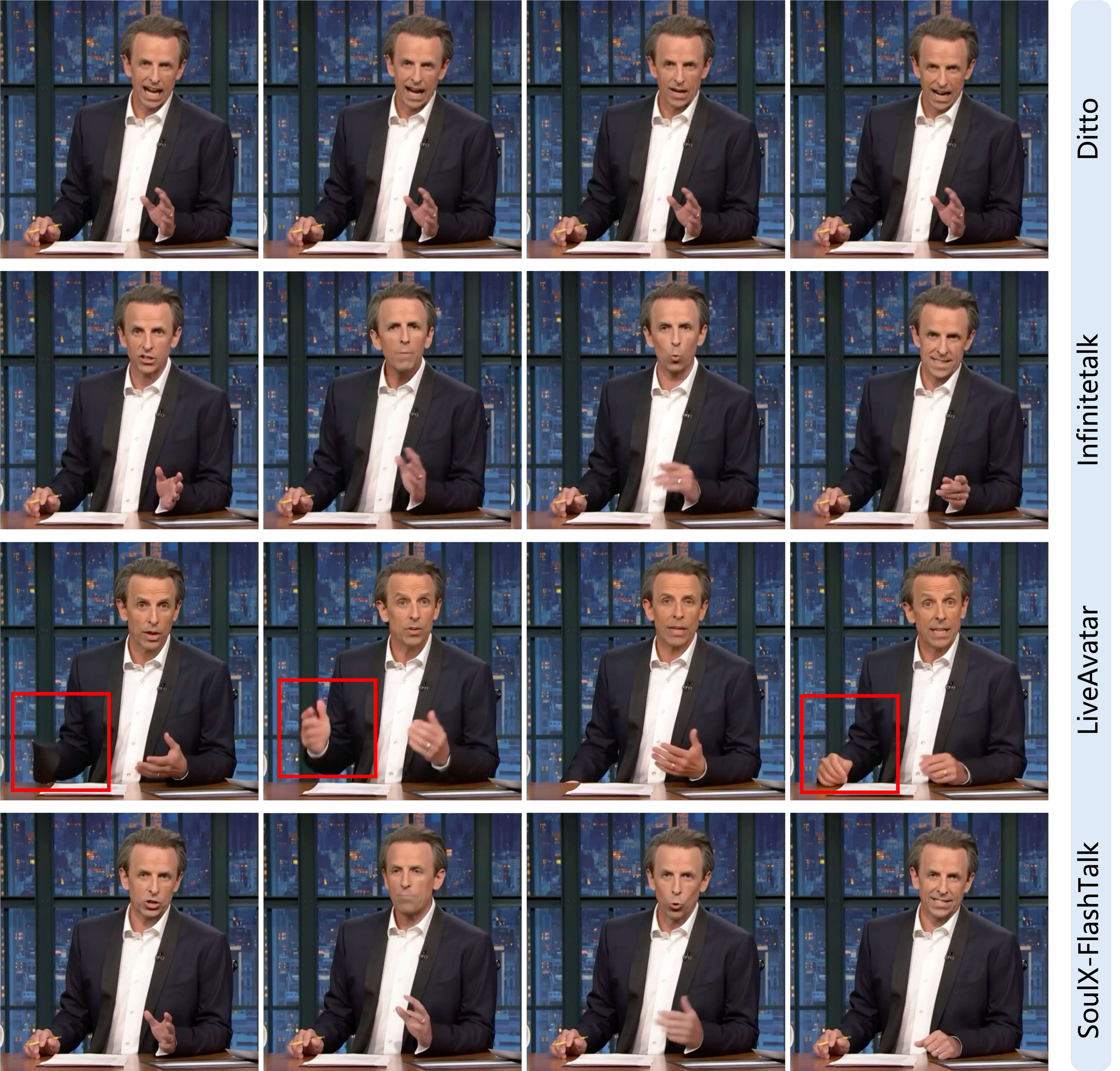}
    \caption{Comparison of videos generated by methods designed for real-time talking head generation. Existing methods generally preserve identity consistency well, but failure cases remain: Ditto tends to produce limited facial motion, while LiveAvatar may introduce local factual inconsistencies or artifacts (highlighted in red). Figure courtesy of~\cite{shen2025soulx}.}
    \label{fig:talkinghead}
\end{figure}
\section{More Related Work}
\label{sec:more_related}
\subsection{Interactive Talking Head Generation}
Recent advances in talking head generation have increasingly focused on interaction, streaming, and long-horizon conversation rather than traditional offline portrait synthesis. 

An early representative method, INFP~\cite{zhu2024infp}, explicitly models speaker–listener interaction in dyadic conversations, capturing speaking and listening behaviors within a shared motion latent space. In parallel, efficiency-oriented methods aim to reduce diffusion cost rather than redesign generation paradigms. For instance, Ditto~\cite{li2025ditto} performs diffusion in a compact motion space to achieve controllable real-time synthesis, while OSA-LCM~\cite{guo2024real} compresses multi-step diffusion into a one-step latent consistency model to further accelerate expressive portrait generation. More recent work extends beyond portrait-level interaction toward streaming diffusion frameworks. InfiniteTalk~\cite{yang2025infinitetalkaudiodrivenvideogeneration} adopts a sparse-frame paradigm, where key reference frames anchor identity and motion style, while context frames enable stable long-horizon synthesis. Similarly, SoulX-FlashTalk~\cite{shen2025soulx} adopts self-correcting bidirectional distillation to preserve temporal coherence during long-form avatar streaming. At the system and architecture level, LiveAvatar~\cite{huang2025live} demonstrates that algorithm–system co-design by exploiting timestep-forcing pipeline parallelism and the rolling sink mechanism can enable large-scale diffusion models to operate in real-time streaming settings, while StreamAvatar~\cite{sun2025streamavatarstreamingdiffusionmodels} proposes a two-stage autoregressive adaptation framework that converts non-causal human video diffusion models into block-causal streaming generators with improved long-term stability.

As illustrated in Figure~\ref{fig:talkinghead}, recent advances mark a clear paradigm shift from traditional offline portrait synthesis to causal and streaming real-time talking head generation, although challenges such as limited motion diversity or local artifacts still remain in existing approaches.

\subsection{Interactive Content Creation}
For AI content creation~\cite{guoedit}, creators often iterate rapidly by repeatedly refining prompts or other input to rewrite structure, swap characters, or adjust shot pacing, making efficient video generation crucial for shifting from offline processing to interactive workflows. 

For video editing, Edit-Your-Interest~\cite{zuo2025edit} caches and dynamically updates spatial attention feature tokens from previous frames, enabling cross-frame information utilization without explicit temporal modeling, thereby effectively reducing computational cost and memory consumption. DiTCtrl~\cite{cai2025ditctrl} performs tuning-free controllable editing using mask-guided KV sharing and latent blending for smooth transitions across semantic segments. For story generation, TaleCrafter~\cite{gong2023talecrafter} modularizes story visualization into four interconnected components—story-to-prompt generation, text-to-layout generation, controllable text-to-image generation, and image-to-video animation—enabling interactive edits on intermediate representations and avoiding repeated end-to-end resampling. Animate-A-Story~\cite{he2023animate} adopts retrieval-augmented narrative synthesis to offload complex motion structure to retrieved priors. For controllable production such as virtual try-on, ViViD~\cite{fang2024vivid} extends diffusion to video with garment/pose encoders and hierarchical temporal modules to strengthen spatiotemporal coherence. PlayerOne~\cite{tu2025PlayerOne} formulates egocentric video generation as motion-conditioned world modeling, introducing a joint reconstruction framework for 4D scenes and video frames that supports real-time first-person exploration while ensuring scene consistency and temporal continuity.

\subsection{Video-Driven Scene Generation}
Video-driven scene generation methods leverage the spatial priors embedded in video generation models to synthesize more coherent and realistic 3D/4D environments.

Some approaches decompose the pipeline into two stages: video generation and 3D optimization. They first use a video model to synthesize a reference video or multi-view sequence, and then recover scene structure via techniques such as 4D Gaussians. VividDream~\cite{lee2024vividdreamgenerating3dscene} introduces a novel pipeline that first constructs and expands a static 3D scene according to an input image, then generates dynamic multi-view videos with a video diffusion model, and finally makes use of them to optimize an explorable 4D scene. Similarly, 4Real~\cite{yu2024real} and Free4D~\cite{liu2025free4dtuningfree4dscene} first generate a temporally consistent reference video and then expand the viewpoint range through frame-conditioned video generation. These methods benefit from a stable modular pipeline; however, because video generation and geometric reconstruction are decoupled, errors can accumulate progressively.

Other approaches aim to jointly model spatiotemporal information within a unified model, directly generating representations that are consistent across time and viewpoints. A promising direction is to combine the geometric priors of feed-forward 3D reconstruction models with the generative capability of video diffusion models. Gen3R~\cite{huang2026gen3r3dscenegeneration} unifies feed-forward 3D reconstruction with video diffusion in a shared latent space, enabling the joint generation of temporally consistent RGB videos and their 3D geometry within a single framework. Other approaches, such as CAT4D~\cite{wu2024cat4dcreate4dmultiview}, address dynamic scene generation by first expanding a monocular video into dynamic multi-view videos with a video diffusion model and then optimizing a deformable 3D Gaussian representation to recover the final 4D scene. StarGen~\cite{zhai2025stargenspatiotemporalautoregressionframework} conditions each temporal sliding window on the overlapping frame from the preceding sliding window to maintain temporal consistency, and on images covering the largest common spatial area in the scene with the current sliding window to improve spatial consistency during long-range generation. Due to their potential advantages in consistency and generation efficiency, these unified approaches are increasingly becoming an important direction for video-driven scene generation.

\section{Conclusions}
\label{sec:conclusions}
In this paper, we provide a comprehensive and systematic review of the critical intersection between efficiency improvement techniques and video-based world models. We explore how efficiency-oriented designs empower video-based world simulators in three primary dimensions: efficient modeling paradigms, efficient architectures, and efficient algorithms. In addition, we investigate how these efficient video generation frameworks directly enhance downstream applications such as autonomous driving, embodied AI, and gaming. Based on this review, we summarize challenges and future opportunities in this rapidly developing field, offering potential solutions for next-generation models facing increasingly complex physical dynamics and substantial computational demands. 


\bibliographystyle{IEEEtran}
\bibliography{main}

@article{goodfellow2014generative,
  title={Generative adversarial nets},
  author={Goodfellow, Ian J and Pouget-Abadie, Jean and Mirza, Mehdi and Xu, Bing and Warde-Farley, David and Ozair, Sherjil and Courville, Aaron and Bengio, Yoshua},
  journal={Advances in neural information processing systems},
  volume={27},
  year={2014}
}

@inproceedings{saito2017temporal,
  title={Temporal generative adversarial nets with singular value clipping},
  author={Saito, Masaki and Matsumoto, Eiichi and Saito, Shunta},
  booktitle={Proceedings of the IEEE international conference on computer vision},
  pages={2830--2839},
  year={2017}
}

@inproceedings{esser2021taming,
  title={Taming transformers for high-resolution image synthesis},
  author={Esser, Patrick and Rombach, Robin and Ommer, Bjorn},
  booktitle={Proceedings of the IEEE/CVF conference on computer vision and pattern recognition},
  pages={12873--12883},
  year={2021}
}

@inproceedings{kalchbrenner2017video,
  title={Video pixel networks},
  author={Kalchbrenner, Nal and Oord, A{\"a}ron and Simonyan, Karen and Danihelka, Ivo and Vinyals, Oriol and Graves, Alex and Kavukcuoglu, Koray},
  booktitle={International Conference on Machine Learning},
  pages={1771--1779},
  year={2017},
  organization={PMLR}
}

@article{yan2021videogpt,
  title={Videogpt: Video generation using vq-vae and transformers},
  author={Yan, Wilson and Zhang, Yunzhi and Abbeel, Pieter and Srinivas, Aravind},
  journal={arXiv preprint arXiv:2104.10157},
  year={2021}
}

@article{ho2022video,
  title={Video diffusion models},
  author={Ho, Jonathan and Salimans, Tim and Gritsenko, Alexey and Chan, William and Norouzi, Mohammad and Fleet, David J},
  journal={Advances in neural information processing systems},
  volume={35},
  pages={8633--8646},
  year={2022}
}

@inproceedings{blattmann2023align,
  title={Align your latents: High-resolution video synthesis with latent diffusion models},
  author={Blattmann, Andreas and Rombach, Robin and Ling, Huan and Dockhorn, Tim and Kim, Seung Wook and Fidler, Sanja and Kreis, Karsten},
  booktitle={Proceedings of the IEEE/CVF conference on computer vision and pattern recognition},
  pages={22563--22575},
  year={2023}
}

@inproceedings{yin2025slow,
  title={From slow bidirectional to fast autoregressive video diffusion models},
  author={Yin, Tianwei and Zhang, Qiang and Zhang, Richard and Freeman, William T and Durand, Fredo and Shechtman, Eli and Huang, Xun},
  booktitle={Proceedings of the Computer Vision and Pattern Recognition Conference},
  pages={22963--22974},
  year={2025}
}

@article{ho2020denoising,
  title={Denoising diffusion probabilistic models},
  author={Ho, Jonathan and Jain, Ajay and Abbeel, Pieter},
  journal={Advances in neural information processing systems},
  volume={33},
  pages={6840--6851},
  year={2020}
}

@inproceedings{
song2020score,
title={Score-Based Generative Modeling through Stochastic Differential Equations},
author={Yang Song and Jascha Sohl-Dickstein and Diederik P Kingma and Abhishek Kumar and Stefano Ermon and Ben Poole},
booktitle={International Conference on Learning Representations},
year={2021},
url={https://openreview.net/forum?id=PxTIG12RRHS}
}

@inproceedings{
lipman2023flow,
title={Flow Matching for Generative Modeling},
author={Yaron Lipman and Ricky T. Q. Chen and Heli Ben-Hamu and Maximilian Nickel and Matthew Le},
booktitle={The Eleventh International Conference on Learning Representations },
year={2023},
url={https://openreview.net/forum?id=PqvMRDCJT9t}
}

@inproceedings{
liu2024instaflow,
title={InstaFlow: One Step is Enough for High-Quality Diffusion-Based Text-to-Image Generation},
author={Xingchao Liu and Xiwen Zhang and Jianzhu Ma and Jian Peng and qiang liu},
booktitle={The Twelfth International Conference on Learning Representations},
year={2024},
url={https://openreview.net/forum?id=1k4yZbbDqX}
}

@inproceedings{rombach2022high,
  title={High-resolution image synthesis with latent diffusion models},
  author={Rombach, Robin and Blattmann, Andreas and Lorenz, Dominik and Esser, Patrick and Ommer, Bj{\"o}rn},
  booktitle={Proceedings of the IEEE/CVF conference on computer vision and pattern recognition},
  pages={10684--10695},
  year={2022}
}

@inproceedings{yu2023magvit,
  title={Magvit: Masked generative video transformer},
  author={Yu, Lijun and Cheng, Yong and Sohn, Kihyuk and Lezama, Jos{\'e} and Zhang, Han and Chang, Huiwen and Hauptmann, Alexander G and Yang, Ming-Hsuan and Hao, Yuan and Essa, Irfan and others},
  booktitle={Proceedings of the IEEE/CVF Conference on Computer Vision and Pattern Recognition},
  pages={10459--10469},
  year={2023}
}

@inproceedings{
yang2025cogvideox,
title={CogVideoX: Text-to-Video Diffusion Models with An Expert Transformer},
author={Zhuoyi Yang and Jiayan Teng and Wendi Zheng and Ming Ding and Shiyu Huang and Jiazheng Xu and Yuanming Yang and Wenyi Hong and Xiaohan Zhang and Guanyu Feng and Da Yin and Yuxuan.Zhang and Weihan Wang and Yean Cheng and Bin Xu and Xiaotao Gu and Yuxiao Dong and Jie Tang},
booktitle={The Thirteenth International Conference on Learning Representations},
year={2025},
url={https://openreview.net/forum?id=LQzN6TRFg9}
}

@inproceedings{ronneberger2015u,
  title={U-net: Convolutional networks for biomedical image segmentation},
  author={Ronneberger, Olaf and Fischer, Philipp and Brox, Thomas},
  booktitle={International Conference on Medical image computing and computer-assisted intervention},
  pages={234--241},
  year={2015},
  organization={Springer}
}

@inproceedings{peebles2023scalable,
  title={Scalable diffusion models with transformers},
  author={Peebles, William and Xie, Saining},
  booktitle={Proceedings of the IEEE/CVF international conference on computer vision},
  pages={4195--4205},
  year={2023}
}

@article{videoworldsimulators2024,
  title={Video generation models as world simulators},
  author={Tim Brooks and Bill Peebles and Connor Holmes and Will DePue and Yufei Guo and Li Jing and David Schnurr and Joe Taylor and Troy Luhman and Eric Luhman and Clarence Ng and Ricky Wang and Aditya Ramesh},
  year={2024},
  url={https://openai.com/research/video-generation-models-as-world-simulators},
}

@article{zheng2024open,
  title={Open-sora: Democratizing efficient video production for all},
  author={Zheng, Zangwei and Peng, Xiangyu and Yang, Tianji and Shen, Chenhui and Li, Shenggui and Liu, Hongxin and Zhou, Yukun and Li, Tianyi and You, Yang},
  journal={arXiv preprint arXiv:2412.20404},
  year={2024}
}

@article{
ma2025latte,
title={Latte: Latent Diffusion Transformer for Video Generation},
author={Xin Ma and Yaohui Wang and Xinyuan Chen and Gengyun Jia and Ziwei Liu and Yuan-Fang Li and Cunjian Chen and Yu Qiao},
journal={Transactions on Machine Learning Research},
issn={2835-8856},
year={2025},
url={https://openreview.net/forum?id=ntGPYNUF3t},
note={}
}

@inproceedings{radford2021learning,
  title={Learning transferable visual models from natural language supervision},
  author={Radford, Alec and Kim, Jong Wook and Hallacy, Chris and Ramesh, Aditya and Goh, Gabriel and Agarwal, Sandhini and Sastry, Girish and Askell, Amanda and Mishkin, Pamela and Clark, Jack and others},
  booktitle={International conference on machine learning},
  pages={8748--8763},
  year={2021},
  organization={PmLR}
}

@article{raffel2020exploring,
  title={Exploring the limits of transfer learning with a unified text-to-text transformer},
  author={Raffel, Colin and Shazeer, Noam and Roberts, Adam and Lee, Katherine and Narang, Sharan and Matena, Michael and Zhou, Yanqi and Li, Wei and Liu, Peter J},
  journal={Journal of machine learning research},
  volume={21},
  number={140},
  pages={1--67},
  year={2020}
}

@article{kong2024hunyuanvideo,
  title={Hunyuanvideo: A systematic framework for large video generative models},
  author={Kong, Weijie and Tian, Qi and Zhang, Zijian and Min, Rox and Dai, Zuozhuo and Zhou, Jin and Xiong, Jiangfeng and Li, Xin and Wu, Bo and Zhang, Jianwei and others},
  journal={arXiv preprint arXiv:2412.03603},
  year={2024}
}

@inproceedings{
yang2024learning,
title={Learning Interactive Real-World Simulators},
author={Sherry Yang and Yilun Du and Seyed Kamyar Seyed Ghasemipour and Jonathan Tompson and Leslie Pack Kaelbling and Dale Schuurmans and Pieter Abbeel},
booktitle={The Twelfth International Conference on Learning Representations},
year={2024},
url={https://openreview.net/forum?id=sFyTZEqmUY}
}

@article{ha2018world,
  title={World models},
  author={Ha, David and Schmidhuber, J{\"u}rgen},
  journal={arXiv preprint arXiv:1803.10122},
  volume={2},
  number={3},
  pages={440},
  year={2018}
}

@article{hu2023gaia,
  title={Gaia-1: A generative world model for autonomous driving},
  author={Hu, Anthony and Russell, Lloyd and Yeo, Hudson and Murez, Zak and Fedoseev, George and Kendall, Alex and Shotton, Jamie and Corrado, Gianluca},
  journal={arXiv preprint arXiv:2309.17080},
  year={2023}
}

@inproceedings{wang2024drivedreamer,
  title={Drivedreamer: Towards real-world-drive world models for autonomous driving},
  author={Wang, Xiaofeng and Zhu, Zheng and Huang, Guan and Chen, Xinze and Zhu, Jiagang and Lu, Jiwen},
  booktitle={European conference on computer vision},
  pages={55--72},
  year={2024},
  organization={Springer}
}

@article{du2023learning,
  title={Learning universal policies via text-guided video generation},
  author={Du, Yilun and Yang, Sherry and Dai, Bo and Dai, Hanjun and Nachum, Ofir and Tenenbaum, Josh and Schuurmans, Dale and Abbeel, Pieter},
  journal={Advances in neural information processing systems},
  volume={36},
  pages={9156--9172},
  year={2023}
}

@article{ho2022classifier,
  title={Classifier-free diffusion guidance},
  author={Ho, Jonathan and Salimans, Tim},
  journal={arXiv preprint arXiv:2207.12598},
  year={2022}
}

@inproceedings{
jin2025pyramidal,
title={Pyramidal Flow Matching for Efficient Video Generative Modeling},
author={Yang Jin and Zhicheng Sun and Ningyuan Li and Kun Xu and Kun Xu and Hao Jiang and Nan Zhuang and Quzhe Huang and Yang Song and Yadong MU and Zhouchen Lin},
booktitle={The Thirteenth International Conference on Learning Representations},
year={2025},
url={https://openreview.net/forum?id=66NzcRQuOq}
}

@article{li2026rollingsink,
  title={Rolling Sink: Bridging Limited-Horizon Training and Open-Ended Testing in Autoregressive Video Diffusion},
  author={Li, Haodong and Liu, Shaoteng and Lin, Zhe and Chandraker, Manmohan},
  journal={arXiv preprint arXiv:2602.07775},
  year={2026},
  url={https://arxiv.org/abs/2602.07775}
}

@inproceedings{
ran2025tpdiff,
title={{TPD}iff: Temporal Pyramid Video Diffusion Model},
author={Lingmin Ran and Mike Zheng Shou},
booktitle={The Fourteenth International Conference on Learning Representations},
year={2026},
url={https://openreview.net/forum?id=Eg3KqoI9tS}
}

@inproceedings{hu2025ultragen,
  title={UltraGen: High-Resolution Video Generation with Hierarchical Attention},
  author={Hu, Teng and Zhang, Jiangning and Su, Zihan and Yi, Ran},
  booktitle={Proceedings of the AAAI Conference on Artificial Intelligence},
  volume={40},
  number={6},
  pages={4923--4931},
  year={2026}
}

@article{zhang2025waver,
  title={Waver: Wave your way to lifelike video generation},
  author={Zhang, Yifu and Yang, Hao and Zhang, Yuqi and Hu, Yifei and Zhu, Fengda and Lin, Chuang and Mei, Xiaofeng and Jiang, Yi and Peng, Bingyue and Yuan, Zehuan},
  journal={arXiv preprint arXiv:2508.15761},
  year={2025}
}

@article{chen2025dc,
  title={Dc-videogen: Efficient video generation with deep compression video autoencoder},
  author={Chen, Junyu and He, Wenkun and Gu, Yuchao and Zhao, Yuyang and Yu, Jincheng and Chen, Junsong and Zou, Dongyun and Lin, Yujun and Zhang, Zhekai and Li, Muyang and others},
  journal={arXiv preprint arXiv:2509.25182},
  year={2025}
}

@InProceedings{zhang2025regen,
    author    = {Zhang, Yitian and Mai, Long and Mahapatra, Aniruddha and Bourgin, David and Hong, Yicong and Casebeer, Jonah and Liu, Feng and Fu, Yun},
    title     = {REGEN: Learning Compact Video Embedding with (Re-)Generative Decoder},
    booktitle = {Proceedings of the IEEE/CVF International Conference on Computer Vision (ICCV)},
    month     = {October},
    year      = {2025},
    pages     = {18453-18462}
}

@inproceedings{zou2025turbo,
  title={Turbo-vaed: Fast and stable transfer of video-vaes to mobile devices},
  author={Zou, Ya and Yao, Jingfeng and Yu, Siyuan and Zhang, Shuai and Liu, Wenyu and Wang, Xinggang},
  booktitle={Proceedings of the AAAI Conference on Artificial Intelligence},
  volume={40},
  number={16},
  pages={14086--14094},
  year={2026}
}

@inproceedings{zhang2026flashvideo,
  title={Flashvideo: Flowing fidelity to detail for efficient high-resolution video generation},
  author={Zhang, Shilong and Li, Wenbo and Chen, Shoufa and Ge, Chongjian and Sun, Peize and Zhang, Yifu and Jiang, Yi and Yuan, Zehuan and Peng, Bingyue and Luo, Ping},
  booktitle={Proceedings of the AAAI Conference on Artificial Intelligence},
  volume={40},
  number={15},
  pages={12735--12743},
  year={2026}
}

@article{cheng2025srdiffusion,
  title={SRDiffusion: Accelerate Video Diffusion Inference via Sketching-Rendering Cooperation},
  author={Cheng, Shenggan and Wei, Yuanxin and Diao, Lansong and Liu, Yong and Chen, Bujiao and Huang, Lianghua and Liu, Yu and Yu, Wenyuan and Du, Jiangsu and Lin, Wei and others},
  journal={arXiv preprint arXiv:2505.19151},
  year={2025}
}

@article{ye2025supergen,
  title={SuperGen: An Efficient Ultra-high-resolution Video Generation System with Sketching and Tiling},
  author={Ye, Fanjiang and Zhao, Zepeng and Mu, Yi and Shen, Jucheng and Li, Renjie and Wang, Kaijian and Sun, Desen and Agarwal, Saurabh and Lee, Myungjin and Cao, Triston and others},
  journal={arXiv preprint arXiv:2508.17756},
  year={2025}
}

@InProceedings{yuan2025vgdfr,
    author    = {Yuan, Zhihang and Xie, Rui and Shang, Yuzhang and Zhang, Hanling and Wang, Siyuan and Yan, Shengen and Dai, Guohao and Wang, Yu},
    title     = {DLFR-Gen: Diffusion-based Video Generation with Dynamic Latent Frame Rate},
    booktitle = {Proceedings of the IEEE/CVF International Conference on Computer Vision (ICCV)},
    month     = {October},
    year      = {2025},
    pages     = {16410-16419}
}

@inproceedings{yuan2025dlfr,
  title={Dlfr-vae: Dynamic latent frame rate vae for video generation},
  author={Yuan, Zhihang and Wang, Siyuan and Shang, Yuzhang and Zhang, Hanling and Fang, Tongcheng and Xie, Rui and Yan, Shengen and Dai, Guohao and Wang, Yu},
  booktitle={Proceedings of the 33rd ACM International Conference on Multimedia},
  pages={10388--10397},
  year={2025}
}

@article{wang2025evoworld,
  title={EvoWorld: Evolving Panoramic World Generation with Explicit 3D Memory},
  author={Wang, Jiahao and Ye, Luoxin and Lu, TaiMing and Xiao, Junfei and Zhang, Jiahan and Guo, Yuxiang and Liu, Xijun and Chellappa, Rama and Peng, Cheng and Yuille, Alan and others},
  journal={arXiv preprint arXiv:2510.01183},
  year={2025}
}

@InProceedings{li2025vmem,
    author    = {Li, Runjia and Torr, Philip and Vedaldi, Andrea and Jakab, Tomas},
    title     = {VMem: Consistent Interactive Video Scene Generation with Surfel-Indexed View Memory},
    booktitle = {Proceedings of the IEEE/CVF International Conference on Computer Vision (ICCV)},
    month     = {October},
    year      = {2025},
    pages     = {25690-25699}
}

@inproceedings{
wu2025video,
title={Video World Models with Long-term Spatial Memory},
author={Tong Wu and Shuai Yang and Ryan Po and Yinghao Xu and Ziwei Liu and Dahua Lin and Gordon Wetzstein},
booktitle={The Thirty-ninth Annual Conference on Neural Information Processing Systems},
year={2025},
url={https://openreview.net/forum?id=HbTxc6U1fO}
}

@article{huang2025voyager,
  title={Voyager: Long-range and world-consistent video diffusion for explorable 3d scene generation},
  author={Huang, Tianyu and Zheng, Wangguandong and Wang, Tengfei and Liu, Yuhao and Wang, Zhenwei and Wu, Junta and Jiang, Jie and Li, Hui and Lau, Rynson and Zuo, Wangmeng and others},
  journal={ACM Transactions on Graphics (TOG)},
  volume={44},
  number={6},
  pages={1--15},
  year={2025},
  publisher={ACM New York, NY, USA}
}

@inproceedings{
li2025stable,
title={Stable Video Infinity: Infinite-Length Video Generation with Error Recycling},
author={Wuyang Li and Wentao Pan and Po-Chien Luan and Yang Gao and Alexandre Alahi},
booktitle={The Fourteenth International Conference on Learning Representations},
year={2026},
url={https://openreview.net/forum?id=X96Ei9n34a}
}

@article{oshima2025worldpack,
  title={WorldPack: Compressed Memory Improves Spatial Consistency in Video World Modeling},
  author={Oshima, Yuta and Iwasawa, Yusuke and Suzuki, Masahiro and Matsuo, Yutaka and Furuta, Hiroki},
  journal={arXiv preprint arXiv:2512.02473},
  year={2025}
}

@article{zhang2025storymem,
  title={StoryMem: Multi-shot Long Video Storytelling with Memory},
  author={Zhang, Kaiwen and Jiang, Liming and Wang, Angtian and Fang, Jacob Zhiyuan and Zhi, Tiancheng and Yan, Qing and Kang, Hao and Lu, Xin and Pan, Xingang},
  journal={arXiv preprint arXiv:2512.19539},
  year={2025}
}

@inproceedings{yu2025context,
  title={Context as memory: Scene-consistent interactive long video generation with memory retrieval},
  author={Yu, Jiwen and Bai, Jianhong and Qin, Yiran and Liu, Quande and Wang, Xintao and Wan, Pengfei and Zhang, Di and Liu, Xihui},
  booktitle={Proceedings of the SIGGRAPH Asia 2025 Conference Papers},
  pages={1--11},
  year={2025}
}

@article{jiang2025lovic,
  title={Lovic: Efficient long video generation with context compression},
  author={Jiang, Jiaxiu and Li, Wenbo and Ren, Jingjing and Qiu, Yuping and Guo, Yong and Xu, Xiaogang and Wu, Han and Zuo, Wangmeng},
  journal={arXiv preprint arXiv:2507.12952},
  year={2025}
}

@inproceedings{
cai2025mixture,
title={Mixture of Contexts for Long Video Generation},
author={Shengqu Cai and Ceyuan Yang and Lvmin Zhang and Yuwei Guo and Junfei Xiao and Ziyan Yang and Yinghao Xu and Zhenheng Yang and Alan Yuille and Leonidas Guibas and Maneesh Agrawala and Lu Jiang and Gordon Wetzstein},
booktitle={The Fourteenth International Conference on Learning Representations},
year={2026},
url={https://openreview.net/forum?id=y6XJZlEC2x}
}

@inproceedings{
yang2025longlive,
title={LongLive: Real-time Interactive Long Video Generation},
author={Shuai Yang and Wei Huang and Ruihang Chu and Yicheng Xiao and Yuyang Zhao and Xianbang Wang and Muyang Li and Enze Xie and Ying-Cong Chen and Yao Lu and Song Han and Yukang Chen},
booktitle={The Fourteenth International Conference on Learning Representations},
year={2026},
url={https://openreview.net/forum?id=nCAODkpsPJ}
}

@article{zhang2026spargeattention2,
  title={SpargeAttention2: Trainable Sparse Attention via Hybrid Top-k+ Top-p Masking and Distillation Fine-Tuning},
  author={Zhang, Jintao and Jiang, Kai and Xiang, Chendong and Feng, Weiqi and Hu, Yuezhou and Xi, Haocheng and Chen, Jianfei and Zhu, Jun},
  journal={arXiv preprint arXiv:2602.13515},
  year={2026}
}

@inproceedings{
chen2026sanavideo,
title={{SANA}-Video: Efficient Video Generation with Block Linear Diffusion Transformer},
author={Junsong Chen and Yuyang Zhao and Jincheng YU and Ruihang Chu and Junyu Chen and Shuai Yang and Xianbang Wang and Yicheng Pan and Daquan Zhou and Huan Ling and Haozhe Liu and Hongwei Yi and Hao Zhang and Muyang Li and Yukang Chen and Han Cai and Sanja Fidler and Ping Luo and Song Han and Enze Xie},
booktitle={The Fourteenth International Conference on Learning Representations},
year={2026},
url={https://openreview.net/forum?id=mzAchylAtf}
}

@article{huang2025linvideo,
  title={Linvideo: A post-training framework towards o (n) attention in efficient video generation},
  author={Huang, Yushi and Ge, Xingtong and Gong, Ruihao and Lv, Chengtao and Zhang, Jun},
  journal={arXiv preprint arXiv:2510.08318},
  year={2025}
}

@inproceedings{
gu2024mamba,
title={Mamba: Linear-Time Sequence Modeling with Selective State Spaces},
author={Albert Gu and Tri Dao},
booktitle={First Conference on Language Modeling},
year={2024},
url={https://openreview.net/forum?id=tEYskw1VY2}
}

@InProceedings{Wang_2025_CVPR,
    author    = {Wang, Hongjie and Ma, Chih-Yao and Liu, Yen-Cheng and Hou, Ji and Xu, Tao and Wang, Jialiang and Juefei-Xu, Felix and Luo, Yaqiao and Zhang, Peizhao and Hou, Tingbo and Vajda, Peter and Jha, Niraj K. and Dai, Xiaoliang},
    title     = {LinGen: Towards High-Resolution Minute-Length Text-to-Video Generation with Linear Computational Complexity},
    booktitle = {Proceedings of the IEEE/CVF Conference on Computer Vision and Pattern Recognition (CVPR)},
    month     = {June},
    year      = {2025},
    pages     = {2578-2588}
}

@inproceedings{
dao2024transformers,
title={Transformers are {SSM}s: Generalized Models and Efficient Algorithms Through Structured State Space Duality},
author={Tri Dao and Albert Gu},
booktitle={Forty-first International Conference on Machine Learning},
year={2024},
url={https://openreview.net/forum?id=ztn8FCR1td}
}

@inproceedings{
zhang2025frame,
title={Frame Context Packing and Drift Prevention in Next-Frame-Prediction Video Diffusion Models},
author={Lvmin Zhang and Shengqu Cai and Muyang Li and Gordon Wetzstein and Maneesh Agrawala},
booktitle={The Thirty-ninth Annual Conference on Neural Information Processing Systems},
year={2025},
url={https://openreview.net/forum?id=J8JCF64aEn}
}

@inproceedings{du2025patchvsr,
  title={PatchVSR: Breaking Video Diffusion Resolution Limits with Patch-wise Video Super-Resolution},
  author={Du, Shian and Xia, Menghan and Liu, Chang and Wang, Xintao and Wang, Jing and Wan, Pengfei and Zhang, Di and Ji, Xiangyang},
  booktitle={Proceedings of the Computer Vision and Pattern Recognition Conference},
  pages={17799--17809},
  year={2025}
}

@article{gao2025seedance,
  title={Seedance 1.0: Exploring the Boundaries of Video Generation Models},
  author={Gao, Yu and Guo, Haoyuan and Hoang, Tuyen and Huang, Weilin and Jiang, Lu and Kong, Fangyuan and Li, Huixia and Li, Jiashi and Li, Liang and Li, Xiaojie and others},
  journal={arXiv preprint arXiv:2506.09113},
  year={2025}
}

@inproceedings{dalal2025one,
  title={One-minute video generation with test-time training},
  author={Dalal, Karan and Koceja, Daniel and Xu, Jiarui and Zhao, Yue and Han, Shihao and Cheung, Ka Chun and Kautz, Jan and Choi, Yejin and Sun, Yu and Wang, Xiaolong},
  booktitle={Proceedings of the Computer Vision and Pattern Recognition Conference},
  pages={17702--17711},
  year={2025}
}

@inproceedings{
sun2025learning,
title={Learning to (Learn at Test Time): {RNN}s with Expressive Hidden States},
author={Yu Sun and Xinhao Li and Karan Dalal and Jiarui Xu and Arjun Vikram and Genghan Zhang and Yann Dubois and Xinlei Chen and Xiaolong Wang and Sanmi Koyejo and Tatsunori Hashimoto and Carlos Guestrin},
booktitle={Forty-second International Conference on Machine Learning},
year={2025},
url={https://openreview.net/forum?id=wXfuOj9C7L}
}

@inproceedings{
zhang2025test,
title={Test-Time Training Done Right},
author={Tianyuan Zhang and Sai Bi and Yicong Hong and Kai Zhang and Fujun Luan and Songlin Yang and Kalyan Sunkavalli and William T. Freeman and Hao Tan},
booktitle={The Fourteenth International Conference on Learning Representations},
year={2026},
url={https://openreview.net/forum?id=Tb9qAxT3xv}
}

@article{huang2025memory,
  title={Memory forcing: Spatio-temporal memory for consistent scene generation on minecraft},
  author={Huang, Junchao and Hu, Xinting and Han, Boyao and Shi, Shaoshuai and Tian, Zhuotao and He, Tianyu and Jiang, Li},
  journal={arXiv preprint arXiv:2510.03198},
  year={2025}
}

@inproceedings{
zhu2026astra,
title={Astra: General Interactive World Model with Autoregressive Denoising},
author={Yixuan Zhu and Feng Jiaqi and Wenzhao Zheng and Yuan Gao and Xin Tao and Pengfei Wan and Jiwen Lu and Jie Zhou},
booktitle={The Fourteenth International Conference on Learning Representations},
year={2026},
url={https://openreview.net/forum?id=8UZpmrxoLG}
}

@inproceedings{
xi2025sparse,
title={Sparse Video-Gen: Accelerating Video Diffusion Transformers with Spatial-Temporal Sparsity},
author={Haocheng Xi and Shuo Yang and Yilong Zhao and Chenfeng Xu and Muyang Li and Xiuyu Li and Yujun Lin and Han Cai and Jintao Zhang and Dacheng Li and Jianfei Chen and Ion Stoica and Kurt Keutzer and Song Han},
booktitle={Forty-second International Conference on Machine Learning},
year={2025},
url={https://openreview.net/forum?id=u8CA3qIS0V}
}

@inproceedings{
zhang2023ho,
title={H2O: Heavy-Hitter Oracle for Efficient Generative Inference of Large Language Models},
author={Zhenyu Zhang and Ying Sheng and Tianyi Zhou and Tianlong Chen and Lianmin Zheng and Ruisi Cai and Zhao Song and Yuandong Tian and Christopher Re and Clark Barrett and Zhangyang Wang and Beidi Chen},
booktitle={Thirty-seventh Conference on Neural Information Processing Systems},
year={2023},
url={https://openreview.net/forum?id=RkRrPp7GKO}
}

@inproceedings{
yang2025sparse,
title={Sparse VideoGen2: Accelerate Video Generation with  Sparse Attention via Semantic-Aware Permutation},
author={Shuo Yang and Haocheng Xi and Yilong Zhao and Muyang Li and Jintao Zhang and Han Cai and Yujun Lin and Xiuyu Li and Chenfeng Xu and Kelly Peng and Jianfei Chen and Song Han and Kurt Keutzer and Ion Stoica},
booktitle={The Thirty-ninth Annual Conference on Neural Information Processing Systems},
year={2025},
url={https://openreview.net/forum?id=WPU17d1l7R}
}

@inproceedings{Fu_2025_BMVC,
author    = {Yunxiang Fu and Chaoqi Chen and Yizhou Yu},
title     = {LaMamba-Diff: Linear-Time High-Fidelity Diffusion Models Based on Local Attention and Mamba},
booktitle = {36th British Machine Vision Conference 2025, {BMVC} 2025, Sheffield, UK, November 24-27, 2025},
publisher = {BMVA},
year      = {2025},
url       = {https://bmva-archive.org.uk/bmvc/2025/assets/papers/Paper_962/paper.pdf}
}

@inproceedings{
sun2025vorta,
title={{VORTA}: Efficient Video Diffusion via Routing Sparse Attention},
author={Wenhao Sun and Rong-Cheng Tu and Yifu Ding and Jingyi Liao and Zhao Jin and Shunyu Liu and Dacheng Tao},
booktitle={The Thirty-ninth Annual Conference on Neural Information Processing Systems},
year={2025},
url={https://openreview.net/forum?id=gY9yOGYB48}
}

@inproceedings{
wu2025vmoba,
title={{VM}o{BA}: Mixture-of-Block Attention for Video Diffusion Models},
author={Jianzong Wu and Liang Hou and Haotian Yang and Ye Tian and Pengfei Wan and Di ZHANG and Yunhai Tong},
booktitle={The Fourteenth International Conference on Learning Representations},
year={2026},
url={https://openreview.net/forum?id=oQaRElUdmh}
}

@inproceedings{
lu2025moba,
title={Mo{BA}: Mixture of Block Attention for Long-Context {LLM}s},
author={Enzhe Lu and Zhejun Jiang and Jingyuan Liu and Yulun Du and Tao Jiang and Chao Hong and Shaowei Liu and Weiran He and Enming Yuan and Yuzhi Wang and Zhiqi Huang and Huan Yuan and Suting Xu and Xinran Xu and Guokun Lai and Yanru Chen and Huabin Zheng and Junjie Yan and Jianlin Su and Yuxin Wu and Yutao Zhang and Zhilin Yang and Xinyu Zhou and Mingxing Zhang and Jiezhong Qiu},
booktitle={The Thirty-ninth Annual Conference on Neural Information Processing Systems},
year={2025},
url={https://openreview.net/forum?id=RlqYCpTu1P}
}

@article{durvasula2025fg,
  title={FG-Attn: Leveraging Fine-Grained Sparsity In Diffusion Transformers},
  author={Durvasula, Sankeerth and Sreedhar, Kavya and Moustafa, Zain and Kothawade, Suraj and Gondimalla, Ashish and Subramanian, Suvinay and Shahidi, Narges and Vijaykumar, Nandita},
  journal={arXiv preprint arXiv:2509.16518},
  year={2025}
}

@inproceedings{chen2025sparse,
  title={Sparse-vdit: Unleashing the power of sparse attention to accelerate video diffusion transformers},
  author={Chen, Pengtao and Zeng, Xianfang and Zhao, Maosen and Shen, Mingzhu and Cheng, Wei and Yu, Gang and Chen, Tao},
  booktitle={Proceedings of the AAAI Conference on Artificial Intelligence},
  volume={40},
  number={4},
  pages={2957--2965},
  year={2026}
}

@inproceedings{
zhang2025fast,
title={Fast Video Generation with Sliding Tile Attention},
author={Peiyuan Zhang and Yongqi Chen and Runlong Su and Hangliang Ding and Ion Stoica and Zhengzhong Liu and Hao Zhang},
booktitle={Forty-second International Conference on Machine Learning},
year={2025},
url={https://openreview.net/forum?id=U74MOXPEJd}
}

@inproceedings{
li2025radial,
title={Radial Attention: \${\textbackslash}mathcal O(n {\textbackslash}log n)\$ Sparse Attention for Long Video Generation},
author={Xingyang Li and Muyang Li and Tianle Cai and Haocheng Xi and Shuo Yang and Yujun Lin and Lvmin Zhang and Songlin Yang and Jinbo Hu and Kelly Peng and Maneesh Agrawala and Ion Stoica and Kurt Keutzer and Song Han},
booktitle={The Thirty-ninth Annual Conference on Neural Information Processing Systems},
year={2025},
url={https://openreview.net/forum?id=hYovE4nHTt}
}

@article{li2025compact,
  title={Compact attention: Exploiting structured spatio-temporal sparsity for fast video generation},
  author={Li, Qirui and Zheng, Guangcong and Zhao, Qi and Li, Jie and Dong, Bin and Yao, Yiwu and Li, Xi},
  journal={arXiv preprint arXiv:2508.12969},
  year={2025}
}

@article{zhan2025bidirectional,
  title={Bidirectional sparse attention for faster video diffusion training},
  author={Zhan, Chenlu and Li, Wen and Shen, Chuyu and Zhang, Jun and Wu, Suhui and Zhang, Hao},
  journal={arXiv preprint arXiv:2509.01085},
  year={2025}
}

@inproceedings{
yuan2024ditfastattn,
title={Di{TF}astAttn: Attention Compression for Diffusion Transformer Models},
author={Zhihang Yuan and Hanling Zhang and Lu Pu and Xuefei Ning and Linfeng Zhang and Tianchen Zhao and Shengen Yan and Guohao Dai and Yu Wang},
booktitle={The Thirty-eighth Annual Conference on Neural Information Processing Systems},
year={2024},
url={https://openreview.net/forum?id=51HQpkQy3t}
}

@inproceedings{
zhang2025sla,
title={{SLA}: Beyond Sparsity in Diffusion Transformers via Fine-Tunable Sparse{\textendash}Linear Attention},
author={Jintao Zhang and Haoxu Wang and Kai Jiang and Shuo Yang and Kaiwen Zheng and Haocheng Xi and Ziteng Wang and Hongzhou Zhu and Min Zhao and Ion Stoica and Joseph E. Gonzalez and Jun Zhu and Jianfei Chen},
booktitle={The Fourteenth International Conference on Learning Representations},
year={2026},
url={https://openreview.net/forum?id=eD8IPvNoZB}
}

@inproceedings{
sun2025asymrnr,
title={AsymRnR: Video Diffusion Transformers Acceleration with Asymmetric Reduction and Restoration},
author={Wenhao Sun and Rong-Cheng Tu and Jingyi Liao and Zhao Jin and Dacheng Tao},
booktitle={Forty-second International Conference on Machine Learning},
year={2025},
url={https://openreview.net/forum?id=5PiZevq9fY}
}

@inproceedings{
gu2025blade,
title={{BLADE}: Block-Sparse Attention Meets Step Distillation for Efficient Video Generation},
author={Youping Gu and XIAOLONG LI and Yuhao Hu and Chen Minqi and Bohan Zhuang},
booktitle={The Fourteenth International Conference on Learning Representations},
year={2026},
url={https://openreview.net/forum?id=O9J20MsmRl}
}

@article{mao2025yume,
  title={Yume-1.5: A Text-Controlled Interactive World Generation Model},
  author={Mao, Xiaofeng and Li, Zhen and Li, Chuanhao and Xu, Xiaojie and Ying, Kaining and He, Tong and Pang, Jiangmiao and Qiao, Yu and Zhang, Kaipeng},
  journal={arXiv preprint arXiv:2512.22096},
  year={2025}
}

@inproceedings{
song2025videonsa,
title={Video{NSA}: Native Sparse Attention Scales Video Understanding},
author={Enxin Song and Wenhao Chai and Shusheng Yang and Ethan J. Armand and Xiaojun Shan and Haiyang Xu and Jianwen Xie and Zhuowen Tu},
booktitle={The Fourteenth International Conference on Learning Representations},
year={2026},
url={https://openreview.net/forum?id=zA2LbsUMDd}
}

@inproceedings{
zhang2025spargeattention,
title={SpargeAttention: Accurate and Training-free Sparse Attention Accelerating Any Model Inference},
author={Jintao Zhang and Chendong Xiang and Haofeng Huang and Jia wei and Haocheng Xi and Jun Zhu and Jianfei Chen},
booktitle={Forty-second International Conference on Machine Learning},
year={2025},
url={https://openreview.net/forum?id=74c3Wwk8Tc}
}

@inproceedings{
liu2025astraea,
title={{ASTRAEA}: A Token-wise Acceleration Framework for Video Diffusion Transformers},
author={Haosong Liu and Yuge Cheng and Wenxuan Miao and Zihan Liu and Aiyue Chen and Jing Lin and Yiwu Yao and Chen Chen and Jingwen Leng and Yu Feng and Minyi Guo},
booktitle={The Fourteenth International Conference on Learning Representations},
year={2026},
url={https://openreview.net/forum?id=e8P4Oo8S6U}
}

@article{russell2025gaia,
  title={Gaia-2: A controllable multi-view generative world model for autonomous driving},
  author={Russell, Lloyd and Hu, Anthony and Bertoni, Lorenzo and Fedoseev, George and Shotton, Jamie and Arani, Elahe and Corrado, Gianluca},
  journal={arXiv preprint arXiv:2503.20523},
  year={2025}
}

@online{wayve_gaia3_2025,
  title   = {GAIA-3: Scaling World Models to Power Safety and Evaluation},
  author  = {{Wayve}},
  date    = {2025-12-02},
  url     = {https://wayve.ai/thinking/gaia-3/},
  note    = {Wayve Blog (Research)}
}

@inproceedings{zhao2025drivedreamer4d,
  title={Drivedreamer4d: World models are effective data machines for 4d driving scene representation},
  author={Zhao, Guosheng and Ni, Chaojun and Wang, Xiaofeng and Zhu, Zheng and Zhang, Xueyang and Wang, Yida and Huang, Guan and Chen, Xinze and Wang, Boyuan and Zhang, Youyi and others},
  booktitle={Proceedings of the Computer Vision and Pattern Recognition Conference},
  pages={12015--12026},
  year={2025}
}

@article{guo2024infinitydrive,
  title={Infinitydrive: Breaking time limits in driving world models},
  author={Guo, Xi and Ding, Chenjing and Dou, Haoxuan and Zhang, Xin and Tang, Weixuan and Wu, Wei},
  journal={arXiv preprint arXiv:2412.01522},
  year={2024}
}

@inproceedings{wang2024driving,
  title={Driving into the future: Multiview visual forecasting and planning with world model for autonomous driving},
  author={Wang, Yuqi and He, Jiawei and Fan, Lue and Li, Hongxin and Chen, Yuntao and Zhang, Zhaoxiang},
  booktitle={Proceedings of the IEEE/CVF Conference on Computer Vision and Pattern Recognition},
  pages={14749--14759},
  year={2024}
}

@article{gao2024vista,
  title={Vista: A generalizable driving world model with high fidelity and versatile controllability},
  author={Gao, Shenyuan and Yang, Jiazhi and Chen, Li and Chitta, Kashyap and Qiu, Yihang and Geiger, Andreas and Zhang, Jun and Li, Hongyang},
  journal={Advances in Neural Information Processing Systems},
  volume={37},
  pages={91560--91596},
  year={2024}
}

@article{wang2025mila,
  title={MiLA: Multi-view Intensive-fidelity Long-term Video Generation World Model for Autonomous Driving},
  author={Wang, Haiguang and Liu, Daqi and Xie, Hongwei and Liu, Haisong and Ma, Enhui and Yu, Kaicheng and Wang, Limin and Wang, Bing},
  journal={arXiv preprint arXiv:2503.15875},
  year={2025}
}

@misc{jia2023adriverigeneralworldmodel,
      title={ADriver-I: A General World Model for Autonomous Driving}, 
      author={Fan Jia and Weixin Mao and Yingfei Liu and Yucheng Zhao and Yuqing Wen and Chi Zhang and Xiangyu Zhang and Tiancai Wang},
      year={2023},
      eprint={2311.13549},
      archivePrefix={arXiv},
      primaryClass={cs.CV},
      url={https://arxiv.org/abs/2311.13549}, 
}

@inproceedings{zhang2025epona,
  author = {Zhang, Kaiwen and Tang, Zhenyu and Hu, Xiaotao and Pan, Xingang and Guo, Xiaoyang and Liu, Yuan and Huang,
  Jingwei and Yuan, Li and Zhang, Qian and Long, Xiao-Xiao and Cao, Xun and Yin, Wei},
  title = {Epona: Autoregressive Diffusion World Model for Autonomous Driving},
  booktitle = {Proceedings of the IEEE/CVF International Conference on Computer Vision (ICCV)},
  year = {2025}
}

@InProceedings{yang2024genadgeneralizedpredictivemodel,
    author    = {Yang, Jiazhi and Gao, Shenyuan and Qiu, Yihang and Chen, Li and Li, Tianyu and Dai, Bo and Chitta, Kashyap and Wu, Penghao and Zeng, Jia and Luo, Ping and Zhang, Jun and Geiger, Andreas and Qiao, Yu and Li, Hongyang},
    title     = {Generalized Predictive Model for Autonomous Driving},
    booktitle = {Proceedings of the IEEE/CVF Conference on Computer Vision and Pattern Recognition (CVPR)},
    month     = {June},
    year      = {2024},
    pages     = {14662-14672}
}

@article{feng2025vidar,
  title={Vidar: Embodied video diffusion model for generalist manipulation},
  author={Feng, Yao and Tan, Hengkai and Mao, Xinyi and Xiang, Chendong and Liu, Guodong and Huang, Shuhe and Su, Hang and Zhu, Jun},
  journal={arXiv preprint arXiv:2507.12898},
  year={2025}
}

@inproceedings{
jang2025dreamgen,
title={DreamGen: Unlocking Generalization in Robot Learning through Video World Models},
author={Joel Jang and Seonghyeon Ye and Zongyu Lin and Jiannan Xiang and Johan Bjorck and Yu Fang and Fengyuan Hu and Spencer Huang and Kaushil Kundalia and Yen-Chen Lin and Lo{\"\i}c Magne and Ajay Mandlekar and Avnish Narayan and You Liang Tan and Guanzhi Wang and Jing Wang and Qi Wang and Yinzhen Xu and Xiaohui Zeng and Kaiyuan Zheng and Ruijie Zheng and Ming-Yu Liu and Luke Zettlemoyer and Dieter Fox and Jan Kautz and Scott Reed and Yuke Zhu and Linxi Fan},
booktitle={9th Annual Conference on Robot Learning},
year={2025},
url={https://openreview.net/forum?id=3CnxNqmklv}
}

@article{ali2025world,
  title={World simulation with video foundation models for physical ai},
  author={Ali, Arslan and Bai, Junjie and Bala, Maciej and Balaji, Yogesh and Blakeman, Aaron and Cai, Tiffany and Cao, Jiaxin and Cao, Tianshi and Cha, Elizabeth and Chao, Yu-Wei and others},
  journal={arXiv preprint arXiv:2511.00062},
  year={2025}
}

@article{ni2025generated,
  title={From Generated Human Videos to Physically Plausible Robot Trajectories},
  author={Ni, James and Wang, Zekai and Lin, Wei and Bar, Amir and LeCun, Yann and Darrell, Trevor and Malik, Jitendra and Herzig, Roei},
  journal={arXiv preprint arXiv:2512.05094},
  year={2025}
}

@article{deng2026rethinking,
  title={Rethinking Video Generation Model for the Embodied World},
  author={Deng, Yufan and Pan, Zilin and Zhang, Hongyu and Li, Xiaojie and Hu, Ruoqing and Ding, Yufei and Zou, Yiming and Zeng, Yan and Zhou, Daquan},
  journal={arXiv preprint arXiv:2601.15282},
  year={2026}
}

@article{xiao2025world,
  title={World-env: Leveraging world model as a virtual environment for vla post-training},
  author={Xiao, Junjin and Yang, Yandan and Chang, Xinyuan and Chen, Ronghan and Xiong, Feng and Xu, Mu and Zheng, Wei-Shi and Zhang, Qing},
  journal={arXiv preprint arXiv:2509.24948},
  year={2025}
}

@article{jiang2025enerverse,
  title={Enerverse-ac: Envisioning embodied environments with action condition},
  author={Jiang, Yuxin and Chen, Shengcong and Huang, Siyuan and Chen, Liliang and Zhou, Pengfei and Liao, Yue and He, Xindong and Liu, Chiming and Li, Hongsheng and Yao, Maoqing and others},
  journal={arXiv preprint arXiv:2505.09723},
  year={2025}
}

@inproceedings{
guo2025ctrl,
title={Ctrl-World: A Controllable Generative World Model for Robot Manipulation},
author={Yanjiang Guo and Lucy Xiaoyang Shi and Jianyu Chen and Chelsea Finn},
booktitle={The Fourteenth International Conference on Learning Representations},
year={2026},
url={https://openreview.net/forum?id=748bHL2BAv}
}

@article{xu2025vilp,
  title={VILP: Imitation Learning with Latent Video Planning},
  author={Xu, Zhengtong and Qiu, Qiang and She, Yu},
  journal={IEEE Robotics and Automation Letters},
  year={2025},
  publisher={IEEE}
}

@article{li2025unified,
  title={Unified video action model},
  author={Li, Shuang and Gao, Yihuai and Sadigh, Dorsa and Song, Shuran},
  journal={arXiv preprint arXiv:2503.00200},
  year={2025}
}

@inproceedings{yang2025roboenvision,
  title={Roboenvision: A long-horizon video generation model for multi-task robot manipulation},
  author={Yang, Liudi and Bai, Yang and Eskandar, George and Shen, Fengyi and Altillawi, Mohammad and Chen, Dong and Majumder, Soumajit and Liu, Ziyuan and Kutyniok, Gitta and Valada, Abhinav},
  booktitle={2025 IEEE/RSJ International Conference on Intelligent Robots and Systems (IROS)},
  pages={21281--21288},
  year={2025},
  organization={IEEE}
}

@inproceedings{
che2024gamegen,
title={GameGen-X: Interactive Open-world Game Video Generation},
author={Haoxuan Che and Xuanhua He and Quande Liu and Cheng Jin and Hao Chen},
booktitle={The Thirteenth International Conference on Learning Representations},
year={2025},
url={https://openreview.net/forum?id=8VG8tpPZhe}
}

@InProceedings{yu2025gamefactory,
    author    = {Yu, Jiwen and Qin, Yiran and Wang, Xintao and Wan, Pengfei and Zhang, Di and Liu, Xihui},
    title     = {GameFactory: Creating New Games with Generative Interactive Videos},
    booktitle = {Proceedings of the IEEE/CVF International Conference on Computer Vision (ICCV)},
    month     = {October},
    year      = {2025},
    pages     = {11590-11599}
}

@article{guo2025mineworld,
  title={Mineworld: a real-time and open-source interactive world model on minecraft},
  author={Guo, Junliang and Ye, Yang and He, Tianyu and Wu, Haoyu and Jiang, Yushu and Pearce, Tim and Bian, Jiang},
  journal={arXiv preprint arXiv:2504.08388},
  year={2025}
}

@article{he2025matrix,
  title={Matrix-game 2.0: An open-source real-time and streaming interactive world model},
  author={He, Xianglong and Peng, Chunli and Liu, Zexiang and Wang, Boyang and Zhang, Yifan and Cui, Qi and Kang, Fei and Jiang, Biao and An, Mengyin and Ren, Yangyang and others},
  journal={arXiv preprint arXiv:2508.13009},
  year={2025}
}

@inproceedings{savov2025exploration,
  title={Exploration-Driven Generative Interactive Environments},
  author={Savov, Nedko and Kazemi, Naser and Mahdi, Mohammad and Paudel, Danda Pani and Wang, Xi and Van Gool, Luc},
  booktitle={Proceedings of the Computer Vision and Pattern Recognition Conference},
  pages={27597--27607},
  year={2025}
}

@article{tang2025hunyuan,
  title={Hunyuan-GameCraft-2: Instruction-following Interactive Game World Model},
  author={Tang, Junshu and Liu, Jiacheng and Li, Jiaqi and Wu, Longhuang and Yang, Haoyu and Zhao, Penghao and Gong, Siruis and Yuan, Xiang and Shao, Shuai and Lu, Qinglin},
  journal={arXiv preprint arXiv:2511.23429},
  year={2025}
}

@article{sun2025worldplay,
  title={WorldPlay: Towards Long-Term Geometric Consistency for Real-Time Interactive World Modeling},
  author={Sun, Wenqiang and Zhang, Haiyu and Wang, Haoyuan and Wu, Junta and Wang, Zehan and Wang, Zhenwei and Wang, Yunhong and Zhang, Jun and Wang, Tengfei and Guo, Chunchao},
  journal={arXiv preprint arXiv:2512.14614},
  year={2025}
}

@misc{robbyantteam2026advancingopensourceworldmodels,
      title={Advancing Open-source World Models}, 
      author={Robbyant Team and Zelin Gao and Qiuyu Wang and Yanhong Zeng and Jiapeng Zhu and Ka Leong Cheng and Yixuan Li and Hanlin Wang and Yinghao Xu and Shuailei Ma and Yihang Chen and Jie Liu and Yansong Cheng and Yao Yao and Jiayi Zhu and Yihao Meng and Kecheng Zheng and Qingyan Bai and Jingye Chen and Zehong Shen and Yue Yu and Xing Zhu and Yujun Shen and Hao Ouyang},
      year={2026},
      eprint={2601.20540},
      archivePrefix={arXiv},
      primaryClass={cs.CV},
      url={https://arxiv.org/abs/2601.20540}, 
}

@online{deepmind_genie3_models,
  title        = {Genie 3},
  subtitle     = {Generate and explore interactive worlds},
  author       = {{Google DeepMind}},
  organization = {Google DeepMind},
  url          = {https://deepmind.google/models/genie/},
}

@InProceedings{zhu2024infp,
    author    = {Zhu, Yongming and Zhang, Longhao and Rong, Zhengkun and Hu, Tianshu and Liang, Shuang and Ge, Zhipeng},
    title     = {INFP: Audio-Driven Interactive Head Generation in Dyadic Conversations},
    booktitle = {Proceedings of the IEEE/CVF Conference on Computer Vision and Pattern Recognition (CVPR)},
    month     = {June},
    year      = {2025},
    pages     = {10667-10677}
}

@misc{sun2025streamavatarstreamingdiffusionmodels,
      title={StreamAvatar: Streaming Diffusion Models for Real-Time Interactive Human Avatars}, 
      author={Zhiyao Sun and Ziqiao Peng and Yifeng Ma and Yi Chen and Zhengguang Zhou and Zixiang Zhou and Guozhen Zhang and Youliang Zhang and Yuan Zhou and Qinglin Lu and Yong-Jin Liu},
      year={2025},
      eprint={2512.22065},
      archivePrefix={arXiv},
      primaryClass={cs.CV},
      url={https://arxiv.org/abs/2512.22065}, 
}

@inproceedings{li2025ditto,
  title={Ditto: Motion-space diffusion for controllable realtime talking head synthesis},
  author={Li, Tianqi and Zheng, Ruobing and Yang, Minghui and Chen, Jingdong and Yang, Ming},
  booktitle={Proceedings of the 33rd ACM International Conference on Multimedia},
  pages={9704--9713},
  year={2025}
}

@misc{shen2025soulx,
      title={SoulX-FlashTalk: Real-Time Infinite Streaming of Audio-Driven Avatars via Self-Correcting Bidirectional Distillation}, 
      author={Le Shen and Qian Qiao and Tan Yu and Ke Zhou and Tianhang Yu and Yu Zhan and Zhenjie Wang and Ming Tao and Shunshun Yin and Siyuan Liu},
      year={2025},
      eprint={2512.23379},
      archivePrefix={arXiv},
      primaryClass={cs.CV},
      url={https://arxiv.org/abs/2512.23379}, 
}

@article{guo2024real,
  title={Real-time One-Step Diffusion-based Expressive Portrait Videos Generation},
  author={Guo, Hanzhong and Yi, Hongwei and Zhou, Daquan and Bergman, Alexander William and Lingelbach, Michael and Yu, Yizhou},
  journal={arXiv preprint arXiv:2412.13479},
  year={2024}
}

@article{huang2025live,
  title={Live avatar: Streaming real-time audio-driven avatar generation with infinite length},
  author={Huang, Yubo and Guo, Hailong and Wu, Fangtai and Zhang, Shifeng and Huang, Shijie and Gan, Qijun and Liu, Lin and Zhao, Sirui and Chen, Enhong and Liu, Jiaming and others},
  journal={arXiv preprint arXiv:2512.04677},
  year={2025}
}

@article{zuo2025edit,
  title={Edit-Your-Interest: Efficient Video Editing via Feature Most-Similar Propagation},
  author={Zuo, Yi and Wang, Zitao and Li, Lingling and Liu, Xu and Liu, Fang and Jiao, Licheng},
  journal={arXiv preprint arXiv:2510.13084},
  year={2025}
}

@inproceedings{cai2025ditctrl,
  title={Ditctrl: Exploring attention control in multi-modal diffusion transformer for tuning-free multi-prompt longer video generation},
  author={Cai, Minghong and Cun, Xiaodong and Li, Xiaoyu and Liu, Wenze and Zhang, Zhaoyang and Zhang, Yong and Shan, Ying and Yue, Xiangyu},
  booktitle={Proceedings of the Computer Vision and Pattern Recognition Conference},
  pages={7763--7772},
  year={2025}
}

@article{gong2023talecrafter,
  title={Talecrafter: Interactive story visualization with multiple characters},
  author={Gong, Yuan and Pang, Youxin and Cun, Xiaodong and Xia, Menghan and He, Yingqing and Chen, Haoxin and Wang, Longyue and Zhang, Yong and Wang, Xintao and Shan, Ying and others},
  journal={arXiv preprint arXiv:2305.18247},
  year={2023}
}

@article{he2023animate,
  title={Animate-a-story: Storytelling with retrieval-augmented video generation},
  author={He, Yingqing and Xia, Menghan and Chen, Haoxin and Cun, Xiaodong and Gong, Yuan and Xing, Jinbo and Zhang, Yong and Wang, Xintao and Weng, Chao and Shan, Ying and others},
  journal={arXiv preprint arXiv:2307.06940},
  year={2023}
}

@article{fang2024vivid,
  title={Vivid: Video virtual try-on using diffusion models},
  author={Fang, Zixun and Zhai, Wei and Su, Aimin and Song, Hongliang and Zhu, Kai and Wang, Mao and Chen, Yu and Liu, Zhiheng and Cao, Yang and Zha, Zheng-Jun},
  journal={arXiv preprint arXiv:2405.11794},
  year={2024}
}

@InProceedings{yahia2024mobile,
    author    = {Ben Yahia, Haitam and Korzhenkov, Denis and Lelekas, Ioannis and Ghodrati, Amir and Habibian, Amirhossein},
    title     = {Mobile Video Diffusion},
    booktitle = {Proceedings of the IEEE/CVF International Conference on Computer Vision (ICCV)},
    month     = {October},
    year      = {2025},
    pages     = {19450-19460}
}

@inproceedings{wu2025snapgen,
  title={Snapgen-v: Generating a five-second video within five seconds on a mobile device},
  author={Wu, Yushu and Zhang, Zhixing and Li, Yanyu and Xu, Yanwu and Kag, Anil and Sui, Yang and Coskun, Huseyin and Ma, Ke and Lebedev, Aleksei and Hu, Ju and others},
  booktitle={Proceedings of the Computer Vision and Pattern Recognition Conference},
  pages={2479--2490},
  year={2025}
}

@article{wu2025taming,
  title={Taming Diffusion Transformer for Efficient Mobile Video Generation in Seconds},
  author={Wu, Yushu and Li, Yanyu and Kag, Anil and Skorokhodov, Ivan and Menapace, Willi and Ma, Ke and Sahni, Arpit and Hu, Ju and Siarohin, Aliaksandr and Sagar, Dhritiman and others},
  journal={arXiv preprint arXiv:2507.13343},
  year={2025}
}

@inproceedings{wu2025individual,
  title={Individual Content and Motion Dynamics Preserved Pruning for Video Diffusion Models},
  author={Wu, Yiming and Chen, Zhenghao and Wang, Huan and Xu, Dong},
  booktitle={Proceedings of the 33rd ACM International Conference on Multimedia},
  pages={9714--9723},
  year={2025}
}

@article{kim2025device,
  title={On-device Sora: Enabling Training-Free Diffusion-based Text-to-Video Generation for Mobile Devices},
  author={Kim, Bosung and Lee, Kyuhwan and Jeong, Isu and Cheon, Jungmin and Lee, Yeojin and Lee, Seulki},
  journal={arXiv preprint arXiv:2502.04363},
  year={2025}
}

@inproceedings{
zhao2025riflex,
title={{RIFLE}x: A Free Lunch for Length Extrapolation in Video Diffusion Transformers},
author={Min Zhao and Guande He and Yixiao Chen and Hongzhou Zhu and Chongxuan Li and Jun Zhu},
booktitle={Forty-second International Conference on Machine Learning},
year={2025},
url={https://openreview.net/forum?id=v3B79m7t8Z}
}

@inproceedings{
zhao2025ultravico,
title={UltraViCo: Breaking Extrapolation Limits in Video Diffusion Transformers},
author={Min Zhao and Hongzhou Zhu and Yingze Wang and Bokai Yan and Jintao Zhang and Guande He and Ling Yang and Chongxuan Li and Jun Zhu},
booktitle={The Fourteenth International Conference on Learning Representations},
year={2026},
url={https://openreview.net/forum?id=fLLCmC53u9}
}

@article{yesiltepe2025infinity,
  title={Infinity-RoPE: Action-Controllable Infinite Video Generation Emerges From Autoregressive Self-Rollout},
  author={Yesiltepe, Hidir and Meral, Tuna Han Salih and Akan, Adil Kaan and Oktay, Kaan and Yanardag, Pinar},
  journal={arXiv preprint arXiv:2511.20649},
  year={2025}
}

@inproceedings{
qiu2024freenoise,
title={FreeNoise: Tuning-Free Longer Video Diffusion via Noise Rescheduling},
author={Haonan Qiu and Menghan Xia and Yong Zhang and Yingqing He and Xintao Wang and Ying Shan and Ziwei Liu},
booktitle={The Twelfth International Conference on Learning Representations},
year={2024},
url={https://openreview.net/forum?id=ijoqFqSC7p}
}

@inproceedings{
zhang2025sageattention,
title={SageAttention: Accurate 8-Bit Attention for Plug-and-play Inference Acceleration},
author={Jintao Zhang and Jia wei and Pengle Zhang and Jun Zhu and Jianfei Chen},
booktitle={The Thirteenth International Conference on Learning Representations},
year={2025},
url={https://openreview.net/forum?id=OL44KtasKc}
}

@inproceedings{
liu2025fpsattention,
title={{FPSA}ttention: Training-Aware {FP}8 and Sparsity Co-Design for Fast Video Diffusion},
author={Akide Liu and Zeyu Zhang and Zhexin Li and Xuehai Bai and Yuanjie Xing and Yizeng Han and Jiasheng Tang and Jichao Wu and Mingyang Yang and Weihua Chen and Jiahao He and Yuanyu He and Fan Wang and Gholamreza Haffari and Bohan Zhuang},
booktitle={The Thirty-ninth Annual Conference on Neural Information Processing Systems},
year={2025},
url={https://openreview.net/forum?id=T62TYoF8R3}
}

@inproceedings{
zhang2025sageattention2,
title={SageAttention2: Efficient Attention with Thorough Outlier Smoothing and Per-thread {INT}4 Quantization},
author={Jintao Zhang and Haofeng Huang and Pengle Zhang and Jia wei and Jun Zhu and Jianfei Chen},
booktitle={Forty-second International Conference on Machine Learning},
year={2025},
url={https://openreview.net/forum?id=nC8XliUxeg}
}

@article{zhang2025sageattention2++,
  title={Sageattention2++: A more efficient implementation of sageattention2},
  author={Zhang, Jintao and Xu, Xiaoming and Wei, Jia and Huang, Haofeng and Zhang, Pengle and Xiang, Chendong and Zhu, Jun and Chen, Jianfei},
  journal={arXiv preprint arXiv:2505.21136},
  year={2025}
}

@inproceedings{
zhang2025sageattention3,
title={SageAttention3: Microscaling {FP}4 Attention for Inference and An Exploration of 8-Bit Training},
author={Jintao Zhang and Jia wei and Haoxu Wang and Pengle Zhang and Xiaoming Xu and Haofeng Huang and Kai Jiang and Jun Zhu and Jianfei Chen},
booktitle={The Thirty-ninth Annual Conference on Neural Information Processing Systems},
year={2025},
url={https://openreview.net/forum?id=JbJVWljk7r}
}

@inproceedings{
zhao2025viditq,
title={ViDiT-Q: Efficient and Accurate Quantization of Diffusion Transformers for Image and Video Generation},
author={Tianchen Zhao and Tongcheng Fang and Haofeng Huang and Rui Wan and Widyadewi Soedarmadji and Enshu Liu and Shiyao Li and Zinan Lin and Guohao Dai and Shengen Yan and Huazhong Yang and Xuefei Ning and Yu Wang},
booktitle={The Thirteenth International Conference on Learning Representations},
year={2025},
url={https://openreview.net/forum?id=E1N1oxd63b}
}

@inproceedings{
li2025dvd,
title={{DVD}-Quant: Data-free Video Diffusion Transformers Quantization},
author={Zhiteng Li and Hanxuan Li and Junyi Wu and Kai Liu and Haotong Qin and Linghe Kong and Guihai Chen and Yulun Zhang and Xiaokang Yang},
booktitle={The Fourteenth International Conference on Learning Representations},
year={2026},
url={https://openreview.net/forum?id=3AnRMvlVDw}
}

@article{yang2025lrq,
  title={LRQ-DiT: Log-Rotation Post-Training Quantization of Diffusion Transformers for Image and Video Generation},
  author={Yang, Lianwei and Lin, Haokun and Zhao, Tianchen and Wu, Yichen and Zhu, Hongyu and Xie, Ruiqi and Sun, Zhenan and Wang, Yu and Gu, Qingyi},
  journal={arXiv preprint arXiv:2508.03485},
  year={2025}
}

@article{zhang2026adatsq,
  title={AdaTSQ: Pushing the Pareto Frontier of Diffusion Transformers via Temporal-Sensitivity Quantization},
  author={Zhang, Shaoqiu and Ding, Zizhong and Yang, Kaicheng and Wu, Junyi and Yan, Xianglong and Li, Xi and Duan, Bingnan and Fang, Jianping and Zhang, Yulun},
  journal={arXiv preprint arXiv:2602.09883},
  year={2026}
}

@inproceedings{
huang2026qvgen,
title={{QVG}en: Pushing the Limit of Quantized Video Generative Models},
author={Yushi Huang and Ruihao Gong and Jing Liu and Yifu Ding and Chengtao Lv and Haotong Qin and Jun Zhang},
booktitle={The Fourteenth International Conference on Learning Representations},
year={2026},
url={https://openreview.net/forum?id=XJXZXuTj11}
}

@InProceedings{Wu_2025_ICCV,
    author    = {Wu, Junyi and Li, Zhiteng and Hui, Zheng and Zhang, Yulun and Kong, Linghe and Yang, Xiaokang},
    title     = {QuantCache: Adaptive Importance-Guided Quantization with Hierarchical Latent and Layer Caching for Video Generation},
    booktitle = {Proceedings of the IEEE/CVF International Conference on Computer Vision (ICCV)},
    month     = {October},
    year      = {2025},
    pages     = {15035-15044}
}

@inproceedings{bolya2023token,
  title={Token merging for fast stable diffusion},
  author={Bolya, Daniel and Hoffman, Judy},
  booktitle={Proceedings of the IEEE/CVF conference on computer vision and pattern recognition},
  pages={4599--4603},
  year={2023}
}

@inproceedings{li2024vidtome,
  title={Vidtome: Video token merging for zero-shot video editing},
  author={Li, Xirui and Ma, Chao and Yang, Xiaokang and Yang, Ming-Hsuan},
  booktitle={Proceedings of the IEEE/CVF Conference on Computer Vision and Pattern Recognition},
  pages={7486--7495},
  year={2024}
}

@inproceedings{wu2025importance,
  title={Importance-based token merging for efficient image and video generation},
  author={Wu, Haoyu and Xu, Jingyi and Le, Hieu and Samaras, Dimitris},
  booktitle={Proceedings of the IEEE/CVF International Conference on Computer Vision},
  pages={4983--4995},
  year={2025}
}

@inproceedings{anagnostidis2025flexidit,
  title={Flexidit: Your diffusion transformer can easily generate high-quality samples with less compute},
  author={Anagnostidis, Sotiris and Bachmann, Gregor and Kim, Yeongmin and Kohler, Jonas and Georgopoulos, Markos and Sanakoyeu, Artsiom and Du, Yuming and Pumarola, Albert and Thabet, Ali and Sch{\"o}nfeld, Edgar},
  booktitle={Proceedings of the Computer Vision and Pattern Recognition Conference},
  pages={28316--28326},
  year={2025}
}

@inproceedings{
zhang2025trainingfree,
title={Training-Free Efficient Video Generation via Dynamic Token Carving},
author={Yuechen Zhang and Jinbo Xing and Bin Xia and Shaoteng Liu and Bohao PENG and Xin Tao and Pengfei Wan and Eric Lo and Jiaya Jia},
booktitle={The Thirty-ninth Annual Conference on Neural Information Processing Systems},
year={2025},
url={https://openreview.net/forum?id=CdkFnJSG4G}
}

@article{song2025hero,
  title={Hero: Hierarchical extrapolation and refresh for efficient world models},
  author={Song, Quanjian and Wang, Xinyu and Zhou, Donghao and Lin, Jingyu and Chen, Cunjian and Ma, Yue and Li, Xiu},
  journal={arXiv preprint arXiv:2508.17588},
  year={2025}
}

@inproceedings{sun2025unicp,
  title={Unicp: A unified caching and pruning framework for efficient video generation},
  author={Sun, Wenzhang and Hou, Qirui and Di, Donglin and Yang, Jiahui and Ma, Yongjia and Cui, Jianxun},
  booktitle={Proceedings of the 7th ACM International Conference on Multimedia in Asia},
  pages={1--7},
  year={2025}
}

@inproceedings{
xie2025glad,
title={Glad: A Streaming Scene Generator for Autonomous Driving},
author={Bin Xie and Yingfei Liu and Tiancai Wang and Jiale Cao and Xiangyu Zhang},
booktitle={The Thirteenth International Conference on Learning Representations},
year={2025},
url={https://openreview.net/forum?id=ZFxpclrCCf}
}

@inproceedings{wang2025stage,
  title={STAGE: A stream-centric generative world model for long-horizon driving-scene simulation},
  author={Wang, Jiamin and Yao, Yichen and Feng, Xiang and Wu, Hang and Wang, Yaming and Huang, Qingqiu and Ma, Yuexin and Zhu, Xinge},
  booktitle={2025 IEEE/RSJ International Conference on Intelligent Robots and Systems (IROS)},
  pages={14163--14169},
  year={2025},
  organization={IEEE}
}

@article{rahimi2026mad,
  title={MAD: Motion Appearance Decoupling for efficient Driving World Models},
  author={Rahimi, Ahmad and Gerard, Valentin and Zablocki, Eloi and Cord, Matthieu and Alahi, Alexandre},
  journal={arXiv preprint arXiv:2601.09452},
  year={2026}
}

@inproceedings{li2025uniscene,
  title={Uniscene: Unified occupancy-centric driving scene generation},
  author={Li, Bohan and Guo, Jiazhe and Liu, Hongsi and Zou, Yingshuang and Ding, Yikang and Chen, Xiwu and Zhu, Hu and Tan, Feiyang and Zhang, Chi and Wang, Tiancai and others},
  booktitle={Proceedings of the Computer Vision and Pattern Recognition Conference},
  pages={11971--11981},
  year={2025}
}

@inproceedings{
zhu2025worldsplat,
title={WorldSplat: Gaussian-Centric Feed-Forward 4D Scene Generation for Autonomous Driving},
author={Ziyue Zhu and Zhanqian Wu and Zhenxin Zhu and Lijun Zhou and Haiyang Sun and BING WANG and Kun Ma and Guang Chen and Hangjun Ye and Jin Xie and jian Yang},
booktitle={The Fourteenth International Conference on Learning Representations},
year={2026},
url={https://openreview.net/forum?id=KWeX6tYno6}
}

@inproceedings{zhu2025other,
  title={Other vehicle trajectories are also needed: A driving world model unifies ego-other vehicle trajectories in video latent space},
  author={Zhu, Jian and Jia, Zhengyu and Gao, Tian and Deng, Jiaxin and Li, Shidi and Zhang, Lang and Liu, Fu and Jia, Peng and Lang, Xianpeng},
  booktitle={Proceedings of the AAAI Conference on Artificial Intelligence},
  volume={40},
  number={16},
  pages={13934--13942},
  year={2026}
}

@article{santana2016learning,
  title={Learning a driving simulator},
  author={Santana, Eder and Hotz, George},
  journal={arXiv preprint arXiv:1608.01230},
  year={2016}
}

@inproceedings{gao2025magicdrive,
  title={MagicDrive-V2: High-resolution long video generation for autonomous driving with adaptive control},
  author={Gao, Ruiyuan and Chen, Kai and Xiao, Bo and Hong, Lanqing and Li, Zhenguo and Xu, Qiang},
  booktitle={Proceedings of the IEEE/CVF International Conference on Computer Vision},
  pages={28135--28144},
  year={2025}
}

@inproceedings{yang2025drivearena,
  title={Drivearena: A closed-loop generative simulation platform for autonomous driving},
  author={Yang, Xuemeng and Wen, Licheng and Wei, Tiantian and Ma, Yukai and Mei, Jianbiao and Li, Xin and Lei, Wenjie and Fu, Daocheng and Cai, Pinlong and Dou, Min and others},
  booktitle={Proceedings of the IEEE/CVF International Conference on Computer Vision},
  pages={26933--26943},
  year={2025}
}

@article{xia2025drivelaw,
  title={DriveLaW: Unifying Planning and Video Generation in a Latent Driving World},
  author={Xia, Tianze and Li, Yongkang and Zhou, Lijun and Yao, Jingfeng and Xiong, Kaixin and Sun, Haiyang and Wang, Bing and Ma, Kun and Ye, Hangjun and Liu, Wenyu and others},
  journal={arXiv preprint arXiv:2512.23421},
  year={2025}
}

@inproceedings{lu2024wovogen,
  title={Wovogen: World volume-aware diffusion for controllable multi-camera driving scene generation},
  author={Lu, Jiachen and Huang, Ze and Yang, Zeyu and Zhang, Jiahui and Zhang, Li},
  booktitle={European Conference on Computer Vision},
  pages={329--345},
  year={2024},
  organization={Springer}
}

@inproceedings{chen2025drivinggpt,
  title={Drivinggpt: Unifying driving world modeling and planning with multi-modal autoregressive transformers},
  author={Chen, Yuntao and Wang, Yuqi and Zhang, Zhaoxiang},
  booktitle={Proceedings of the IEEE/CVF International Conference on Computer Vision},
  pages={26890--26900},
  year={2025}
}

@article{bartoccioni2025vavim,
  title={Vavim and vavam: Autonomous driving through video generative modeling},
  author={Bartoccioni, Florent and Ramzi, Elias and Besnier, Victor and Venkataramanan, Shashanka and Vu, Tuan-Hung and Xu, Yihong and Chambon, Loick and Gidaris, Spyros and Odabas, Serkan and Hurych, David and others},
  journal={arXiv preprint arXiv:2502.15672},
  year={2025}
}

@article{team2025gigaworld,
  title={Gigaworld-0: World models as data engine to empower embodied ai},
  author={Team, GigaWorld and Ye, Angen and Wang, Boyuan and Ni, Chaojun and Huang, Guan and Zhao, Guosheng and Li, Haoyun and Zhu, Jiagang and Li, Kerui and Xu, Mengyuan and others},
  journal={arXiv preprint arXiv:2511.19861},
  year={2025}
}

@article{patel2025robotic,
  title={Robotic Manipulation by Imitating Generated Videos Without Physical Demonstrations},
  author={Patel, Shivansh and Mohan, Shraddhaa and Mai, Hanlin and Jain, Unnat and Lazebnik, Svetlana and Li, Yunzhu},
  journal={arXiv preprint arXiv:2507.00990},
  year={2025}
}

@article{qiu2025lucibot,
  title={LuciBot: Automated Robot Policy Learning from Generated Videos},
  author={Qiu, Xiaowen and Wang, Yian and Cai, Jiting and Chen, Zhehuan and Lin, Chunru and Wang, Tsun-Hsuan and Gan, Chuang},
  journal={arXiv preprint arXiv:2503.09871},
  year={2025}
}

@inproceedings{
bharadhwaj2024gen2act,
title={Gen2Act: Human Video Generation in Novel Scenarios enables Generalizable Robot Manipulation},
author={Homanga Bharadhwaj and Debidatta Dwibedi and Abhinav Gupta and Shubham Tulsiani and Carl Doersch and Ted Xiao and Dhruv Shah and Fei Xia and Dorsa Sadigh and Sean Kirmani},
booktitle={9th Annual Conference on Robot Learning},
year={2025},
url={https://openreview.net/forum?id=HprBJupvvM}
}

@article{soni2024videoagent,
  title={Videoagent: Self-improving video generation},
  author={Soni, Achint and Venkataraman, Sreyas and Chandra, Abhranil and Fischmeister, Sebastian and Liang, Percy and Dai, Bo and Yang, Sherry},
  journal={arXiv preprint arXiv:2410.10076},
  year={2024}
}

@inproceedings{
liang2024dreamitate,
title={Dreamitate: Real-World Visuomotor Policy Learning via Video Generation},
author={Junbang Liang and Ruoshi Liu and Ege Ozguroglu and Sruthi Sudhakar and Achal Dave and Pavel Tokmakov and Shuran Song and Carl Vondrick},
booktitle={8th Annual Conference on Robot Learning},
year={2024},
url={https://openreview.net/forum?id=InT87E5sr4}
}

@inproceedings{
zhang2025gevrm,
title={{GEVRM}: Goal-Expressive Video Generation Model For Robust Visual Manipulation},
author={Hongyin Zhang and Pengxiang Ding and Shangke Lyu and Ying Peng and Donglin Wang},
booktitle={The Thirteenth International Conference on Learning Representations},
year={2025},
url={https://openreview.net/forum?id=hPWWXpCaJ7}
}

@article{escontrela2023video,
  title={Video prediction models as rewards for reinforcement learning},
  author={Escontrela, Alejandro and Adeniji, Ademi and Yan, Wilson and Jain, Ajay and Peng, Xue Bin and Goldberg, Ken and Lee, Youngwoon and Hafner, Danijar and Abbeel, Pieter},
  journal={Advances in Neural Information Processing Systems},
  volume={36},
  pages={68760--68783},
  year={2023}
}

@article{sharma2026world,
  title={World-Gymnast: Training Robots with Reinforcement Learning in a World Model},
  author={Sharma, Ansh Kumar and Sun, Yixiang and Lu, Ninghao and Zhang, Yunzhe and Liu, Jiarao and Yang, Sherry},
  journal={arXiv preprint arXiv:2602.02454},
  year={2026}
}

@article{li2025worldeval,
  title={WorldEval: World Model as Real-World Robot Policies Evaluator},
  author={Li, Yaxuan and Zhu, Yichen and Wen, Junjie and Shen, Chaomin and Xu, Yi},
  journal={arXiv preprint arXiv:2505.19017},
  year={2025}
}

@inproceedings{
huang2025enerverseenvisioningembodiedfuture,
title={EnerVerse: Envisioning Embodied Future Space for Robotics Manipulation},
author={Siyuan Huang and Liliang Chen and Pengfei Zhou and Shengcong Chen and Yue Liao and Zhengkai Jiang and Yue Hu and Peng Gao and Hongsheng Li and Maoqing Yao and Guanghui Ren},
booktitle={The Thirty-ninth Annual Conference on Neural Information Processing Systems},
year={2025},
url={https://openreview.net/forum?id=igtjRQfght}
}

@inproceedings{
ge2025,
title={Genie Envisioner: A Unified World Foundation Platform for Robotic Manipulation},
author={Yue Liao and Pengfei Zhou and Siyuan Huang and Donglin Yang and Shengcong Chen and Yuxin Jiang and Yue Hu and Si Liu and Jianlan Luo and Liliang Chen and Shuicheng YAN and Maoqing Yao and Guanghui Ren},
booktitle={The Fourteenth International Conference on Learning Representations},
year={2026},
url={https://openreview.net/forum?id=fHLtSxDFKC}
}

@misc{yang2024playablegamegeneration,
      title={Playable Game Generation}, 
      author={Mingyu Yang and Junyou Li and Zhongbin Fang and Sheng Chen and Yangbin Yu and Qiang Fu and Wei Yang and Deheng Ye},
      year={2024},
      eprint={2412.00887},
      archivePrefix={arXiv},
      primaryClass={cs.AI},
      url={https://arxiv.org/abs/2412.00887}
}

@article{lingbotva2026,
  title={Causal World Modeling for Robot Control},
  author={Li, Lin and Zhang, Qihang and Luo, Yiming and Yang, Shuai and Wang, Ruilin and Han, Fei and Yu, Mingrui and Gao, Zelin and Xue, Nan and Zhu, Xing and Shen, Yujun and Xu, Yinghao},
  journal={arXiv preprint arXiv:2601.21998},
  year={2026}
}

@misc{li2025hunyuangamecrafthighdynamicinteractivegame,
    title={Hunyuan-GameCraft: High-dynamic Interactive Game Video Generation with Hybrid History Condition}, 
    author={Jiaqi Li and Junshu Tang and Zhiyong Xu and Longhuang Wu and Yuan Zhou and Shuai Shao and Tianbao Yu and Zhiguo Cao and Qinglin Lu},
    year={2025},
    eprint={2506.17201},
    archivePrefix={arXiv},
    primaryClass={cs.CV},
    url={https://arxiv.org/abs/2506.17201}, 
}

@article{zhang2025matrixgame,
  title     = {Matrix-Game: Interactive World Foundation Model},
  author    = {Yifan Zhang and Chunli Peng and Boyang Wang and Puyi Wang and Qingcheng Zhu and Zedong Gao and Eric Li and Yang Liu and Yahui Zhou},
  journal   = {arXiv preprint arXiv:2506.18701},
  year      = {2025}
}

@article{hafner2025dreamerv4,
    title   = {Training Agents Inside of Scalable World Models},
    author  = {Hafner, Danijar and Yan, Wilson and Lillicrap, Timothy},
    journal = {arXiv preprint arXiv:2509.24527},
    year    = {2025},
    url     = {https://arxiv.org/abs/2509.24527}
}

@misc{ren2025cosmosdrivedreamsscalablesyntheticdriving,
      title={Cosmos-Drive-Dreams: Scalable Synthetic Driving Data Generation with World Foundation Models}, 
      author={Xuanchi Ren and Yifan Lu and Tianshi Cao and Ruiyuan Gao and Shengyu Huang and Amirmojtaba Sabour and Tianchang Shen and Tobias Pfaff and Jay Zhangjie Wu and Runjian Chen and Seung Wook Kim and Jun Gao and Laura Leal-Taixe and Mike Chen and Sanja Fidler and Huan Ling},
      year={2025},
      eprint={2506.09042},
      archivePrefix={arXiv},
      primaryClass={cs.CV},
      url={https://arxiv.org/abs/2506.09042}, 
}

@misc{gao2026dreamdojogeneralistrobotworld,
      title={DreamDojo: A Generalist Robot World Model from Large-Scale Human Videos}, 
      author={Shenyuan Gao and William Liang and Kaiyuan Zheng and Ayaan Malik and Seonghyeon Ye and Sihyun Yu and Wei-Cheng Tseng and Yuzhu Dong and Kaichun Mo and Chen-Hsuan Lin and Qianli Ma and Seungjun Nah and Loic Magne and Jiannan Xiang and Yuqi Xie and Ruijie Zheng and Dantong Niu and You Liang Tan and K. R. Zentner and George Kurian and Suneel Indupuru and Pooya Jannaty and Jinwei Gu and Jun Zhang and Jitendra Malik and Pieter Abbeel and Ming-Yu Liu and Yuke Zhu and Joel Jang and Linxi "Jim" Fan},
      year={2026},
      eprint={2602.06949},
      archivePrefix={arXiv},
      primaryClass={cs.RO},
      url={https://arxiv.org/abs/2602.06949}, 
}

@inproceedings{
kim2026cosmospolicyfinetuningvideo,
title={Cosmos Policy: Fine-Tuning Video Models for Visuomotor Control and Planning},
author={Moo Jin Kim and Yihuai Gao and Tsung-Yi Lin and Yen-Chen Lin and Yunhao Ge and Grace Lam and Percy Liang and Shuran Song and Ming-Yu Liu and Chelsea Finn and Jinwei Gu},
booktitle={The Fourteenth International Conference on Learning Representations},
year={2026},
url={https://openreview.net/forum?id=wPEIStHxYH}
}

@misc{lee2024vividdreamgenerating3dscene,
      title={VividDream: Generating 3D Scene with Ambient Dynamics}, 
      author={Yao-Chih Lee and Yi-Ting Chen and Andrew Wang and Ting-Hsuan Liao and Brandon Y. Feng and Jia-Bin Huang},
      year={2024},
      eprint={2405.20334},
      archivePrefix={arXiv},
      primaryClass={cs.CV},
      url={https://arxiv.org/abs/2405.20334}, 
}

@inproceedings{
yu2024real,
title={4Real: Towards Photorealistic 4D Scene Generation via Video Diffusion Models},
author={Heng Yu and Chaoyang Wang and Peiye Zhuang and Willi Menapace and Aliaksandr Siarohin and Junli Cao and Laszlo Attila Jeni and Sergey Tulyakov and Hsin-Ying Lee},
booktitle={The Thirty-eighth Annual Conference on Neural Information Processing Systems},
year={2024},
url={https://openreview.net/forum?id=SO1aRpwVLk}
}

@InProceedings{liu2025free4dtuningfree4dscene,
    author    = {Liu, Tianqi and Huang, Zihao and Chen, Zhaoxi and Wang, Guangcong and Hu, Shoukang and Shen, Liao and Sun, Huiqiang and Cao, Zhiguo and Li, Wei and Liu, Ziwei},
    title     = {Free4D: Tuning-free 4D Scene Generation with Spatial-Temporal Consistency},
    booktitle = {Proceedings of the IEEE/CVF International Conference on Computer Vision (ICCV)},
    month     = {October},
    year      = {2025},
    pages     = {25571-25582}
}

@InProceedings{wu2024cat4dcreate4dmultiview,
    author    = {Wu, Rundi and Gao, Ruiqi and Poole, Ben and Trevithick, Alex and Zheng, Changxi and Barron, Jonathan T. and Holynski, Aleksander},
    title     = {CAT4D: Create Anything in 4D with Multi-View Video Diffusion Models},
    booktitle = {Proceedings of the IEEE/CVF Conference on Computer Vision and Pattern Recognition (CVPR)},
    month     = {June},
    year      = {2025},
    pages     = {26057-26068}
}

@InProceedings{zhai2025stargenspatiotemporalautoregressionframework,
    author    = {Zhai, Shangjin and Ye, Zhichao and Liu, Jialin and Xie, Weijian and Hu, Jiaqi and Peng, Zhen and Xue, Hua and Chen, Danpeng and Wang, Xiaomeng and Yang, Lei and Wang, Nan and Liu, Haomin and Zhang, Guofeng},
    title     = {StarGen: A Spatiotemporal Autoregression Framework with Video Diffusion Model for Scalable and Controllable Scene Generation},
    booktitle = {Proceedings of the IEEE/CVF Conference on Computer Vision and Pattern Recognition (CVPR)},
    month     = {June},
    year      = {2025},
    pages     = {26822-26833}
}

@misc{huang2026gen3r3dscenegeneration,
      title={Gen3R: 3D Scene Generation Meets Feed-Forward Reconstruction}, 
      author={Jiaxin Huang and Yuanbo Yang and Bangbang Yang and Lin Ma and Yuewen Ma and Yiyi Liao},
      year={2026},
      eprint={2601.04090},
      archivePrefix={arXiv},
      primaryClass={cs.CV},
      url={https://arxiv.org/abs/2601.04090}, 
}

@misc{yang2025infinitetalkaudiodrivenvideogeneration,
      title={InfiniteTalk: Audio-driven Video Generation for Sparse-Frame Video Dubbing}, 
      author={Shaoshu Yang and Zhe Kong and Feng Gao and Meng Cheng and Xiangyu Liu and Yong Zhang and Zhuoliang Kang and Wenhan Luo and Xunliang Cai and Ran He and Xiaoming Wei},
      year={2025},
      eprint={2508.14033},
      archivePrefix={arXiv},
      primaryClass={cs.CV},
      url={https://arxiv.org/abs/2508.14033}, 
}

@inproceedings{
tu2025PlayerOne,
title={PlayerOne: Egocentric World Simulator},
author={Yuanpeng Tu and Hao Luo and Xi Chen and Xiang Bai and Fan Wang and Hengshuang Zhao},
booktitle={The Thirty-ninth Annual Conference on Neural Information Processing Systems},
year={2025},
url={https://openreview.net/forum?id=Gq4Gay8rDB}
}

@article{yu2025videossm,
  title={Videossm: Autoregressive long video generation with hybrid state-space memory},
  author={Yu, Yifei and Wu, Xiaoshan and Hu, Xinting and Hu, Tao and Sun, Yangtian and Lyu, Xiaoyang and Wang, Bo and Ma, Lin and Ma, Yuewen and Wang, Zhongrui and others},
  journal={arXiv preprint arXiv:2512.04519},
  year={2025}
}

@article{ji2025memflow,
  title={Memflow: Flowing adaptive memory for consistent and efficient long video narratives},
  author={Ji, Sihui and Chen, Xi and Yang, Shuai and Tao, Xin and Wan, Pengfei and Zhao, Hengshuang},
  journal={arXiv preprint arXiv:2512.14699},
  year={2025}
}

@incollection{wang2024animatelcm,
  title={Animatelcm: Computation-efficient personalized style video generation without personalized video data},
  author={Wang, Fu-Yun and Huang, Zhaoyang and Bian, Weikang and Shi, Xiaoyu and Sun, Keqiang and Song, Guanglu and Liu, Yu and Li, Hongsheng},
  booktitle={SIGGRAPH Asia 2024 Technical Communications},
  pages={1--5},
  year={2024}
}

@article{wang2024videolcm,
  title={Videolcm: Video latent consistency model},
  author={Wang, Xiang and Zhang, Shiwei and Zhang, Han and Liu, Yu and Zhang, Yingya and Gao, Changxin and Sang, Nong},
  journal={arXiv preprint arXiv:2312.09109},
  year={2023}
}

@article{zhu2024accelerating,
  title={Accelerating video diffusion models via distribution matching},
  author={Zhu, Yuanzhi and Yan, Hanshu and Yang, Huan and Zhang, Kai and Li, Junnan},
  journal={arXiv preprint arXiv:2412.05899},
  year={2024}
}

@article{yin2024improved,
  title={Improved distribution matching distillation for fast image synthesis},
  author={Yin, Tianwei and Gharbi, Micha{\"e}l and Park, Taesung and Zhang, Richard and Shechtman, Eli and Durand, Fredo and Freeman, Bill},
  journal={Advances in neural information processing systems},
  volume={37},
  pages={47455--47487},
  year={2024}
}

@inproceedings{yin2024one,
  title={One-step diffusion with distribution matching distillation},
  author={Yin, Tianwei and Gharbi, Micha{\"e}l and Zhang, Richard and Shechtman, Eli and Durand, Fredo and Freeman, William T and Park, Taesung},
  booktitle={Proceedings of the IEEE/CVF conference on computer vision and pattern recognition},
  pages={6613--6623},
  year={2024}
}

@inproceedings{
kondratyuk2023videopoet,
title={VideoPoet: A Large Language Model for Zero-Shot Video Generation},
author={Dan Kondratyuk and Lijun Yu and Xiuye Gu and Jose Lezama and Jonathan Huang and Grant Schindler and Rachel Hornung and Vighnesh Birodkar and Jimmy Yan and Ming-Chang Chiu and Krishna Somandepalli and Hassan Akbari and Yair Alon and Yong Cheng and Joshua V. Dillon and Agrim Gupta and Meera Hahn and Anja Hauth and David Hendon and Alonso Martinez and David Minnen and Mikhail Sirotenko and Kihyuk Sohn and Xuan Yang and Hartwig Adam and Ming-Hsuan Yang and Irfan Essa and Huisheng Wang and David A Ross and Bryan Seybold and Lu Jiang},
booktitle={Forty-first International Conference on Machine Learning},
year={2024},
url={https://openreview.net/forum?id=LRkJwPIDuE}
}

@inproceedings{
bruce2024genie,
title={Genie: Generative Interactive Environments},
author={Jake Bruce and Michael D Dennis and Ashley Edwards and Jack Parker-Holder and Yuge Shi and Edward Hughes and Matthew Lai and Aditi Mavalankar and Richie Steigerwald and Chris Apps and Yusuf Aytar and Sarah Maria Elisabeth Bechtle and Feryal Behbahani and Stephanie C.Y. Chan and Nicolas Heess and Lucy Gonzalez and Simon Osindero and Sherjil Ozair and Scott Reed and Jingwei Zhang and Konrad Zolna and Jeff Clune and Nando de Freitas and Satinder Singh and Tim Rockt{\"a}schel},
booktitle={Forty-first International Conference on Machine Learning},
year={2024},
url={https://openreview.net/forum?id=bJbSbJskOS}
}

@inproceedings{
chen2024diffusionforcing,
title={Diffusion Forcing: Next-token Prediction Meets Full-Sequence Diffusion},
author={Boyuan Chen and Diego Mart{\'\i} Mons{\'o} and Yilun Du and Max Simchowitz and Russ Tedrake and Vincent Sitzmann},
booktitle={The Thirty-eighth Annual Conference on Neural Information Processing Systems},
year={2024},
url={https://openreview.net/forum?id=yDo1ynArjj}
}

@inproceedings{
huang2025selfforcing,
title={Self Forcing: Bridging the Train-Test Gap in Autoregressive Video Diffusion},
author={Xun Huang and Zhengqi Li and Guande He and Mingyuan Zhou and Eli Shechtman},
booktitle={The Thirty-ninth Annual Conference on Neural Information Processing Systems},
year={2025},
url={https://openreview.net/forum?id=mSiN7i0BYH}
}

@inproceedings{
rollingforcing2025,
title={Rolling Forcing: Autoregressive Long Video Diffusion in Real Time},
author={Kunhao Liu and Wenbo Hu and Jiale Xu and Ying Shan and Shijian Lu},
booktitle={The Fourteenth International Conference on Learning Representations},
year={2026},
url={https://openreview.net/forum?id=IAyzXjbfwo}
}

@article{chen2024videoinfinity,
  title={Video-infinity: Distributed long video generation},
  author={Tan, Zhenxiong and Yang, Xingyi and Liu, Songhua and Wang, Xinchao},
  journal={arXiv preprint arXiv:2406.16260},
  year={2024}
}

@InProceedings{li2024distrifusion,
    author    = {Li, Muyang and Cai, Tianle and Cao, Jiaxin and Zhang, Qinsheng and Cai, Han and Bai, Junjie and Jia, Yangqing and Li, Kai and Han, Song},
    title     = {DistriFusion: Distributed Parallel Inference for High-Resolution Diffusion Models},
    booktitle = {Proceedings of the IEEE/CVF Conference on Computer Vision and Pattern Recognition (CVPR)},
    month     = {June},
    year      = {2024},
    pages     = {7183-7193}
}

@InProceedings{liu2024teacache,
    author    = {Liu, Feng and Zhang, Shiwei and Wang, Xiaofeng and Wei, Yujie and Qiu, Haonan and Zhao, Yuzhong and Zhang, Yingya and Ye, Qixiang and Wan, Fang},
    title     = {Timestep Embedding Tells: It's Time to Cache for Video Diffusion Model},
    booktitle = {Proceedings of the IEEE/CVF Conference on Computer Vision and Pattern Recognition (CVPR)},
    month     = {June},
    year      = {2025},
    pages     = {7353-7363}
}

@inproceedings{
cui2025selfforcingplus,
title={Self-Forcing++: Towards Minute-Scale High-Quality Video Generation},
author={Justin Cui and Jie Wu and Ming Li and Tao Yang and Xiaojie Li and Rui Wang and Andrew Bai and Yuanhao Ban and Cho-Jui Hsieh},
booktitle={The Fourteenth International Conference on Learning Representations},
year={2026},
url={https://openreview.net/forum?id=DzvPiqh23f}
}

@article{zhu2026causalforcing,
  title={Causal Forcing: Autoregressive Diffusion Distillation Done Right for High-Quality Real-Time Interactive Video Generation},
  author={Zhu, Hongzhou and Zhao, Min and He, Guande and Su, Hang and Li, Chongxuan and Zhu, Jun},
  journal={arXiv preprint arXiv:2602.02214},
  year={2026}
}

@article{lu2025rewardforcing,
  title={Reward forcing: Efficient streaming video generation with rewarded distribution matching distillation},
  author={Lu, Yunhong and Zeng, Yanhong and Li, Haobo and Ouyang, Hao and Wang, Qiuyu and Cheng, Ka Leong and Zhu, Jiapeng and Cao, Hengyuan and Zhang, Zhipeng and Zhu, Xing and others},
  journal={arXiv preprint arXiv:2512.04678},
  year={2025}
}

@inproceedings{
lin2025aapt,
title={Autoregressive Adversarial Post-Training for Real-Time Interactive Video Generation},
author={Shanchuan Lin and Ceyuan Yang and Hao He and Jianwen Jiang and Yuxi Ren and Xin Xia and Yang Zhao and Xuefeng Xiao and Lu Jiang},
booktitle={The Thirty-ninth Annual Conference on Neural Information Processing Systems},
year={2025},
url={https://openreview.net/forum?id=lF6SHARvmG}
}

@article{wan2025wan,
  title={Wan: Open and advanced large-scale video generative models},
  author={Wan, Team and Wang, Ang and Ai, Baole and Wen, Bin and Mao, Chaojie and Xie, Chen-Wei and Chen, Di and Yu, Feiwu and Zhao, Haiming and Yang, Jianxiao and others},
  journal={arXiv preprint arXiv:2503.20314},
  year={2025}
}

@inproceedings{tian2024emo,
  title={Emo: Emote portrait alive generating expressive portrait videos with audio2video diffusion model under weak conditions},
  author={Tian, Linrui and Wang, Qi and Zhang, Bang and Bo, Liefeng},
  booktitle={European Conference on Computer Vision},
  pages={244--260},
  year={2024},
  organization={Springer}
}

@article{yi2025magicinfinite,
  title={Magicinfinite: Generating infinite talking videos with your words and voice},
  author={Yi, Hongwei and Ye, Tian and Shao, Shitong and Yang, Xuancheng and Zhao, Jiantong and Guo, Hanzhong and Wang, Terrance and Yin, Qingyu and Xie, Zeke and Zhu, Lei and others},
  journal={arXiv preprint arXiv:2503.05978},
  year={2025}
}

@inproceedings{
salimans2022progressive,
title={Progressive Distillation for Fast Sampling of Diffusion Models},
author={Tim Salimans and Jonathan Ho},
booktitle={International Conference on Learning Representations},
year={2022},
url={https://openreview.net/forum?id=TIdIXIpzhoI}
}

@article{
zheng2024memo,
title={{MEMO}: Memory-Guided Diffusion for Expressive Talking Video Generation},
author={Longtao Zheng and Yifan Zhang and Hanzhong Guo and Jiachun Pan and Zhenxiong Tan and Jiahao Lu and Chuanxin Tang and Bo An and Shuicheng YAN},
journal={Transactions on Machine Learning Research},
issn={2835-8856},
year={2026},
url={https://openreview.net/forum?id=uBcHcM7Kzi},
note={J2C Certification}
}

@article{wu2025hunyuanvideo,
  title={Hunyuanvideo 1.5 technical report},
  author={Wu, Bing and Zou, Chang and Li, Changlin and Huang, Duojun and Yang, Fang and Tan, Hao and Peng, Jack and Wu, Jianbing and Xiong, Jiangfeng and Jiang, Jie and others},
  journal={arXiv preprint arXiv:2511.18870},
  year={2025}
}

@inproceedings{
lin2025seaweedapt,
title={Diffusion Adversarial Post-Training for One-Step Video Generation},
author={Shanchuan Lin and Xin Xia and Yuxi Ren and Ceyuan Yang and Xuefeng Xiao and Lu Jiang},
booktitle={Forty-second International Conference on Machine Learning},
year={2025},
url={https://openreview.net/forum?id=AAgzsnhc28}
}

@article{zhang2025turbodiffusion,
  title={TurboDiffusion: Accelerating Video Diffusion Models by 100-200 Times},
  author={Zhang, Jintao and Zheng, Kaiwen and Jiang, Kai and Wang, Haoxu and Stoica, Ion and Gonzalez, Joseph E and Chen, Jianfei and Zhu, Jun},
  journal={arXiv preprint arXiv:2512.16093},
  year={2025}
}

@misc{fastvideo2025,
  title={FastVideo},
  author={FastVideo Team},
  year={2025},
  note={\url{https://huggingface.co/FastVideo}}
}

@inproceedings{
wu2023unleashing,
title={Unleashing Large-Scale Video Generative Pre-training for Visual Robot Manipulation},
author={Hongtao Wu and Ya Jing and Chilam Cheang and Guangzeng Chen and Jiafeng Xu and Xinghang Li and Minghuan Liu and Hang Li and Tao Kong},
booktitle={The Twelfth International Conference on Learning Representations},
year={2024},
url={https://openreview.net/forum?id=NxoFmGgWC9}
}

@inproceedings{guoedit,
  author    = {Hanzhong Guo and Jie Wu and Jie Liu and Yu Gao and Zilyu Ye and Linxiao Yuan and Xionghui Wang and Yizhou Yu and Weilin Huang},
  title     = {Leveraging Verifier-Based Reinforcement Learning in Image Editing},
  booktitle = {Proceedings of the IEEE/CVF Conference on Computer Vision and Pattern Recognition (CVPR)},
  year      = {2026},
  address   = {Denver, Colorado},
  month     = {June}
}

@InProceedings{xie2024progressive,
    author    = {Xie, Desai and Xu, Zhan and Hong, Yicong and Tan, Hao and Liu, Difan and Liu, Feng and Kaufman, Arie and Zhou, Yang},
    title     = {Progressive Autoregressive Video Diffusion Models},
    booktitle = {Proceedings of the IEEE/CVF Conference on Computer Vision and Pattern Recognition (CVPR) Workshops},
    month     = {June},
    year      = {2025},
    pages     = {6376-6386}
}

@inproceedings{
wu2024ivideogpt,
title={iVideo{GPT}: Interactive Video{GPT}s are Scalable World Models},
author={Jialong Wu and Shaofeng Yin and Ningya Feng and Xu He and Dong Li and Jianye HAO and Mingsheng Long},
booktitle={The Thirty-eighth Annual Conference on Neural Information Processing Systems},
year={2024},
url={https://openreview.net/forum?id=4TENzBftZR}
}

@article{fang2024xdit,
  title={xDiT: an Inference Engine for Diffusion Transformers (DiTs) with Massive Parallelism},
  author={Fang, Jiarui and Pan, Jinzhe and Sun, Xibo and Li, Aoyu and Wang, Jiannan},
  journal={arXiv preprint arXiv:2411.01738},
  year={2024}
}

@article{wang2024loong,
  title={Loong: Generating minute-level long videos with autoregressive language models},
  author={Wang, Yuqing and Xiong, Tianwei and Zhou, Daquan and Lin, Zhijie and Zhao, Yang and Kang, Bingyi and Feng, Jiashi and Liu, Xihui},
  journal={arXiv preprint arXiv:2410.02757},
  year={2024}
}

@inproceedings{song2023consistency,
  title={Consistency models},
  author={Song, Yang and Dhariwal, Prafulla and Chen, Mark and Sutskever, Ilya},
  booktitle={Proceedings of the 40th International Conference on Machine Learning},
  pages={32211--32252},
  year={2023}
}

@article{luo2023latent,
  title={Latent consistency models: Synthesizing high-resolution images with few-step inference},
  author={Luo, Simian and Tan, Yiqin and Huang, Longbo and Li, Jian and Zhao, Hang},
  journal={arXiv preprint arXiv:2310.04378},
  year={2023}
}

@inproceedings{
frans2024one,
title={One Step Diffusion via Shortcut Models},
author={Kevin Frans and Danijar Hafner and Sergey Levine and Pieter Abbeel},
booktitle={The Thirteenth International Conference on Learning Representations},
year={2025},
url={https://openreview.net/forum?id=OlzB6LnXcS}
}

@article{liang2026gpd,
  title={GPD: Guided Progressive Distillation for Fast and High-Quality Video Generation},
  author={Liang, Xiao and Zhang, Yunzhu and Zhu, Linchao},
  journal={arXiv preprint arXiv:2602.01814},
  year={2026}
}

@article{maes2026leworldmodel,
  title={LeWorldModel: Stable End-to-End Joint-Embedding Predictive Architecture from Pixels},
  author={Maes, Lucas and Lidec, Quentin Le and Scieur, Damien and LeCun, Yann and Balestriero, Randall},
  journal={arXiv preprint arXiv:2603.19312},
  year={2026}
}

@inproceedings{ramesh2021zero,
  title={Zero-shot text-to-image generation},
  author={Ramesh, Aditya and Pavlov, Mikhail and Goh, Gabriel and Gray, Scott and Voss, Chelsea and Radford, Alec and Chen, Mark and Sutskever, Ilya},
  booktitle={International conference on machine learning},
  pages={8821--8831},
  year={2021},
  organization={Pmlr}
}

@article{betker2023improving,
  title={Improving image generation with better captions},
  author={Betker, James and Goh, Gabriel and Jing, Li and Brooks, Tim and Wang, Jianfeng and Li, Linjie and Ouyang, Long and Zhuang, Juntang and Lee, Joyce and Guo, Yufei and others},
  journal={Computer Science. https://cdn. openai. com/papers/dall-e-3. pdf},
  volume={2},
  number={3},
  pages={8},
  year={2023}
}

@inproceedings{esser2024scaling,
  title={Scaling rectified flow transformers for high-resolution image synthesis},
  author={Esser, Patrick and Kulal, Sumith and Blattmann, Andreas and Entezari, Rahim and M{\"u}ller, Jonas and Saini, Harry and Levi, Yam and Lorenz, Dominik and Sauer, Axel and Boesel, Frederic and others},
  booktitle={Forty-first International Conference on Machine Learning},
  year={2024}
}

@inproceedings{podellsdxl,
  title={SDXL: Improving Latent Diffusion Models for High-Resolution Image Synthesis},
  author={Podell, Dustin and English, Zion and Lacey, Kyle and Blattmann, Andreas and Dockhorn, Tim and M{\"u}ller, Jonas and Penna, Joe and Rombach, Robin},
  booktitle={The Twelfth International Conference on Learning Representations},
  year={2024}
}

@article{saharia2022photorealistic,
  title={Photorealistic text-to-image diffusion models with deep language understanding},
  author={Saharia, Chitwan and Chan, William and Saxena, Saurabh and Li, Lala and Whang, Jay and Denton, Emily L and Ghasemipour, Kamyar and Gontijo Lopes, Raphael and Karagol Ayan, Burcu and Salimans, Tim and others},
  journal={Advances in neural information processing systems},
  volume={35},
  pages={36479--36494},
  year={2022}
}

@article{lv2025fastercache,
  title={FasterCache: Training-Free Video Diffusion Model Acceleration with High Quality},
  author={Lv, Zhengyao and Si, Chenyang and Song, Junhao and Yang, Zhenyu and Qiao, Yu and Liu, Ziwei and Wong, Kwan-Yee K.},
  journal={arXiv preprint arXiv:2410.19355},
  year={2025}
}

@article{wang2026precisecache,
  title={PreciseCache: Precise Feature Caching for Efficient and High-fidelity Video Generation},
  author={Wang, Jiangshan and Zhao, Kang and Guo, Jiayi and Wang, Jiayu and Guo, Hang and Zhu, Chenyang and Li, Xiu and Yue, Xiangyu},
  journal={arXiv preprint arXiv:2603.00976},
  year={2026}
}

@inproceedings{
zhao2025pab,
title={Real-Time Video Generation with Pyramid Attention Broadcast},
author={Xuanlei Zhao and Xiaolong Jin and Kai Wang and Yang You},
booktitle={The Thirteenth International Conference on Learning Representations},
year={2025},
url={https://openreview.net/forum?id=hDBrQ4DApF}
}

@misc{yuan2026fastwam,
  title={Fast-WAM: Do World Action Models Need Test-time Future Imagination?},
  author={Tianyuan Yuan and Zibin Dong and Yicheng Liu and Hang Zhao},
  year={2026},
  note={arXiv preprint arXiv:2603.16666}
}

@article{feng2026worldcache,
  title={WorldCache: Accelerating World Models for Free via Heterogeneous Token Caching},
  author={Feng, Weilun and Fan, Guoxin and Qin, Haotong and Yang, Chuanguang and Wu, Mingqiang and Li, Yuqi and Li, Xiangqi and An, Zhulin and Huang, Libo and Wang, Dingrui and Liao, Longlong and Magno, Michele and Xu, Yongjun},
  journal={arXiv preprint arXiv:2603.06331},
  year={2026}
}

@article{lecun2022path,
  title={A path towards autonomous machine intelligence version 0.9. 2, 2022-06-27},
  author={LeCun, Yann},
  journal={Open Review},
  volume={62},
  number={1},
  pages={1--62},
  year={2022}
}

@InProceedings{Huang_2024_VBench,
    author    = {Huang, Ziqi and He, Yinan and Yu, Jiashuo and Zhang, Fan and Si, Chenyang and Jiang, Yuming and Zhang, Yuanhan and Wu, Tianxing and Jin, Qingyang and Chanpaisit, Nattapol and Wang, Yaohui and Chen, Xinyuan and Wang, Limin and Lin, Dahua and Qiao, Yu and Liu, Ziwei},
    title     = {VBench: Comprehensive Benchmark Suite for Video Generative Models},
    booktitle = {Proceedings of the IEEE/CVF Conference on Computer Vision and Pattern Recognition (CVPR)},
    month     = {June},
    year      = {2024},
    pages     = {21807-21818}
}

@article{ye2026dreamzero,
  title={World action models are zero-shot policies},
  author={Ye, Seonghyeon and Ge, Yunhao and Zheng, Kaiyuan and Gao, Shenyuan and Yu, Sihyun and Kurian, George and Indupuru, Suneel and Tan, You Liang and Zhu, Chuning and Xiang, Jiannan and others},
  journal={arXiv preprint arXiv:2602.15922},
  year={2026}
}

@inproceedings{
yi2025deep,
title={Deep Forcing: Training-Free Long Video Generation with Deep Sink and Participative Compression},
author={Jung Yi and Wooseok Jang and Paul Hyunbin Cho and Jisu Nam and Heeji Yoon and Seungryong Kim},
booktitle={Forty-third International Conference on Machine Learning},
year={2026},
url={https://openreview.net/forum?id=gtmyFnvXAW}
}

\vfill

\end{document}